\documentclass[superscriptaddress,nofootinbib,preprint,11pt,eqsecnum]{revtex4-1}

\pdfoutput=1

\usepackage[left=0.6in, right=0.6in, top=0.9in, bottom=0.9in]{geometry}
\usepackage{slashed}
\usepackage{comment}
\usepackage[overload]{empheq}
\usepackage[utf8]{inputenc}
\usepackage[english]{babel}
\usepackage{amsmath,amssymb,amsfonts,bm}
\usepackage{graphicx}
\usepackage{xcolor}
\usepackage{booktabs}
\usepackage{empheq}
\usepackage{braket}
\usepackage{float}
\usepackage[colorlinks=true,linkcolor=red,urlcolor=blue,citecolor=blue]{hyperref}
\usepackage{tikz}
\usetikzlibrary{arrows.meta,positioning,decorations.markings,backgrounds,calc,decorations.pathreplacing}
\allowdisplaybreaks

\newcommand{\mi}{\mathrm{i}}

\newcommand{\dd}{\mathrm{d}}
\newcommand{\dbar}{\mathchar'26\mkern-12mu \mathrm{d}}
\newcommand{\deltabar}{\mathchar'26\mkern-9mu \delta}

\newcommand{\Seff}{S_{\rm eff}}
\newcommand{\SIF}{S_{\rm IF}}
\newcommand{\Os}{O_s}

\usetikzlibrary{decorations.markings}
\newcommand{\midarrow}[3]{%
  \begin{scope}[shift={(#1,#2)}, rotate=#3]
    \fill (-0.10,-0.065) -- (0.12,0) -- (-0.10,0.065) -- cycle;
  \end{scope}%
}

\newcommand{\rarrowhead}[2]{%
  \fill (#1-0.10,#2-0.055) -- (#1+0.12,#2) -- (#1-0.10,#2+0.055) -- cycle;
}
\newcommand{\larrowhead}[2]{%
  \fill (#1+0.10,#2-0.055) -- (#1-0.12,#2) -- (#1+0.10,#2+0.055) -- cycle;
}

\begin{document}
\preprint{CALT-TH/2026-018}

\title{An open EFT for dark matter searches with matter-wave interferometers}

\author{Leonardo Badurina}
\email{badurina@caltech.edu}
\affiliation{Walter Burke Institute for Theoretical Physics and Leinweber Forum for Theoretical Physics, California Institute of Technology, Pasadena, CA 91125, USA}

\author{Kathryn M. Zurek}
\email{kzurek@caltech.edu}
\affiliation{Walter Burke Institute for Theoretical Physics and Leinweber Forum for Theoretical Physics, California Institute of Technology, Pasadena, CA 91125, USA}
\date{\today}

\begin{abstract}
Matter-wave interferometers (MWIs) are a genuine \textit{quantum} probe of dark matter (DM), being sensitive to DM-induced phase shifts and loss of contrast. 
Treating the interferometer as an open quantum system and the dark sector as its environment, we derive an open EFT to study these signatures using the Schwinger--Keldysh formalism.  
Starting from a microscopic theory, we derive the interferometer's effective action, which we then project onto the many-body worldline sectors relevant for atom interferometers and MWIs employing mesoscopic objects, i.e., composite-particle interferometers (CPIs). 
This construction clarifies the heavy-probe limit and organizes phase shifts and decoherence as distinct real-time components of the influence action: the retarded kernel controls the causal response of the environment, while the Keldysh kernel controls environmental fluctuations and hence loss of contrast. 
At leading order, these kernels reduce to: (i) DM bubble diagrams in the case of four-point interactions, and (ii) dressed in-medium mediator propagators in the case of long-range interactions, where the dressing is provided by the DM environment. 
Within this framework, we study the signatures of elastic spin-independent sub-GeV DM scattering off nucleons and find that: (i) owing to Bose enhancements below the DM wave-particle boundary, ambitious CPIs could probe Higgs-portal DM with mass $m_\chi \lesssim 1$~eV in an unexplored region of model space; (ii) for DM coupled to nucleons via a light hadrophilic scalar, the DM background can drive the mediator’s in-medium mass tachyonic over much of parameter space; outside these regions, CPIs could probe provide powerful probes of light DM.
\end{abstract}

\maketitle

\begingroup
\hypersetup{linkcolor=black}
\tableofcontents
\endgroup

\newpage

\section{Introduction}\label{sec:intro}

Matter-wave interferometers (MWIs) are sensitive to dark matter (DM) through observables that are intrinsically quantum. 
Indeed, a DM background need neither produce a resolvable recoil displacement, nor deposit energy, in order to leave a detectable signature.
Rather, DM can modify the reduced density matrix of the probe, by introducing a relative phase between spatially separated wavepackets and suppressing quantum coherence.

Thanks to their exceptional metrological precision, MWIs are poised to become versatile probes of DM across a wide range of masses. 
For instance, atom gradiometers, where pairs of atom interferometers (AIs) are driven by common laser pulses, have been proposed to probe linearly coupled and quadratically coupled ultralight scalar DM~\cite{Arvanitaki:2016fyj,Badurina:2021lwr,Badurina:2023wpk,Banks:2024sli}, as well as spin-2 DM~\cite{Blas:2024kps}, in the $10^{-18}$--$10^{-12}$~eV mass range. 
Space-based versions of these experiments could probe the quadratic axion-photon coupling in uncharted regions of parameter space~\cite{Beadle:2023flm}, while atom interferometers in which atoms are prepared in a superposition of spin states could probe axion-like particle DM via ALP-nucleon
interactions~\cite{Graham:2017ivz}. 
Co-located interferometers employing different isotopes also offer sensitivity to $B-L$ vector DM in the same mass range~\cite{Graham:2015ifn,MAGIS-100:2021etm,
Zhou:2024qdw}. 
Composite-particle interferometers (CPIs), which are MWIs interfering superpositions of mesoscopic objects, could detect collisional decoherence from elastic, spin-independent scattering with sub-GeV DM~\cite{Riedel:2012ur,Riedel:2016acj,Du:2022ceh,Badurina:2024nge,Murgui:2026mlr}. 
Futuristic space-based gradiometers proposed for mid-frequency gravitational wave detection may also probe a DM subcomponent populated by small sub-asteroid mass DM clumps and constrain small ultralight DM overdensities via purely gravitational interactions~\cite{Badurina:2025xwl}. Trapped-ion interferometers originally conceived for rotation sensing could detect small magnetic fields sourced by axion-like particles or kinetically mixed dark photons in otherwise unexplored regions of ultralight dark matter parameter space~\cite{Badurina:2025idj}.

As a result of these disparate phenomenological regimes, the theoretical description of MWIs interacting with a DM background remains fragmented. 
A unified treatment should interpolate between particle-like and wave-like DM, while retaining dependence on the quantum state of the DM, including its occupation number, spin statistics, and coherence time. 
It should also distinguish between two regimes, depending on whether the DM coherence time is shorter or longer than the interferometric time.
In the former case, a Markovian (or Lindblad-type) description is adequate, while in the latter a non-Markovian treatment would be appropriate, since interactions between the MWI and the DM environment are non-local in time~\cite{Breuer:2007juk}. 
Furthermore, the commonly used heavy-probe approximation, which neglects DM-induced displacement recoils, must be revisited once momentum transfer produces a resolvable recoil phase or becomes comparable to the wavepacket width. 
Finally, such a framework should account for the many-body structure of a realistic probe. As shown in the context of elastic spin-independent DM-nucleon scattering~\cite{Badurina:2024nge,Murgui:2025unt}, different many-atom states can give rise to different form factors that parametrically change the expected reach of MWIs to sub-GeV DM. 

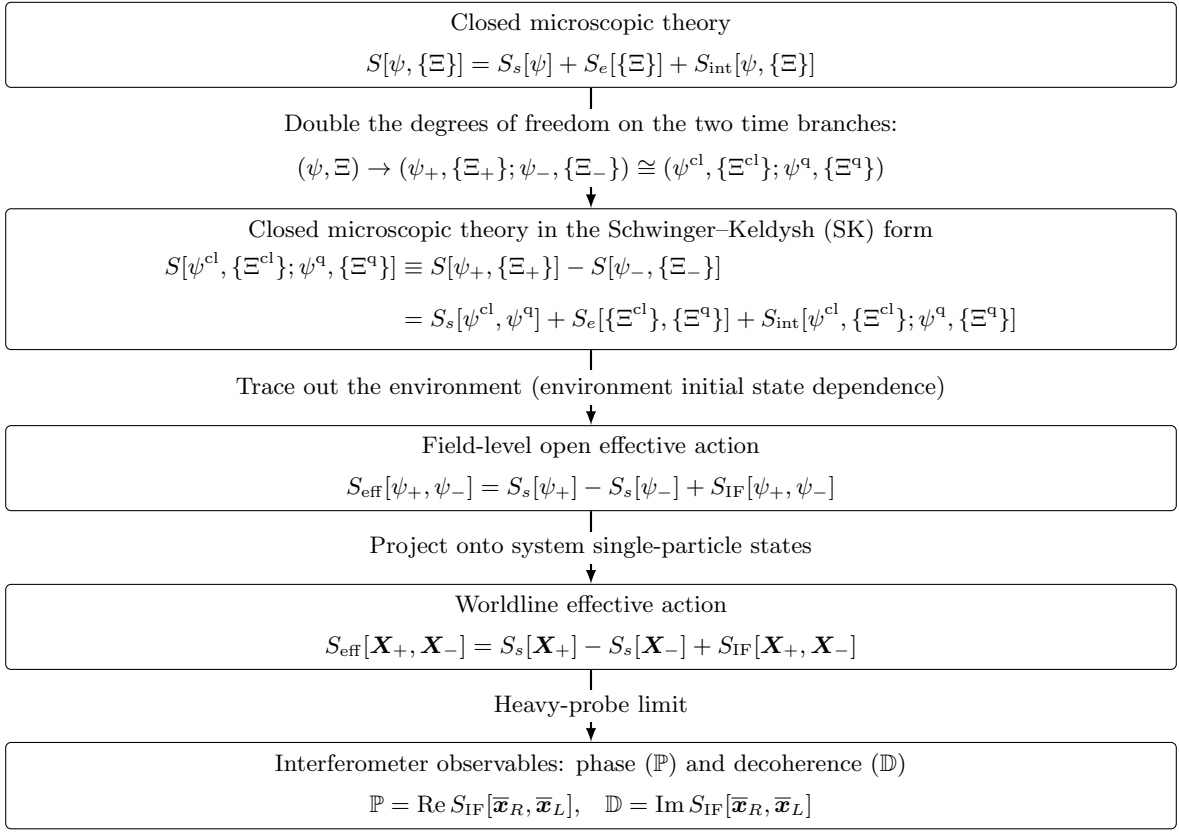
\begin{figure}[t]
\centering
\begin{tikzpicture}[
  box/.style={
    draw,
    rounded corners=2pt,
    align=flush center,
    inner sep=4pt,
    text width=0.82\textwidth,
    minimum height=0.78cm,
    font=\footnotesize
  },
  flowlabel/.style={
    align=flush center,
    font=\footnotesize,
    fill=white,
    inner sep=2pt,
    text width=0.76\textwidth
  },
  arrow/.style={->, >=latex, thick}
]
\node[box] (closed) at (0,0) {
  Closed microscopic theory\\[-1pt]
  $S[\psi,\{\Xi\}]=S_s[\psi]+S_e[\{\Xi\}]+S_{\rm int}[\psi,\{\Xi\}]$
};

\node[box] (inin) at (0,-3.1) {
  Closed microscopic theory in the Schwinger--Keldysh (SK) form\\[-1pt]
  $\begin{aligned}
  S[\psi^\mathrm{cl},\{\Xi^\mathrm{cl}\};\psi^\mathrm{q},\{\Xi^\mathrm{q}\}]
  &\equiv S[\psi_+,\{\Xi_+\}] - S[\psi_-,\{\Xi_-\}]\\
  &= S_s[\psi^\mathrm{cl},\psi^\mathrm{q}]
  +S_e[\{\Xi^\mathrm{cl}\},\{\Xi^\mathrm{q}\}]
  +S_{\rm int}[\psi^\mathrm{cl},\{\Xi^\mathrm{cl}\};
  \psi^\mathrm{q},\{\Xi^\mathrm{q}\}]
  \end{aligned}$
};

\node[box] (open) at (0,-5.6) {
  Field-level open effective action\\[-1pt]
  $\Seff[\psi_+,\psi_-]
  =S_s[\psi_+]-S_s[\psi_-]+\SIF[\psi_+,\psi_-]$
};

\node[box] (wl) at (0,-7.7) {
  Worldline effective action\\[-1pt]
  $\Seff[\boldsymbol{X}_+,\boldsymbol{X}_-]
  =S_s[\boldsymbol{X}_+]-S_s[\boldsymbol{X}_-]
  +\SIF[\boldsymbol{X}_+,\boldsymbol{X}_-]$
};

\node[box] (obs) at (0,-9.8) {
  Interferometer observables: phase ($\mathbb{P}$) and decoherence ($\mathbb{D}$)\\[-1pt]
  $\mathbb{P}=\operatorname{Re}\SIF[\overline{\boldsymbol{x}}_{R},\overline{\boldsymbol{x}}_{L}]$,\quad
  $\mathbb{D}=\operatorname{Im}\SIF[\overline{\boldsymbol{x}}_{R},\overline{\boldsymbol{x}}_{L}]$
};

\draw[arrow] (closed) -- node[flowlabel] {
  Double the degrees of freedom on the two time branches:
  $(\psi,\Xi)\to(\psi_+,\{\Xi_+\};\psi_-,\{\Xi_-\})
  \cong
  (\psi^\mathrm{cl},\{\Xi^\mathrm{cl}\};\psi^\mathrm{q},\{\Xi^\mathrm{q}\})$
} (inin);

\draw[arrow] (inin) -- node[flowlabel] {
  Trace out the environment (environment initial state dependence)
} (open);

\draw[arrow] (open) -- node[flowlabel] {
  Project onto system single-particle states
} (wl);

\draw[arrow] (wl) -- node[flowlabel] {
  Heavy-probe limit
} (obs);

\end{tikzpicture}
\caption{Summary of the logical structure of the open EFT construction presented in this work. We assume that the system and environment are described by fields $\psi$ and $\{\Xi\}$, respectively. 
The system and environment free actions are $S_s$ and $S_e$, respectively, and $S_\mathrm{int}$ depends on interactions. 
In the closed-time path, we double the number of fields: $\psi_+$ and $\{\Xi_+\}$ are defined on the forward time branch, and $\psi_-$ and $\{\Xi_-\}$ are defined on the backward time branch. 
These can also be equivalently defined in the Keldysh basis, which is a linear combination of $+$ and $-$ fields. 
We also compute the action for the full (or closed) theory by computing the difference between the total action along the forward and backward time branches.
By tracing out the environment, we obtain an open effective action that describes the dynamics of the system degrees of freedom.
The initial state of the environment and system-environment interactions are captured by the influence action $S_\mathrm{IF}$, which can have a non-zero pure imaginary contribution. 
By projecting onto single-particle states, we obtain an open effective action defined along system worldlines $\boldsymbol{X}_\pm$.
In the heavy-probe limit, the phase shift and decoherence can be directly extracted from the real and imaginary parts of the influence action evaluated along the unperturbed left and right arm worldlines $\overline{\boldsymbol{x}}_{L,R}$.}
\label{fig:logic}
\end{figure}

In this work, we construct such a framework.  
We model the MWI as an open quantum system coupled to the environment, which we take to be the dark sector, and derive its dynamics from first principles using the Schwinger--Keldysh, or in-in, formalism~\cite{Schwinger:1960qe,Keldysh:1964ud} (for a modern review, see Ref.~\cite{Haehl2025}). 
This formalism preserves causality and unitarity at the level of the full theory and implements real-time evolution along a closed-time path (CTP). 
The evolution of kets and bras is captured by the forward ($``+"$) and backward ($``-"$) branches, respectively, with fields defined on both time branches.
Tracing out the environment yields an \textit{open} EFT for MWIs, as summarized in Fig.~\ref{fig:logic}. 
The effective action for a MWI then depends on the influence action $\SIF$, which is fixed by the system-environment couplings and by real-time correlators, or kernels, of the environment fields.
Crucially, the influence action's real and imaginary parts determine the DM-induced phase shift and decoherence, respectively.

The Schwinger--Keldysh approach has been used in closely related interferometric settings, including gravitationally induced decoherence in interferometers~\cite{Blencowe:2012mp,Parikh:2020fhy,Kanno:2020usf}, observable effects due to soft graviton~\cite{Zurek:2020ukz,Wilson-Gerow:2024ljx} and soft photon~\cite{DeLisle:2022pjo} bremsstrahlung, decoherence and phase shifts induced by chameleon fields in atom interferometers~\cite{Burrage:2018pyg,Kading:2023mdk,Kading:2024jqe}, and more generally to study the decay of the interference term between two Gaussian packets caused by a weakly coupled nonlinear bath of oscillators~\cite{Hu:1991di,Hu:1993vs}.
For a review on the use of this formalism in the context of open quantum systems more generally, see Ref.~\cite{Sieberer_2016}. 
More broadly, open EFT methods have recently found applications in cosmology and astrophysics~\cite{Chen:2017ryl,Salcedo:2024smn,Burgess:2025dwm,Colas:2025ind,Salcedo:2026cqb,Li:2025azq,Cespedes:2026fdp,Li:2026lwl,Modrekiladze:2024htc}, gravity~\cite{Salcedo:2025ezu,Modrekiladze:2026twz} and gauge theories~\cite{Salcedo:2024nex,Kaplanek:2025moq,Kaplanek:2026kpp}. EFTs have been widely employed to study the signatures of DM in other direct detection experiments, e.g., Refs.~\cite{Fan:2010gt,Fitzpatrick:2012ix,Trickle:2020oki,Mitridate:2023izi,Krnjaic:2024bdd,Berghaus:2026kmj}. 

Our framework cleanly separates the phenomenology of contact and long-range DM-MWI interactions. 
For contact DM-nucleon interactions, the environment kernel is the DM matter response, which depends on the finite density of the DM background. 
Because the interferometer observables depend on these kernels, our formalism immediately identifies MWI observables that inherit Bose enhancement or Pauli blocking, and those that remain at most linear in the DM occupation number.
For long-range interactions, our construction also shows how the finite DM background density dresses the mediator propagator, which in turn gives rise to screening or anti-screening effects.
For attractive interactions, this correction can give rise to a tachyonic mediator mass, which signals the onset of DM condensation and the breakdown of the plane wave approximation underpinning our calculations. 
These in-medium effects are closely analogous to those discussed in the context of BSM physics in dense astrophysical media~\cite{Hardy:2016kme,Lasenby:2020rlf,Hook:2017psm,Balkin:2020dsr,DeRocco:2022rze} and nuclear physics at finite density~\cite{KAPLAN1988273}.

A concise account of this construction in the case of contact interactions appeared in Ref.~\cite{Badurina:2026owr}. 
Here, we give the detailed EFT derivation of those results and extend the analysis in several directions. 
Firstly, we present a unified many-body treatment spanning single atoms, dilute clouds, and bound mesoscopic objects, making clear where coherent enhancements enter. 
Secondly, we discuss corrections beyond the heavy-probe limit. 
Thirdly, we consider the phenomenology of contact interactions at and below the DM wave-particle boundary, and show that ambitious CPIs may be able to probe unexplored regions of Higgs-portal DM parameter space. 
Fourthly, we discuss the case of long-range interactions, where we show how the DM environment dresses the mediator propagator appearing in the observables. 
This is especially relevant in light of the anticipated sensitivity of AIs to sub-keV DM through elastic DM-nucleon scattering via a light hadrophilic mediator~\cite{Du:2022ceh,Murgui:2026mlr}. 

The paper is organized as follows. 
In section~\ref{sec:openEFT} we derive the effective action starting from a microscopic theory. 
We then connect the influence action to interferometer observables by projecting the system fields to worldlines. 
For pedagogical reasons, we first focus in section~\ref{sec:observables} on the single-atom case (already presented in Ref.~\cite{Badurina:2026owr}) and in section~\ref{sec:oEFT_multi} we generalize our results to multi-atom platforms, such as AIs and CPIs.
In section~\ref{sec:DM-kernels} we show how the tools developed in previous sections can be used to compute the signatures of sub-GeV elastic spin-independent DM-nucleon scattering, distinguishing between contact and long-range interactions.
We summarize our results in section~\ref{sec:discussion}. 
Appendices~\ref{app:finite-mass}--\ref{app:DM_mod} provide additional details.

\noindent \textbf{Conventions and notation} \\
We use natural units and the Minkowski metric $\eta_{\mu\nu}=\mathrm{diag}(+1,-1,-1,-1)$. 
The inner product of a four-vector $x^\mu = (t, \mathbf{x})$ is defined as $x_\mu p^\mu = x\cdot p$. 
We use the notation $\int \dbar^n p = \int \mathrm{d}^n p/(2\pi)^n$, $\deltabar^{(n)} (\cdot) = (2\pi)^n\delta^{(n)}(\cdot)$, and $\int_{t_i}^{t_f} d^4x \equiv  \int_{t_i}^{t_f} \mathrm{d}t \int_{-\infty}^{\infty} d^3 x $.
The Fourier transform (FT) of a function $h(x)$ and its inverse FT are, respectively, defined as
\begin{equation}\label{eq:FT}
\begin{aligned}
\widetilde{h}(p) = \int \mathrm{d}^4 x \, h(x) e^{\mi p\cdot x} \, \quad , \quad h(x) = \int \dbar^4 p \, \widetilde{h}(p) e^{-\mi p\cdot x} \, .
\end{aligned}
\end{equation}
We will also use the 3-momentum FT and its inverse defined as  
\begin{equation}\label{eq:FT_3}
\begin{aligned}
\widetilde{h}(\mathbf{p},t) = \int \mathrm{d}^3 x \, h(x) e^{-\mi \mathbf{p}\cdot \mathbf{x}} \, \quad , \quad h(x) = \int \dbar^3 p \, \widetilde{h}(\mathbf{p},t) e^{\mi \mathbf{p}\cdot \mathbf{x}} \, ,
\end{aligned}
\end{equation}
respectively. 

Unless otherwise stated, we will assume throughout the Standard Halo Model (SHM) in the lab frame with a finite escape velocity, which is described by the velocity distribution
\begin{equation}\label{eq:SHM}
\begin{aligned}
f_\chi(\mathbf{v}) &= \frac{1}{N_0} e^{-|\mathbf{v}+\mathbf{v}_\oplus |^2/v_0^2} \, \Theta(v_\mathrm{esc}-|\mathbf{v}+\mathbf{v}_\oplus |) \, , \\
N_0 & =  \pi^{3/2}v_0^2 \left [ v_0 \mathrm{erf}(v_\mathrm{esc}/v_0) - \frac{2v_\mathrm{esc}}{\sqrt{\pi}} \exp \left ( - v_\mathrm{esc}^2/v_0^2\right )\right ] \, ,
\end{aligned}
\end{equation} 
with $v_0 \approx 230$~km/s the DM velocity dispersion, $v_\mathrm{esc} \approx 600$~km/s the escape velocity, and $|\mathbf{v}_\oplus| \approx 240$~km/s the Earth's velocity in the galactic rest frame.
We define the spin-summed DM occupation number per mode in terms of the DM energy density $\rho_\chi$ and mass $m_\chi$ as 
\begin{equation}
n_\chi(\mathbf{p}) \equiv \frac{\rho_\chi}{m_\chi} f_\chi(\mathbf{p}) \, ,
\end{equation}
where $f_\chi(\mathbf{p}) = (2\pi)^3 f_\chi(\mathbf{v})/m_\chi^3$ is the DM 3-momentum distribution with normalization $\int \dbar^3 p f_\chi(\mathbf{p}) = 1$.

\section{The open EFT from the closed theory}\label{sec:openEFT}

Laboratory probes used to search for dark matter are inherently \emph{open} systems: they interact continuously, albeit weakly, with an external environment constituted by dark matter particles. 
Even when the total system, i.e., probe plus environment, evolves unitarily, the dynamics of the probe alone is generically non-unitary once the environmental degrees of freedom are traced out. 
This loss of unitarity manifests physically as decoherence, dissipation, and stochastic fluctuations, and can be naturally captured by the Schwinger--Keldysh (or closed time-path) formalism~\cite{Keldysh:1964ud,Schwinger:1960qe}, which implements real-time evolution while preserving causality and unitarity at the level of the full system. 
In this section we introduce the key ingredients of this formalism to the unfamiliar reader, and set the conventions and notation used throughout.

\subsection{Open quantum systems and the Schwinger--Keldysh formalism}\label{sec:OQS_SK}

Consider a quantum system (the laboratory probe, labeled by ``s'') coupled to an environment (the dark matter background, labeled by ``e''). 
Let $\psi$ and $\{\Xi\}$ denote the system and (set of) environmental degrees of freedom, respectively, which we take to be quantum fields. 
For DM-nucleon interactions, the system degrees of freedom include nucleons, while the environment degrees of freedom include DM fields $\chi$ and, for long-range interactions, a mediator, which we assume to be the scalar $\phi$. 
The Hamiltonian for the system plus environment is then
\begin{equation}
\hat{H}[\psi,\{\Xi\}]=\hat{H}_s[\psi]+\hat{H}_e[\{\Xi\}]+\hat{H}_{\rm int}[\psi,\{\Xi\}]  \, ,
\label{eq:hamiltonian}
\end{equation}
with $\hat{H}_s[\psi]$ and $\hat{H}_e[\{\Xi\}]$ the free Hamiltonians for the system and environment, respectively, and $\hat{H}_{\rm int}[\psi,\{\Xi\}]$ the interaction Hamiltonian.  
Since the environment cannot be directly measured, the object of interest is the reduced density matrix of the system at time $t_f$, which can be computed by first evolving the system and environment from initial time $t_i$ to $t_f$ using the unitary operator $\hat{U}(t_f,t_i)=\mathrm{T} \exp \left [-\mi \int_{t_i}^{t_f} \mathrm{d}t \, \hat{H}(t) \right ]$, and subsequently by tracing over the environment degrees of freedom, which we represent by $\mathrm{Tr}_e[\cdot]$. 
Assuming a factorized initial state $\hat{\rho}_i=\hat{\rho}_{s}(t_i)\otimes\hat{\rho}_e$ at time $t_i$ and a stationary environment state $\hat{\rho}_e$, which follows from the Born approximation~\cite{Weiss:2021uhm}, the reduced density matrix of the system at time $t_f$ reads as
\begin{equation}\label{eq:reduced_rho}
\hat{\rho}_s(t_f)=\mathrm{Tr}_e\!\left[\hat{U}(t_f,t_{i})\,\hat{\rho}_{i}\,\hat{U}^\dagger(t_f,t_{i})\right] = \mathrm{Tr}_e\!\left[\hat{U}(t_f,t_{i})\,\hat{\rho}_{i}\,\hat{U}(t_{i},t_f)\right] \,  ,
\end{equation} 

In the Schwinger--Keldysh (SK) formulation, also known as the \textit{in-in} formalism, the RHS of Eq.~\eqref{eq:reduced_rho} admits a path-integral representation along a closed time contour extending from time $t_i$ to time $t_f$ and back. The forward ($``+"$) and backward ($``-"$) branches of the contour correspond to the time evolution of kets and bras, respectively, and together encode the unitary evolution of the full system. 
The degrees of freedom are doubled: $(\psi, \{\Xi\}) \rightarrow (\psi_+,\psi_-; \{\Xi_+\},\{\Xi_-\})$, where the $``+"$ and $``-"$ fields live on the forward and backward branches of the closed-time path. 
As a result, the total classical action takes the form
\begin{equation}
\begin{aligned}
S[\psi_+,\psi_-; \{\Xi_+\},\{\Xi_-\}] & = S[\psi_+,\{\Xi_+\}] - S[\psi_-,\{\Xi_-\}]  \, , \\
S[\psi_\pm,\{\Xi_\pm\}] &=  S_s[\psi_\pm]+S_e[\{\Xi_\pm\}] + S_{\rm int}[\psi_\pm,\{\Xi_\pm\}]  \, .
\label{eq:closed-action}
\end{aligned}
\end{equation}
The doubling of degrees of freedom is a direct consequence of enforcing unitarity and causality in real-time evolution, and provides a powerful tool for distinguishing between genuine non-unitary and unitary evolution. 
For example, in the absence of system-environment interactions, or if the evolution of the system can be expressed in terms of a unitary operator, the evolution along the forward and backward time branches is identical up to complex conjugation. 
Non-unitary effects emerge whenever the trace over the environment gives rise to terms that mix system degrees of freedom evaluated along the forward and backward time branches. 
After the environment is traced out, the system evolves with the \textit{open} effective action
\begin{equation}
S[\psi_+,\psi_-; \{\Xi_+\},\{\Xi_-\}] \rightarrow \Seff[\psi_+,\psi_-]=S_s[\psi_+]-S_s[\psi_-]+\SIF[\psi_+,\psi_-].
\label{eq:open-action}
\end{equation}
The influence action is thus the field-level object that depends on the initial state of the environment, the field content of the environment and the interactions between system and environment.
As we shall explicitly see in the next sections, all dependence on the DM model and on the DM state is packaged into retarded and symmetric correlation functions, which preserve path ordering. 
Correlation functions of $``+"$ fields are time ordered, while correlation functions of  $``-"$ fields are anti-time ordered. 
In correlators involving both $``+"$ and $``-"$ fields, all $``+"$ fields are placed to the right of all $``-"$ fields. For a pedagogical and modern introduction, see Refs.~\cite{Kamenev_2011,Haehl2025}.

It is especially convenient to introduce the Keldysh basis, which facilitates the rewriting of all degrees of freedom in terms of quantum (``q") and classical (``cl") fields.\footnote{In the literature, the quantum and classical fields are respectively also known as anti-symmetric and symmetric, or as advanced and retarded fields.}
For bosonic fields,
\begin{equation} \label{eq:cl/k}
\begin{aligned}
O^\mathrm{cl}  & =\frac{1}{2}(O_{_+}+O_{_-}), \qquad
O^\mathrm{q} =O_{_+}-O_{_-} \, .
\end{aligned}
\end{equation}
In this basis, the effective action satisfies the following non-equilibrium constraints~\cite{Crossley:2015evo,Liu:2018kfw}:
\begin{equation}\label{eq:non-eq_constraints}
S_{_\mathrm{eff}}[\psi^{_\mathrm{cl}},\psi^{_\mathrm{q}}=0] = 0 \quad , \quad S_{_\mathrm{eff}}[\psi^{_\mathrm{cl}},\psi^{_\mathrm{q}}] = -S^*_{_\mathrm{eff}}[\psi^{_\mathrm{cl}},-\psi^{_\mathrm{q}}] \quad , \quad 
\mathrm{Im} \left \{ S_{_\mathrm{eff}}[\psi^{_\mathrm{cl}},\psi^{_\mathrm{q}}] \right \}  \geq 0 \, .
\end{equation}
Crucially, this implies that (i) the effective action is at least linear in $\psi^q$, and (ii) that terms in the effective action that have odd or even powers of $\psi^q$ are purely real or purely imaginary, respectively~\cite{Colas:2025app}. 

\subsection{From the microscopic theory to the influence action}\label{sec:operators}

Having laid out the logic behind the Schwinger--Keldysh construction, we now show how to obtain the influence action for a laboratory probe, whose degrees of freedom consist of atoms. 

For simplicity, we focus on monopole-monopole interactions, i.e., we shall consider interaction Lagrangians of the type 
\begin{equation}\label{eq:interaction_L}
\mathcal{L}_{\mathrm{int}} = \frac{g}{\Lambda^D} O_s[\psi] O_e[\Xi]
\end{equation}
where $O_s$ and $O_e$ are functionals of system and environment degrees of freedom, $g\ll1$ is a dimensionless coupling, $\Lambda$ is some high-energy scale and $D$ ensures that the mass dimension of the interaction term is dimension four. 
The notation of the environment operator is deliberately agnostic about the DM model. 
As we shall see explicitly in later sections, $O_e = O_\chi$  in the case of contact interactions, while $O_e = \phi$ in the case of long-range Yukawa interactions. 

At energies relevant for DM direct detection with MWIs, atoms, although built from nucleons and electrons, can be treated as a single heavy non-relativistic (NR) complex field $\psi$. 
Hence, the atom's free dynamics are governed by
\begin{equation}\label{eq:S0}
  \mathcal{L}_s
  = \psi^\dagger\!\left(\mi\partial_t + \frac{\nabla^2}{2m_\mathrm{at}}\right)\psi \,,
\end{equation}
where $m_\mathrm{at}$ is the total atomic mass. 
Furthermore, our EFT will automatically inherit a global $U(1)$ symmetry that forbids operators with an odd number of $\psi$ fields. 
Hence, the lowest dimension system current is given by 
\begin{equation}
\begin{aligned}
\Os(x)&=\int\dd^3r\,\varrho(\mathbf r)\,\psi^\dagger(t,\mathbf x-\mathbf r)\psi(t,\mathbf x-\mathbf r) \, , \\
F_A(\mathbf q)&=\int\dd^3r\,e^{-\mi\mathbf q\cdot\mathbf r}\varrho(\mathbf r) \, ,
\end{aligned}
\label{eq:nuclear-ff}
\end{equation}
where $\varrho(\mathbf{r})$ is the atom's charge density profile and $F_{A}(\mathbf{q})$ its form factor~\cite{Trickle:2019nya}.\footnote{We emphasize that $\psi$ represents the atom as a whole, and not, e.g., the individual nucleons, which have already been integrated out. Their effect is encoded entirely in $m_\mathrm{at}$ and in the composite operator built from $\psi$.} 

Using Eqs.~\eqref{eq:cl/k} and \eqref{eq:interaction_L}, and introducing the notation $\langle\!\langle \cdot \rangle\!\rangle \equiv \mathrm{Tr}[\cdot \hat{\rho}_e]$, the influence action through $\mathcal{O}(g^2)$ is related to the Feynman--Vernon influence functional~\cite{Feynman:1963fq}:
\begin{equation}\label{eq:S_IF}
\begin{aligned}
\exp \left \{ \mi S_{\mathrm{IF}} \right \} &= \left\langle\!\!\!\left\langle \exp \left \{ \mi \frac{g}{\Lambda^D}  \int_{t_i}^{t_f} d^4 x \left [ O^\mathrm{cl}_{s} (x) O^\mathrm{q}_{e}  (x) + O^\mathrm{q}_{s}  (x) O^\mathrm{cl}_{e} (x) \right ] \right \} \right\rangle\!\!\!\right\rangle \\
& = \exp \Bigg \{ \mi \frac{g}{\Lambda^D}  \int_{t_i}^{t_f}\mathrm{d}^4x \,\, O_s^\mathrm{q}(x)\, \langle \!\langle O_e^\mathrm{cl}(x) \rangle \!\rangle 
- \frac{\mi}{2} \left (\frac{g}{\Lambda^D} \right)^2 \!\int_{t_i}^{t_f}\!\mathrm{d}^4x \! \int_{t_i}^{t_f} \,\mathrm{d}^4 x'\,   \vec{O}_s^\mathsf{T}(x) \mathsf{K}(x,x')
 \vec{O}_s(x') \, \\
 & \qquad  \qquad +\mathcal{O}(g^3)  \Bigg \} \, ,
\end{aligned}
\end{equation}
where we introduced the system current vector 
\begin{equation}
 \vec{O}_s^\mathsf{T}(x) = \left [ O_s^\mathrm{cl}(x) \quad  O_s^\mathrm{q}(x) \right ] \, ,
\end{equation}
and the environment matrix kernel 
\begin{equation}\label{eq:K_matrix}
\mathsf{K}(x,x') = \begin{bmatrix} 0 & K_{\mathrm{a}}(x,x')  \\ K_{\mathrm{r}}(x,x') & K_{\mathrm{k}}(x,x') \end{bmatrix} \, ,
\end{equation}
whose entries depend explicitly on the connected correlation functions of environment currents:
\begin{align}
K_{\mathrm{r}}(x,x')&= -\mi \, \langle\!\langle O_e^\mathrm{cl}(x) O_e^\mathrm{q}(x')\rangle\!\rangle_c = -\mi \, \theta(t-t')\langle\!\langle[\hat{O}_e(x),\hat{O}_e(x')]\rangle\!\rangle_c \, , \label{eq:retarded}\\
K_{\mathrm{a}}(x,x')&=  -\mi \, \langle\!\langle  O_e^\mathrm{q}(x) O_e^\mathrm{cl}(x') \rangle\!\rangle_c = +\mi \, \theta(t'-t)\langle\!\langle[\hat{O}_e(x),\hat{O}_e(x')]\rangle\!\rangle_c \, , \label{eq:advanced}\\
K_{\mathrm{k}}(x,x')&=  -\mi \, \langle\!\langle O_e^\mathrm{cl}(x) O_e^\mathrm{cl}(x') \rangle\!\rangle_c = -\frac{\mi}{2}\langle\!\langle\{\hat{O}_e(x),\hat{O}_e(x')\}\rangle\!\rangle_c \, . \label{eq:noise} 
\end{align}
Note that the correlation function $\langle\!\langle O_e^\mathrm{q}(x) O_e^\mathrm{q}(x')\rangle\!\rangle_c$ vanishes. 
Furthermore, for real environment functionals $O_e(x)$, $K_{\mathrm{k}}(x,x')$ is pure imaginary, while $K_{\mathrm{a}}(x,x')$ and $K_{\mathrm{r}}(x,x')$ are real and satisfy the relation 
\begin{equation}\label{eq:rel_a_r}
K_{\mathrm{r}}(x,x') = K^*_{\mathrm{a}}(x',x) \, .
\end{equation} 

Crucially, the explicit form of the environment kernels depends on the properties of the environment. 
In the remainder of this work we specialize to a stationary and homogeneous DM background, whose three-momentum distribution is given by a boosted Maxwell--Boltzmann distribution $f_{\chi}(\mathbf{p})$, with normalization $\int \dbar^3 p \, f_{\chi}(\mathbf{p}) = 1$, as predicted by the Standard Halo Model in the lab frame~\cite{Drukier:1986tm}. 
This assumption is valid whenever the DM phase-space distribution varies negligibly over the interferometric time and length scales, which requires that: 
(i) DM-SM interactions are sufficiently weak that screening~\cite{Emken:2018run} and direct deflection/anti-screening~\cite{Berlin:2019uco} of DM by the local environment are negligible; 
for quadratic couplings in particular, the detector environment can suppress or distort the local DM field amplitude, potentially violating both the homogeneity and stationarity assumptions~\cite{Burrage:2024tdm,Burrage:2025cavity}; 
(ii) any daily or annual modulations of the local DM density, arising for instance from gravitational focusing by the Earth or Sun, occur on timescales much longer than the interferometric time~\cite{Griest:1987vc,Kim:2021yyo}; 
and (iii) the DM distribution is smooth on experimental scales, as supported by $N$-body simulations of the solar neighborhood~\cite{Helmi:2002qp,Vogelsberger:2008qb}. 
Under these assumptions, $\mathsf{K}(x,x') \rightarrow \mathsf{K}(x-x')$.

Furthermore, we assume that the initial DM state is Gaussian, as argued in Ref.~\cite{Cheong:2024ose}. In the case of long-range DM-probe interactions, we take the initial state of the mediator to be the vacuum, which is Gaussian by definition. 
As a result, Wick's theorem applies and the environmental kernels $\mathsf{K}(x-x')$ will depend on products of two-point functions of environmental degrees of freedom. 
Working in the quantum/classical field basis, these two-point functions are either retarded/advanced Green's functions or the Keldysh propagator~\cite{Calzetta:2008iqa}: for real bosonic $\Xi$ the free-theory propagators are 
\begin{equation}\label{eq:GreenBoson}
\begin{aligned}
G_\mathrm{r}^\Xi(x-x') &\equiv -\mi \left\langle\!\!\left\langle \Xi^\mathrm{cl}(x)\Xi^\mathrm{q}(x') \right\rangle\!\!\right\rangle = -\mi\,\theta(t-t') \left\langle\!\!\left\langle \left[ \hat\Xi(x),\hat\Xi(x') \right] \right\rangle\!\!\right\rangle \, ,\\
G_\mathrm{a}^\Xi(x-x') &\equiv -\mi \left\langle\!\!\left\langle \Xi^\mathrm{q}(x)\Xi^\mathrm{cl}(x') \right\rangle\!\!\right\rangle = + \mi\,\theta(t'-t) \left\langle\!\!\left\langle \left[ \hat\Xi(x),\hat\Xi(x') \right] \right\rangle\!\!\right\rangle \, , \\
G_\mathrm{k}^\Xi(x-x') &\equiv -\mi \left\langle\!\!\left\langle \Xi^\mathrm{cl}(x)\Xi^\mathrm{cl}(x') \right\rangle\!\!\right\rangle = -\frac{\mi}{2} \left\langle\!\!\left\langle \left\{ \hat\Xi(x),\hat\Xi(x') \right\} \right\rangle\!\!\right\rangle \, ,
\end{aligned}
\end{equation}
while for fermionic environmental fields the free-theory propagators read as
\begin{equation}\label{eq:GreenFermion}
\begin{aligned}
G_\mathrm{r}^\Xi(x-x') &\equiv -\mi\left\langle\!\!\left\langle \Xi^\mathrm{cl}(x) \overline{\Xi}^\mathrm{cl}(x')  \right\rangle\!\!\right\rangle = -\mi \, \theta(t-t')\left\langle\!\!\left\langle \left \{ \hat{\Xi}(x) , \hat{\overline{\Xi}}(x') \right \}\right\rangle\!\!\right\rangle \, , \\
G_\mathrm{a}^\Xi(x-x') &\equiv -\mi \left\langle\!\!\left\langle \Xi^\mathrm{q}(x) \overline{\Xi}^\mathrm{q}(x')  \right\rangle\!\!\right\rangle = + \mi \, \theta(t'-t) \left\langle\!\!\left\langle \left \{ \hat{\Xi}(x) , \hat{\overline{\Xi}}(x') \right \} \right\rangle\!\!\right\rangle \, , \\
G_\mathrm{k}^\Xi(x-x') &\equiv -\mi \left\langle\!\!\left\langle \Xi^\mathrm{cl}(x) \overline{\Xi}^\mathrm{q}(x')  \right\rangle\!\!\right\rangle = -\mi \left\langle\!\!\left\langle \left [ \hat{\Xi}(x) , \hat{\overline{\Xi}}(x') \right ] \right\rangle\!\!\right\rangle \, ,
\end{aligned}
\end{equation}
where in the latter case we used the Larkin-Ovchinnikov convention~\cite{Larkin_Ovchinnikov_1975}
\begin{equation}\label{eq:Larkin}
\begin{aligned}
\Xi^\mathrm{cl} & =\frac{1}{\sqrt{2}}(\Xi_{_+}+\Xi_{_-}) \,, \,\,\,
\Xi^\mathrm{q} =\frac{1}{\sqrt{2}}(\Xi_{_+}-\Xi_{_-}) \, , \,\,\, 
\overline{\Xi}^\mathrm{cl}  =\frac{1}{\sqrt{2}}(\overline{\Xi}_{_+}-\overline{\Xi}_{_-}) \, , \,\,\,
\overline{\Xi}^\mathrm{q} =\frac{1}{\sqrt{2}}(\overline{\Xi}_{_+}+\overline{\Xi}_{_-}) \, ,
\end{aligned}
\end{equation}
with $\overline{\Xi}=\Xi^\dagger \gamma^0$ the conjugate field. 
As a result of these assumptions, and as we shall see explicitly in section~\ref{sec:DM-kernels}, the entries of the matrix $\mathsf{K}(x-x')$ will be proportional to: (i) DM propagators, in the case of linearly-coupled DM; (ii) bubble diagrams of DM particles, in the case of quadratically-coupled DM; and (iii) in-medium (or DM-dressed) mediator propagators, for long-range DM-nucleon interactions. 

Taking the limit $t_f=-t_i\rightarrow \infty$, using the definition of the FT in Eq.~\eqref{eq:FT}, and noting that $\widetilde{O}^*_s(q) = \widetilde{O}_s(-q)$ and $\widetilde{K}_{\mathrm{a}}(q) = \widetilde{K}_{\mathrm{r}}^*(q)$ (see Eq.~\eqref{eq:rel_a_r}), the influence action may be expressed in four-momentum representation as 
\begin{equation}\label{eq:Seff_gaussian_momentum}
\begin{aligned}
S_{_{\rm IF}}[\psi_{_+},\psi_{_-}]
& = \frac{g}{\Lambda^D} \int \!  \dbar^4 q \, \,  \widetilde{O}_s^\mathrm{q}(q) \langle \!\langle \widetilde{O}_e^\mathrm{cl}(-q) \rangle \!\rangle \\
&\quad  - \left (\frac{g}{\Lambda^D} \right)^2  \!\int\! \dbar^4 q \left \{  \mathrm{Re}\left [ \widetilde{O}_s^\mathrm{cl}(q) \widetilde{O}_s^\mathrm{q}(-q)  \right ] \mathrm{Re} \left [\widetilde{K}_{\mathrm{r}}(q)\right ]  -  \mathrm{Im}\left [ \widetilde{O}_s^\mathrm{cl}(q) \widetilde{O}_s^\mathrm{q}(-q)  \right ] \mathrm{Im} \left [\widetilde{K}_{\mathrm{r}}(q) \right ]  \right \} \\
& \quad - \frac{\mi}{2} \left (\frac{g}{\Lambda^D} \right)^2 \!\int\! \dbar^4 q \, \, \left |\widetilde{O}_s^\mathrm{q}(q) \right |^2 \mathrm{Im} \left [ \widetilde{K}_{\mathrm{k}}(q)\right ] \\
& \quad + \mathcal{O}(g^3) \, ,
\end{aligned}
\end{equation}
where $q$ is the four-momentum transferred by the environment to the system, and the first two lines are real while the third is pure imaginary. 
As we shall see explicitly in section~\ref{sec:observables}, the first line will not lead to observable effects when the environment is homogeneous. 
The second line corresponds to phase shifts, or unitary effects, while the third captures decoherence, or non-unitary processes.

We conclude by noting that, because the environment is described by quantum fields, the kernels $K_{\mathrm{r,a,k}}$ do not generally vanish in vacuum. 
Indeed, the pure vacuum part of $K_{\mathrm{k}}$ reflects the presence of quantum fluctuations, which persist even when the bath occupation number is zero. 

\section{From the influence action to interferometer observables}\label{sec:observables}

Having established the field-level open EFT for an arbitrary system, we now focus on MWIs. 
In particular, we will show explicitly how the reduced density matrix of the MWI can be mapped directly to the \textit{worldline} influence action.

For pedagogical reasons, we start with the single-atom case, which was already discussed in Ref.~\cite{Badurina:2026owr} and contains all the essential structure.
The generalization to an ensemble of $N_\mathrm{at}$ atoms, for example atom interferometers (AIs) and composite-particle interferometers (CPIs), is treated in section~\ref{sec:oEFT_multi}, where many-body effects emerge.

\subsection{A toy example: a single-atom interferometer}\label{sec:toy}

In a single-atom MWI, the atom as a whole  is prepared in a superposition of states that are sent along different spacetime trajectories. 
In practice, this is achieved by entangling the c.o.m. degree of freedom with an internal degree of freedom, e.g., the atomic spin or energy level, which labels the trajectory. 
To make this structure manifest in the field theory, we introduce a discrete path label $\alpha\in\{L,R\}$ and describe the atom as a non-relativistic complex quantum field doublet $\Psi = (\psi_{_L},\psi_{_R})^T$, whose components correspond to the two interferometer arms. 
As a result, for the case of elastic monopole-monopole interactions, the atom's free and interaction Lagrangian take the form displayed in Eqs.~\eqref{eq:interaction_L}-\eqref{eq:nuclear-ff} under the replacement $\psi(t,\mathbf{x-r})\rightarrow \Psi(t,\mathbf{x-r})$.

Since atoms are neither created nor destroyed at sub-GeV scales, the system dynamics and measurements are restricted to the one-particle sector of the Fock space of $\hat{\Psi}$, namely $ \ket{\alpha,\mathbf{y}}\equiv \hat{\psi}_{\alpha}^\dagger(\mathbf{y})\ket{0}_s$, where $\ket{0}_s$ is the system's vacuum state. 
Hence, in second quantization, the action of the system operator on the single atom state is given by
\begin{equation}\label{eq:varrho}
 \hat{O}_s(x) \hat \psi_\alpha^\dagger(\mathbf{y})\ket{0}_s = 
 \varrho (\mathbf{x}-\mathbf{y}) \hat \psi_\alpha^\dagger(\mathbf{y})\ket{0}_s \, .
\end{equation}
This projection then implies that $O_s \to \varrho$ at the level of the influence action.

We define the initial state of the system at time $t_i$ to be given by a linear superposition of co-located $L$ and $R$ wavepackets 
\begin{equation}\label{eq:initial_state}
  \hat{\rho}_{{s}}(t_i)
  =
    \ket{\Phi}\!\bra{\Phi} \, \, , \, \, 
  \ket{\Phi}
  = \frac{1}{\sqrt{2}}\int\!\mathrm{d}^3y\;
    \Phi(\mathbf{y}) \left ( \hat{\psi}_{L}^\dagger(\mathbf{y}) +\hat{\psi}_{R}^\dagger(\mathbf{y}) \right ) \ket{0}_s \, ,
\end{equation} 
where $\Phi(\mathbf{y})$ is the atomic wavepacket of width $\sigma$ centered at the origin.
These wavepackets are subsequently separated and recombined over a finite time interval by state-dependent external forces, described by the operator $\hat{J}$ that is diagonal in the $\{L,R\}$ basis and models the momentum impulses delivered by the beamsplitter and mirror pulses in an atom interferometer~\cite{Glick:2024xlq}. As a result, components of the time-evolved reduced density matrix in the $\alpha, \beta \in \{L,R\}$ and $\{\mathbf{x}_{f,+},\mathbf{x}_{f,-}\}$ bases read as 
\begin{equation}
\begin{aligned}
   \bra{\mathbf{x}_{f,+}}\hat{\rho}^{\alpha\beta}_s(t_f)\ket{\mathbf{x}_{f,-}} & \equiv \bra{\alpha, \mathbf{x}_{f,+}}\hat{\rho}_s(t_f)\ket{\beta, \mathbf{x}_{f,-}} \\
   & =  \bra{\alpha, \mathbf{x}_{f,+}}\mathrm{Tr}_e \left [ \hat{U} \left (t_f,t_{i};\hat{J} \right )\,\hat{\rho}_{i}\,\hat{U} \left (t_{i},t_f;\hat{J} \right ) \right ] \ket{\beta, \mathbf{x}_{f,-}} \, .
\end{aligned}
\end{equation}

In what follows, we will focus on interferometric readouts that depend on the path label. 
This is standard in MWI protocols with internal-state labeling~\cite{Cronin:2009zz}, which entangle the probe's internal degree of freedom with the interferometer arms. 
As a result, the arm-dependent history is encoded in the coherence between the $L$ and $R$ states.
Hence, we shall compute the reduced density matrix where the final position is traced over: 
\begin{equation}\label{eq:op_rho}
   \rho_s^{\alpha\beta}(t_f) \equiv \int d^3x_{f,+} \int d^3x_{f,-} \delta^{(3)}(\mathbf{x}_{f,+}-\mathbf{x}_{f,-})\bra{\mathbf{x}_{f,+}}\hat{\rho}^{\alpha\beta}_s(t_f)\ket{\mathbf{x}_{f,-}} \, .
\end{equation}
Because the dynamics is confined to the one-particle sector and the initial state is given by Eq.~\eqref{eq:initial_state}, Eq.~\eqref{eq:op_rho} admits a path-integral representation over worldlines
$\boldsymbol{X}_\pm$~\cite{Hu:1991di,Hu:1993vs} driven by external impulsive, path-dependent currents $\boldsymbol{J}_{\alpha}\propto m_\mathrm{at}$ given by the eigenvalues of $\hat{J}$.  
Explicitly, setting $\mathbf{x}_{f,+}=\mathbf{x}_{f}$ and performing the integrals over final positions, Eq.~\eqref{eq:op_rho} reads as
\begin{equation}
\begin{aligned}\label{eq:rho_PI_worldline}
\rho_s^{\alpha\beta}(t_f) & = 
\int\!\!\mathrm{d}^3 x_{i,+}\int \mathrm{d}^3 x_{i,-} \!\!\, \bra{\alpha,\mathbf{x}_{i,+}}\hat{\rho}_{{s}}(t_i)\ket{\beta,\mathbf{x}_{i,-}}  \\ & \qquad \qquad \times 
\int\!\!\mathrm{d}^3 x_f \int_{\mathbf{x}_{i,+}}^{\mathbf{x}_f}\!\!\!\!\mathcal{D}\boldsymbol{X}_+\!\!\int_{\mathbf{x}_{i,-}}^{\mathbf{x}_f}\!\!\mathcal{D}\boldsymbol{X}_-
e^{\mi S_s[\boldsymbol{X}_+,\boldsymbol{J}_{\alpha,+}]-\mi S_s[\boldsymbol{X}_-,\boldsymbol{J}_{\beta,-}]}\, 
e^{\mi S_{_\mathrm{IF}}[\boldsymbol{X}_+,\boldsymbol{X}_-]}  \, ,
\end{aligned}
\end{equation}
with boundary conditions $\boldsymbol{X}_\pm(t_i)=\mathbf{x}_{i,\pm}$ and $\boldsymbol{X}_\pm(t_f)=\mathbf{x}_f$, and with the free system action 
\begin{equation}\label{eq:Ssys}
S_s[\boldsymbol{X}_\pm,\boldsymbol{J}_{\alpha,\pm}]\!=\!\!\int_{t_i}^{t_f}\!\mathrm{d}t\Bigl[\tfrac{m_\mathrm{at}}{2}|\dot{\boldsymbol{X}}_\pm|^2\!+\!\boldsymbol{J}_{\alpha,\pm}\!\cdot\!\boldsymbol{X}_\pm\Bigr]\, .
\end{equation}

\begin{figure}[t]
\centering

\begin{tikzpicture}[
  >=Stealth,
  line cap=round,
  line join=round,
  font=\footnotesize,
  x=0.62cm,
  y=0.62cm,
  axis/.style={line width=0.8pt, draw=black!65},
  tubeOuter/.style={draw=black!65, line width=25pt, line cap=butt},
  tubeInner/.style={draw=white, line width=24pt, line cap=butt},
  mainR/.style={line width=1.6pt, draw=orange!90!black},
  mainL/.style={line width=1.6pt, draw=green!55!black, dash pattern=on 4pt off 5pt},
  pathLabel/.style={
    sloped,
    fill=orange!10!white,
    fill opacity=0.9,
    text opacity=1,
    inner sep=1.2pt
  }
]

\def\TopY{2.55}
\def\BotY{-2.55}
\def\ArmX{1.95}
\def\HalfTY{1.8214}
\def\AxisBot{-4.4}
\def\AxisTop{4.5}

\node[text=black!85] at (0,5.0) {Split and hold geometry};

\draw[axis,-{Stealth[length=2.2mm,width=1.4mm]}]
  (-4.0,\AxisBot) -- (-4.0,\AxisTop);
\draw[axis,-{Stealth[length=2.2mm,width=1.4mm]}]
  (-4.0,\AxisBot) -- (4.4,\AxisBot);

\draw[axis] (-4.0,\TopY) -- ++(-0.13,0);
\node[left=4pt,text=black!75] at (-4.18,\TopY)
  {$\frac{T+T'}{2}$};

\draw[axis] (-4.0,\HalfTY) -- ++(-0.13,0);
\node[left=4pt,text=black!75] at (-4.18,\HalfTY)
  {$\frac{T}{2}$};

\draw[axis] (-4.0,0) -- ++(-0.13,0);
\node[left=4pt,text=black!75] at (-4.18,0)
  {$0$};

\draw[axis] (-4.0,-\HalfTY) -- ++(-0.13,0);
\node[left=4pt,text=black!75] at (-4.18,-\HalfTY)
  {$-\frac{T}{2}$};

\draw[axis] (-4.0,\BotY) -- ++(-0.13,0);
\node[left=4pt,text=black!75] at (-4.18,\BotY)
  {$-\frac{T+T'}{2}$};

\foreach \x/\lbl in {
  -1.95/{$-\frac{\Delta x}{2}$},
   0/{$0$},
   1.95/{$\frac{\Delta x}{2}$}
}{
  \draw[axis] (\x,\AxisBot) -- ++(0,-0.13);
  \node[below=3pt,text=black!75] at (\x,-4.55) {\lbl};
}

\coordinate (Top) at (0,\TopY);
\coordinate (UR) at (\ArmX,\HalfTY);
\coordinate (LR) at (\ArmX,-\HalfTY);
\coordinate (Bot) at (0,\BotY);
\coordinate (LL) at (-\ArmX,-\HalfTY);
\coordinate (UL) at (-\ArmX,\HalfTY);

\draw[
  decorate,
  decoration={brace, amplitude=5pt},
  line width=0.7pt,
  draw=black!70
]
  (-6.45,\BotY) -- (-6.45,\TopY)
  node[midway,left=7pt,text=black!80] {$T_s$};

\draw[tubeOuter]
  (0,4.2) -- (Top) -- (UR) -- (LR) -- (Bot) -- (0,-4.2);
\draw[tubeOuter]
  (0,4.2) -- (Top) -- (UL) -- (LL) -- (Bot) -- (0,-4.2);

\draw[tubeInner]
  (0,4.2) -- (Top) -- (UR) -- (LR) -- (Bot) -- (0,-4.2);
\draw[tubeInner]
  (0,4.2) -- (Top) -- (UL) -- (LL) -- (Bot) -- (0,-4.2);

\draw[orange!10!white,line width=24pt,line cap=butt]
  (0,4.2) -- (Top) -- (UR) -- (LR) -- (Bot) -- (0,-4.2);
\draw[orange!10!white,line width=24pt,line cap=butt]
  (0,4.2) -- (Top) -- (UL) -- (LL) -- (Bot) -- (0,-4.2);

\draw[mainR]
  (0,4.2) -- (Top) -- (UR) -- (LR) -- (Bot) -- (0,-4.2);
\draw[mainL]
  (0,4.2) -- (Top) -- (UL) -- (LL) -- (Bot) -- (0,-4.2);

\path (UL) -- (Top)
  node[pos=0.48,pathLabel,text=green!55!black]
  {$\overline{\boldsymbol{x}}_L$};

\path (Top) -- (UR)
  node[pos=0.52,pathLabel,text=orange!90!black]
  {$\overline{\boldsymbol{x}}_R$};

\draw[axis,<->,line width=0.6pt]
  (-0.63,-3.55) -- (0.63,-3.55);
\node[text=black!80] at (1.05,-3.4) {$\sigma$};

\end{tikzpicture}
\caption{Spacetime diagram for the interferometric geometry considered in this work. 
The total interferometric time is $T_s$ and the maximal spatial separation between the two paths is $\Delta x$. 
The unperturbed classical worldlines $\overline{\boldsymbol{x}}_{L,R}(t)$, which correspond to the particular solutions of Eq.~\eqref{eq:unperturbed}, defined in Eq.~\eqref{eq:Xbar}, are shown in solid orange and dashed green, together with the initial wavepacket of size $\sigma$ shown as the shaded tube. 
For simplicity, we will consider the regime $T'\ll T$ with $|\boldsymbol{\Delta x}|$ fixed.}
\label{fig:paths}
\end{figure}
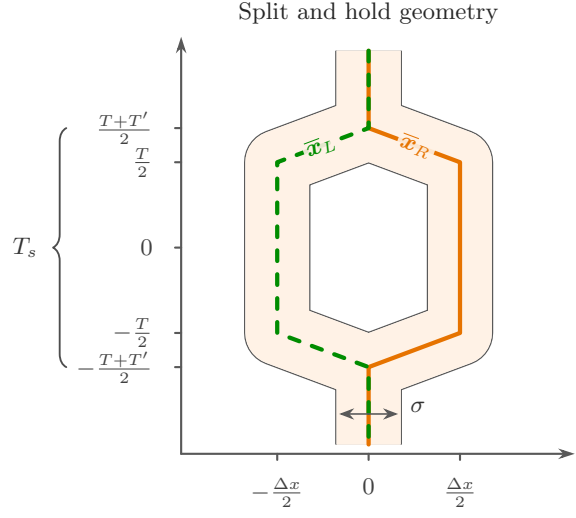

We assume that both arms couple to the environment with the same charge. 
We also assume that the interferometer closes, as shown in Fig.~\ref{fig:paths}.
 As a result, the $\mathcal{O}(g)$ contribution to the influence functional vanishes (see Eq.~\eqref{eq:S_IF}). Introducing the Keldysh representation of the atom's charge density profile, 
\begin{align}
\varrho^\mathrm{cl}(x;\boldsymbol{X}_\pm)&=\frac{1}{2}\bigl[\varrho(\mathbf{x}\!-\!\boldsymbol{X}_+(t))+\varrho(\mathbf{x}\!-\!\boldsymbol{X}_-(t))\bigr]\, , \label{eq:rhocl} \\
\varrho^\mathrm{q}(x;\boldsymbol{X}_\pm)&=\varrho(\mathbf{x}\!-\!\boldsymbol{X}_+(t))-\varrho(\mathbf{x}\!-\!\boldsymbol{X}_-(t))\, , \label{eq:rhoq}
\end{align} 
and the system current vector
\begin{align}
\vec{\varrho}(x)^{\mathsf T} &=
\begin{pmatrix}
\varrho^{\mathrm{cl}}(x;\boldsymbol{X}_\pm) & \varrho^{\mathrm{q}}(x;\boldsymbol{X}_\pm)
\end{pmatrix} \label{eq:Os} \, ,
\end{align}
the leading-order contribution to the influence action takes the compact form
\begin{equation}\label{eq:Sif}
\begin{aligned}
S_{_{\rm IF}}[\boldsymbol{X}_+,\boldsymbol{X}_-] =& 
-\tfrac{1}{2}\!\left(\tfrac{g}{\Lambda^D}\right)^{\!2}\!\!
\int_{t_{i}}^{t_f}\!\! \mathrm{d}^4x\!\! 
\int_{t_{i}}^{t_f}\!\! \mathrm{d}^4x' \vec{\varrho}(x;\boldsymbol{X}_\pm)^{\mathsf T}
\,\mathsf{K}(x - x')\,
\vec{\varrho}(x';\boldsymbol{X}_\pm) + \mathcal{O}(g^3)\, ,
\end{aligned}
\end{equation}
with $\mathsf{K}(x - x')$ the environment matrix kernel (see Eq.~\eqref{eq:K_matrix}).

To evaluate the path integral over worldlines, we decompose the worldlines $\boldsymbol{X}_\pm$ into two distinct contributions: (i) the \textit{classical (driven) trajectory} $\overline{\boldsymbol{X}}_{\alpha,\pm}$ sourced by the external interferometer currents $\boldsymbol{J}_{\alpha,\pm}$ and satisfying the boundary conditions of the path integral, i.e.,
\begin{equation}\label{eq:unperturbed}
m_\mathrm{at} \ddot{\overline{\boldsymbol{X}}}_{\alpha,\pm} = \boldsymbol{J}_{\alpha,\pm} \,, \qquad 
\overline{\boldsymbol{X}}_{\alpha,\pm}(t_i) = \mathbf{x}_{i,\pm} \,, \qquad 
\overline{\boldsymbol{X}}_{\alpha,\pm}(t_f) = \mathbf{x}_f \,,
\end{equation}
which admits the solution
\begin{equation}\label{eq:Xbar}
\overline{\boldsymbol{X}}_{\alpha,\pm}(t) \;=\; 
\mathbf{x}_{i,\pm}\,\frac{t_f-t}{t_f-t_i} \,+\, \mathbf{x}_f\,\frac{t-t_i}{t_f-t_i} \,+\, 
\underbrace{\int_{t_i}^{t_f}\!\mathrm{d}t'\;G_\mathrm{d}(t,t')\,\boldsymbol{J}_{\alpha,\pm}(t')}_{\overline{\boldsymbol{x}}_{\alpha,\pm}(t)} \,,
\end{equation}
in terms of the Dirichlet Green’s function~\cite{Corradini:2015tik}
\begin{equation}\label{eq:Dirichlet_prop}
G_\mathrm{d}(t,t') = \frac{(t_{<}-t_{i})(t_{>}-t_f)}{m_\mathrm{at}(t_f-t_i)} \,, \quad \, t_{<} = \min(t,t')\, , \quad   t_{>} = \max(t,t') \, , 
\end{equation} 
and the (interferometer geometry-dependent but boundary condition-independent) \textit{particular solution} $\overline{\boldsymbol{x}}_{\alpha,\pm}(t)$; 
and (ii) \textit{worldline fluctuations} $\boldsymbol{\delta X}_\pm$ obeying the Dirichlet boundary conditions
\begin{equation}\label{eq:dX_BC}
\boldsymbol{\delta X}_\pm(t_i) = \boldsymbol{\delta X}_\pm(t_f) = \mathbf{0} \,.
\end{equation}
As a result, recalling from Eq.~\eqref{eq:initial_state} that $\bra{\alpha,\mathbf{x}_{i,+}}\hat{\rho}_{{s}}(t_i)\ket{\beta,\mathbf{x}_{i,-}} = \tfrac{1}{2}\Phi(\mathbf{x}_{i,+})\Phi^*(\mathbf{x}_{i,-})$, Eq.~\eqref{eq:rho_PI_worldline} can be rewritten as a path integral over worldline fluctuations:
\begin{equation}\label{eq:rhos}
\begin{aligned}
\rho_s^{\alpha \beta}(t_f) & =\!
\tfrac{1}{2}\int\!\!\mathrm{d}^3 x_{i,+}\int \mathrm{d}^3 x_{i,-} \!\!\, \Phi(\mathbf{x}_{i,+})\Phi^*(\mathbf{x}_{i,-})
\int\!\!\mathrm{d}^3 x_f \,  e^{\mi S_s[\overline{\boldsymbol{X}}_{\alpha,+},\boldsymbol{J}_{\alpha,+}]-\mi S_s[\overline{\boldsymbol{X}}_{\beta,-},\boldsymbol{J}_{\beta,-}]} \\
 & \qquad \qquad \qquad \qquad \times \int_{\mathbf{0}}^{\mathbf{0}}\!\!\!\!\mathcal{D}\boldsymbol{\delta X}_+\!\!\int_{\mathbf{0}}^{\mathbf{0}}\!\!\mathcal{D}\boldsymbol{\delta X}_-
e^{\mi S_s[\delta \boldsymbol{X}_+]-\mi S_s[\delta \boldsymbol{X}_-]}\, 
e^{\mi S_{_\mathrm{IF}}[\overline{\boldsymbol{X}}_{\alpha,+}+\boldsymbol{\delta X}_+,\overline{\boldsymbol{X}}_{\beta,-}+\boldsymbol{\delta X}_-]}  \, . 
\end{aligned}
\end{equation}
Following Ref.~\cite{Badurina:2026owr}, one can solve the path integral perturbatively by: 
(i) introducing fictitious currents $\boldsymbol{j}_\pm$ that couple linearly to the worldline fluctuations, such that $S_s[\delta \boldsymbol{X}_\pm] \rightarrow S_s[\delta \boldsymbol{X}_\pm, \boldsymbol{j}_\pm] $ analogously to Eq.~\eqref{eq:Ssys};
(ii) rewriting $\boldsymbol{\delta X_\pm} \rightarrow \mp \mi \, \boldsymbol{\delta}/\boldsymbol{\delta j}_\pm$ in the influence action in order to pull $S_\mathrm{IF}$ out of the path integral over fluctuations, and 
(iii) performing the exact Gaussian path integral over worldline fluctuations. 

We further focus on symmetric interferometric sequences, such that $\boldsymbol{J}_\alpha = -\boldsymbol{J}_\beta$ for $\alpha \neq \beta$ and $t_f = -t_i = \mathcal{T}$. 
Integrating Eq.~\eqref{eq:Ssys} by parts and substituting the classical trajectories of Eq.~\eqref{eq:Xbar}, the classical contribution to the system phase reduces to
\begin{equation}\label{eq:cl_action}
e^{\mi S_s[\overline{\boldsymbol{X}}_{\alpha,+},\boldsymbol{J}_{\alpha,+}]-\mi S_s[\overline{\boldsymbol{X}}_{\beta,-},\boldsymbol{J}_{\beta,-}]}
= \exp\!\left[\frac{\mi m_\mathrm{at}}{4\mathcal{T}}\left(|\mathbf{x}_f-\mathbf{x}_{i,+}|^{2}-|\mathbf{x}_f-\mathbf{x}_{i,-}|^{2}\right)\right].
\end{equation}
The contribution from the worldline fluctuations is captured by the generating functional~\cite{Corradini:2015tik}
\begin{equation}\label{eq:gen_func}
\begin{aligned}
Z_0[\boldsymbol{j}_+,\boldsymbol{j}_-]
&\equiv \left(\frac{m_\mathrm{at}}{4\pi\mathcal{T}}\right)^{-3}
\int_{\mathbf{0}}^{\mathbf{0}}\!\mathcal{D}\delta\boldsymbol{X}_+
\int_{\mathbf{0}}^{\mathbf{0}}\!\mathcal{D}\delta\boldsymbol{X}_-\,
e^{\mi S_s[\delta\boldsymbol{X}_+,\boldsymbol{j}_+]-\mi S_s[\delta\boldsymbol{X}_-,\boldsymbol{j}_-]} \\
&= \exp\left\{\frac{\mi}{2}\int_{-\mathcal{T}}^{\mathcal{T}}\!\mathrm{d}t\int_{-\mathcal{T}}^{\mathcal{T}}\!\mathrm{d}t'\,
G_\mathrm{d}(t,t')\left[\boldsymbol{j}_+(t)\cdot\boldsymbol{j}_+(t')-\boldsymbol{j}_-(t)\cdot\boldsymbol{j}_-(t')\right]\right\},
\end{aligned}
\end{equation}
with $Z_0[\mathbf{0},\mathbf{0}] = 1$. Using Eqs.~\eqref{eq:cl_action} and \eqref{eq:gen_func},  Eq.~\eqref{eq:rhos} reads as 
\begin{equation}\label{eq:rhos_step}
\begin{aligned}
\rho_s^{\alpha \beta}(\mathcal{T}) & = \tfrac{1}{2} \left ( \tfrac{m_\mathrm{at}}{4\pi\mathcal{T}} \right )^3 \int\!\!\mathrm{d}^3 x_{i,+}\int \mathrm{d}^3 x_{i,-} \!\!\,  \Phi(\mathbf{x}_{i,+})\Phi^*(\mathbf{x}_{i,-})
\\ & \qquad \times
\int\!\!\mathrm{d}^3 x_f \, e^{\tfrac{\mi m_\mathrm{at}}{4\mathcal{T}} \left (|\mathbf{x}_f-\mathbf{x}_{i,+}|^{2}-|\mathbf{x}_f-\mathbf{x}_{i,-}|^{2}\right)} e^{\mi S_{_\mathrm{IF}}\left [\overline{\boldsymbol{X}}_{\alpha,+} - \mi \tfrac{\boldsymbol{\delta}}{\boldsymbol{\delta j}_+},\overline{\boldsymbol{X}}_{\beta,-}+ \mi \tfrac{\boldsymbol{\delta}}{\boldsymbol{\delta j}_-}\right ]}  Z_0[\boldsymbol{j}_+,\boldsymbol{j}_-]\Big |_{\boldsymbol{j}_\pm = 0} \\
& \equiv \tfrac{1}{2} +  \tfrac{\mi}{2} \left \langle S_{_\mathrm{IF}}\left [\overline{\boldsymbol{X}}_{\alpha,+} - \mi \tfrac{\boldsymbol{\delta}}{\boldsymbol{\delta j}_+},\overline{\boldsymbol{X}}_{\beta,-}+ \mi \tfrac{\boldsymbol{\delta}}{\boldsymbol{\delta j}_-}\right ] \right \rangle Z_0[\boldsymbol{j}_+,\boldsymbol{j}_-]\Big |_{\boldsymbol{j}_\pm = 0} +\mathcal{O}(g^3) \, ,
\end{aligned}
\end{equation}
where in going to the last line we linearized the influence functional and defined $\langle \cdot \rangle$ as the average over the boundary conditions.

Henceforth, we will work in the heavy-probe limit, corresponding to sending $m_\mathrm{at}\rightarrow \infty$ while $\boldsymbol{J}_\pm/m_\mathrm{at}$ and $S_\mathrm{IF}$ are held fixed. 
In this limit the atom is treated as a prescribed source with fixed charge density: the environment can phase-shift and decohere the two histories while leaving the system worldlines unaffected.
\footnote{Importantly, since $\boldsymbol{J}_{\alpha,\pm} \propto m_\mathrm{at}$, the limit $m_\mathrm{at}\to\infty$ does not change the interferometer geometry.}
Since $G_{\rm d}\propto m_\mathrm{at}^{-1}$, in the heavy-probe limit Eq.~\eqref{eq:rho_PI_worldline} reduces to the influence functional evaluated on the driven classical paths. 
Furthermore, as we carefully derive in Appendix~\ref{app:finite-mass}, in this limit the external currents $\boldsymbol{J}_\pm$ and the final worldline configuration $\mathbf{x}_f$ fix the initial worldline configurations to be identical, i.e., $\mathbf{x}_{i,+}=\mathbf{x}_{i,-}$. 
Because we assumed a homogeneous environment, the initial worldline configurations drop out of the influence action, so that the influence action exclusively depends on the \textit{particular solution} $\overline{\boldsymbol{x}}_{\alpha,\pm}(t)$ of the classical worldlines (see~Eq.~\eqref{eq:Xbar}), i.e., 
\begin{equation}\label{eq:h.p.}
\lim_{m_\mathrm{at}\rightarrow \infty} \left \langle S_{_\mathrm{IF}}\left [\overline{\boldsymbol{X}}_{\alpha,+} - \mi \tfrac{\boldsymbol{\delta}}{\boldsymbol{\delta j}_+},\overline{\boldsymbol{X}}_{\beta,-}+ \mi \tfrac{\boldsymbol{\delta}}{\boldsymbol{\delta j}_-}\right ] \right \rangle Z_0[\boldsymbol{j}_+,\boldsymbol{j}_-]\Big |_{\boldsymbol{j}_\pm = 0} = S_{_\mathrm{IF}}\left [\overline{\boldsymbol{x}}_{\alpha,+},\overline{\boldsymbol{x}}_{\beta,-}\right ] \, .
\end{equation}
Introducing the 3-momentum FTs of the system currents
\begin{equation}\label{eq:rho_FT}
\widetilde\varrho^{\mathrm{cl},\mathrm{q}}(\mathbf{q},t;\boldsymbol{\overline{x}}_\pm) = \int d^3 x \, \varrho^{\mathrm{cl},\mathrm{q}}(x;\boldsymbol{\overline{x}}_\pm(t)) e^{-\mi \mathbf{q}\cdot \mathbf{x}} = F_A (\mathbf{q})\,\widetilde{\mathcal{W}}^{\mathrm{cl},\mathrm{q}}(\mathbf{q},t;\boldsymbol{\overline{x}}_\pm)\,,
\end{equation}
where $F_A(\mathbf{q})$ is the atomic form factor (see Eq.~\eqref{eq:nuclear-ff}) and
\begin{equation}\label{eq:W}
\begin{aligned}
\widetilde{\mathcal{W}}^\mathrm{cl}(\mathbf{q},t;\boldsymbol{\overline{x}}_{\pm})\!&=\!\tfrac{1}{2} \left[ e^{-\mi\mathbf{q}\cdot\boldsymbol{\overline{x}}_{\alpha,+}(t)}\!+\!e^{-\mi\mathbf{q}\cdot\boldsymbol{\overline{x}}_{\beta,-} (t)}
\right] 
\, , \\
\widetilde{\mathcal{W}}^\mathrm{q}(\mathbf{q},t;\boldsymbol{\overline{x}}_{\pm})\!&=\! e^{-\mi\mathbf{q}\cdot\boldsymbol{\overline{x}}_{\alpha,+}(t)}\!-\!e^{-\mi\mathbf{q}\cdot\boldsymbol{\overline{x}}_{\beta,-} (t)} 
\, , 
\end{aligned}
\end{equation}
capture the worldline histories, the influence action evaluated along the particular solution of the classical worldlines reads as
\begin{equation}\label{eq:Sif_lim}
\begin{aligned}
\!\!\!\! S_{_{\rm IF}}[\boldsymbol{\overline{x}}_{\alpha,+},\boldsymbol{\overline{x}}_{\beta,-}] &= - \left (\tfrac{g}{\Lambda^D} \right)^2  \!\int\! dt \!\int\! dt' \!\int\! \dbar^3 q \,|F_A(\mathbf{q})|^2 \, \mathrm{Re}\left [ \mathcal{\widetilde{W}}^\mathrm{q}(-\mathbf{q},t;\overline{\boldsymbol{x}}_\pm) \mathcal{\widetilde{W}}^\mathrm{cl}(\mathbf{q},t';\overline{\boldsymbol{x}}_\pm)  \right ] \mathrm{Re} \left [\widetilde{K}_{\mathrm{r}}(\mathbf{q},t-t')\right ]  \\
& \quad  + \left (\tfrac{g}{\Lambda^D} \right)^2  \!\int\! dt \!\int\! dt' \!\int\! \dbar^3 q \,|F_A(\mathbf{q})|^2 \, \mathrm{Im}\left [ \mathcal{\widetilde{W}}^\mathrm{q}(-\mathbf{q},t;\overline{\boldsymbol{x}}_\pm) \mathcal{\widetilde{W}}^\mathrm{cl}(\mathbf{q},t';\overline{\boldsymbol{x}}_\pm)  \right ] \mathrm{Im} \left [\widetilde{K}_{\mathrm{r}}(\mathbf{q},t-t') \right ]   \\
& \quad - \tfrac{\mi}{2} \left (\tfrac{g}{\Lambda^D} \right)^2 \!\int\! dt \!\int\! dt' \!\int\! \dbar^3 q \,|F_A(\mathbf{q})|^2 \, \mathcal{\widetilde{W}}^\mathrm{q}(-\mathbf{q},t;\overline{\boldsymbol{x}}_\pm) \mathcal{\widetilde{W}}^\mathrm{q}(\mathbf{q},t';\overline{\boldsymbol{x}}_\pm)  \mathrm{Im} \left [ \widetilde{K}_{\mathrm{k}}(\mathbf{q},t-t')\right ] \\
&  \quad  +\mathcal{O}(g^3) \, .
\end{aligned}
\end{equation}
Hence, substituting Eq.~\eqref{eq:h.p.} into Eq.~\eqref{eq:rhos_step} and re-exponentiating the influence functional, the heavy-probe and long-time limit of the system's reduced density matrix takes the form 
\begin{equation}\label{eq:rho_red_matrix}
\boxed{
\lim_{\mathcal{T}\rightarrow\infty}
\lim_{m_\mathrm{at}\rightarrow \infty} 
\hat{\rho}_s(\mathcal{T}) \simeq  \frac{1} {2}\begin{pmatrix} e^{\mi S_\mathrm{IF}[\overline{\boldsymbol{x}}_{L},\overline{\boldsymbol{x}}_{L}]} & e^{\mi S_\mathrm{IF}[\overline{\boldsymbol{x}}_{L},\overline{\boldsymbol{x}}_{R}]} \\ e^{\mi S_\mathrm{IF}[\overline{\boldsymbol{x}}_{R},\overline{\boldsymbol{x}}_{L}]} & e^{\mi S_\mathrm{IF}[\overline{\boldsymbol{x}}_{R},\overline{\boldsymbol{x}}_{R}]} \end{pmatrix} \equiv \frac{1}{2} \begin{pmatrix} 1 & e^{-\mathbb{D} - \mi \mathbb{P}} \\ e^{-\mathbb{D} +\mi \mathbb{P}} & 1 \end{pmatrix} \, ,}
\end{equation}
with
\begin{align}
\boxed{
\mathbb{D} = \mathrm{Im} \{ S_{_\mathrm{IF}}[\overline{\boldsymbol{x}}_L,\overline{\boldsymbol{x}}_R] \} \, , \quad \mathbb{P} =- \mathrm{Re} \{ S_{_\mathrm{IF}}[\overline{\boldsymbol{x}}_L, \overline{\boldsymbol{x}}_R] \} \, ,} \label{eq:decoherence_phase}
\end{align}
decoherence and phase shift, respectively.

These quantities can then be related to the expectation value of the projective measurements $\hat{P}_{\pm;\theta} = \ket{\pm;\theta}\bra{\pm;\theta}$, where $\ket{\pm;\theta} = \tfrac{1}{\sqrt{2}}(\ket{L} \pm e^{\mi \theta}\ket{R})$:
\begin{equation} \label{eq:proj_meas}
\lim_{\mathcal{T}\rightarrow\infty} \lim_{m_\mathrm{at}\rightarrow \infty}\mathrm{Tr}\left [ \hat{\rho}_s(\mathcal{T}) \hat{P}_{\pm;\theta} \right ] = \tfrac{1}{2} \left [ 1 \pm e^{-\mathbb{D}}\cos(\theta - \mathbb{P} )\right ] \, .
\end{equation} 
At leading order, these quantities can then be extracted from the total number of clicks in the plus or minus ports, e.g., 
\begin{equation}
\text{Expected number of counts in the `$-$' port} = \begin{cases}
\frac{1}{2}\mathbb{D}  + \mathcal{O}(g^3) & \text{for $\theta = 0$,} \\
\frac{1}{2}\left (1-\mathbb{P} \right)  + \mathcal{O}(g^3) & \text{for $\theta = \pi/2$,}
\end{cases}
\end{equation}
As a result, the shot-noise error on the count is Poissonian and, by repeating the experiment $N_\mathrm{exp}$ times, the error scales with $1/\sqrt{N_\mathrm{exp}}$.

Several features are worth emphasizing before specializing to a particular interferometer geometry. 
Firstly, in the off-diagonal components of the system's reduced density matrix, the two Schwinger--Keldysh branches are identified with the two arms of the interferometer. 
Secondly, $\mathbb{D}$ and $\mathbb{P}$ depend on different components of the system currents in the Keldysh basis. 
$\mathbb{D}$ only depends on $\varrho^\mathrm{q}$, while $\mathbb{P}$ depends on both $\varrho^\mathrm{cl}$ and $\varrho^\mathrm{q}$.  
This asymmetry between phase and contrast on the system side explains why, for a system of $N_\mathrm{at}$ initially uncorrelated atoms, only the phase shift can be enhanced by a factor of $N_\mathrm{at}$, as discussed in section~\ref{sec:oEFT_multi}.

The derivation above also clarifies how a MWI could detect interactions with the environment that would be invisible to classical detectors, which rely on recoil displacements or internal-state transitions.
In the heavy-probe limit, the atom's c.o.m. motion is frozen onto its classical trajectory. 
As a result, wavepacket spreading and recoil are both neglected at this order. 
Nevertheless, the reduced density matrix is modified via $\mathbb{D}$, which suppresses its off-diagonal elements, and $\mathbb{P}$, which introduces a phase shift. 
By contrast, a classical probe prepared in either the $L$ or $R$ arm would, at the same order, follow its unperturbed trajectory and register nothing at all, since its density matrix stays diagonal. 
It is precisely this sensitivity to phase shifts and loss of contrast, rather than to deposited energy or recoil displacements, that makes MWIs exquisitely sensitive probes of DM through ultrasoft scattering.

To find explicit expressions for Eq.~\eqref{eq:decoherence_phase}, it is necessary to specify a concrete interferometric geometry. 
Here, we limit our attention to the interferometric geometry depicted in Fig.~\ref{fig:paths}. 
In this case, $\boldsymbol{J}_R\!=\!-\boldsymbol{J}_L$ and
\begin{align}\label{eq:current}
\boldsymbol{J}_L(t)\!=\!\tfrac{m_\mathrm{at}\mathbf{v}}{2}\!&\Big[\!-\!\delta\left(t\!+\!\tfrac{T+T'}{2}\right)\!+\!\delta\left(t\!+\!\tfrac{T}{2}\right)
\!+\!\delta\left (t\!-\!\tfrac{T}{2}\right)\!-\!\delta \left (t\!-\!\tfrac{T+T'}{2}\right)\Big],
\end{align}
where $\mathbf{v}$ is the kick-induced atom velocity and $T_s=T+T'$ is the total interferometric time. 
The maximal spatial separation between the arms is $|\boldsymbol{\Delta x}|=|\mathbf{v}|T'/2$. 
In the limit $T\gg T'$, we obtain the split and hold geometry, used in trapped MWIs~\cite{Hughes:2009suspension,Xu2019}, Bordé interferometers~\cite{Borde1989} and other atom interferometer geometries~\cite{Ufrecht:2020ugr,Roura:2020qci,DiPumpo:2021redshift}. 

In the asymptotic time limit, corresponding to $\mathcal{T}\rightarrow \infty$, one can express the influence action (see Eq.~\eqref{eq:Sif_lim}) as an integral over the four momentum transfer $q$ (e.g., Eq.~\eqref{eq:Seff_gaussian_momentum}).  
In this limit, the left/right particular solutions (see Eq.~\eqref{eq:Xbar}), defined by the external currents of Eq.~\eqref{eq:current} in the limit $T\gg T'$, read as
\begin{equation}
\overline{\boldsymbol{x}}_{{\alpha}}(t) = \int dt' \frac{|t-t'|}{2m_\mathrm{at}} \boldsymbol{J}_{\alpha}(t') \approx
\begin{cases} - \boldsymbol{\Delta x}/2  &~\text{for $\alpha = L$ and $|t|\leq T/2$,}  \\ 
+ \boldsymbol{\Delta x}/2  &~\text{for $\alpha = R$ and $|t|\leq T/2$,} \\
0  &~\text{otherwise.}\
\end{cases}
\end{equation}
The 4-momentum FT of the system charge density for this geometry evaluates to
\begin{equation}\label{app:eq:syscurrent}
\widetilde\varrho(q;\overline{\boldsymbol{x}}_{{\alpha,\pm}})=\int\!\mathrm{d}^4x\,\varrho(\mathbf{x}\!-\!\overline{\boldsymbol{x}}_{{\alpha,\pm}}(t)) \, e^{\mi q\cdot x}\!\approx \!F_A(\mathbf{q}) 
\widetilde{\mathcal{W}}(q;\overline{\boldsymbol{x}}_{{\alpha,\pm}})
\, , 
\end{equation}
with 
\begin{equation}
\begin{aligned}
\widetilde{\mathcal{W}}(q;\overline{\boldsymbol{x}}_{{\alpha,\pm}}) = \begin{cases}
\deltabar(q_0)+T\,\mathrm{sinc}(q_0T/2)(e^{\mi\mathbf{q}\cdot\boldsymbol{\Delta x}/2}-1) &~\text{for $\alpha = L$,} \\
\deltabar(q_0)+T\,\mathrm{sinc}(q_0T/2)(e^{-\mi\mathbf{q}\cdot\boldsymbol{\Delta x}/2}-1) 
&~\text{for $\alpha = R$,} 
\end{cases}
\end{aligned}
\end{equation}
such that the decoherence and phase shift observables may be expressed as
\begin{equation}
\boxed{
\mathbb{D}=-\!\left(\frac{g}{\Lambda^D}\right)^{\!2}\!\!\int\!\dbar^4 q |F_A(\mathbf{q})|^2  \; T^2 \mathrm{sinc}^2\!\Bigl(\tfrac{q_0T}{2}\Bigr)\bigl[1-\cos(\mathbf{q}\!\cdot\!\boldsymbol{\Delta x})\bigr]\mathrm{Im}\bigl[\widetilde{K}_\mathrm{k}(q)\bigr]
+ \mathcal{O}(g^3) \, ,}
\label{eq:D_final}
\end{equation}
and
\begin{equation}
\boxed{
\mathbb{P}=\!\left(\frac{g}{\Lambda^D}\right)^{\!2}\!\!\int\! \dbar^4 q |F_A(\mathbf{q})|^2  \sin \Bigl(\tfrac{\mathbf{q} \cdot \boldsymbol{\Delta x}}{2}\Bigr)  \!\left[2T\deltabar(q_0)-4T^2\,\mathrm{sinc}^2\!\Bigl(\tfrac{q_0T}{2}\Bigr)\sin^2\!\Bigl(\tfrac{\mathbf{q} \cdot \boldsymbol{\Delta x}}{4}\Bigr)\right]\!\mathrm{Im}\bigl[\widetilde{K}_\mathrm{r}(q)\bigr]+ \mathcal{O}(g^3)\,,
}
\label{eq:P_final}
\end{equation}
respectively (see Appendix~\ref{app:sys_operators} for a derivation and for a discussion on other interferometric geometries).

Eqs.~\eqref{eq:D_final} and \eqref{eq:P_final} are the key result of this subsection. 
Decoherence is governed entirely by fluctuations of the environment, encoded in the noise kernel $\widetilde{K}_{\mathrm{k}}$ (see~Eq.~\eqref{eq:noise}), while $\mathbb{P}$ depends on the environment's response, captured by the retarded kernel $\widetilde{K}_{\mathrm{r}}$ (see~Eq.~\eqref{eq:retarded}). 
The remaining factors, which depend upon the interferometer geometry via the external currents $\boldsymbol{J}_{L,R}$, and whose time-dependence makes the system-environment interaction sensitive to the entire history of the system, have familiar interpretations. 
The term $1-\cos(\mathbf q\cdot\boldsymbol{\Delta x})$ in $\mathbb{D}$ is the which-path resolution factor~\cite{PhysRevA.42.38,Hornberger:2003umw}. 
For momentum transfers too small to resolve the arm separation, this factor goes to zero, since the environment cannot extract which-path information. 
In $\mathbb{P}$, $\sin(\mathbf q\cdot\boldsymbol{\Delta x}/2)$ encodes the odd, directional part of the response, which is only non-zero for anisotropic environments.
The function $\mathrm{sinc}^2\!\Bigl(\tfrac{q_0T}{2}\Bigr)$ captures the finite-time energy-resolution of the interferometer. 
It enforces energy conservation up to $1/T$, and reduces to an energy-conserving delta function in the Markovian limit, i.e., when $T$ is the longest time scale in the problem. 
As a result, this function governs the transition from the Markovian to the non-Markovian regime. 

\subsection{Beyond the heavy-probe limit}\label{sec:heavy-probe}

Before studying multi-atom platforms, we conclude by commenting on corrections to the heavy-probe limit, which can be organized as a series expansion in $1/m_\mathrm{at}$. 
To build physical intuition for these corrections, consider the case in which the environment transfers momentum $\mathbf{q}$ to the atom. 
Let $\mathbf{P}$ denote the atom's center-of-mass momentum, distributed with variance $\sigma_p^2$, where $\sigma_p$ is the momentum-space width of the system's wavepacket. 
The corresponding change in the atom's kinetic energy is
\begin{equation}
\Delta E_\mathrm{at} = \frac{|\mathbf{P}+\mathbf{q}|^2}{2m_\mathrm{at}} - \frac{|\mathbf{P}|^2}{2m_\mathrm{at}} = \frac{|\mathbf{q}|^2}{2m_\mathrm{at}} + \frac{\mathbf{P}\cdot\mathbf{q}}{m_\mathrm{at}} \, ,
\end{equation}
and is equivalent to the energy deposited by DM. 
At the level of the influence action (see Eq.~\eqref{eq:Sif_lim}), and therefore of the interferometric observables, the kinetic-energy shift $|\mathbf q|^2/2m_{\rm at}$ contributes a phase of order $|\mathbf q|^2|t-t'|/2m_{\rm at}$, while the momentum fluctuations contribute a typical phase of order $|\mathbf q|\sigma_p|t-t'|/m_\mathrm{at}$. 
Since the system currents restrict $|t-t'|\lesssim T$, while correlations in the environment kernel are suppressed for $|t-t'|\gtrsim\tau_e$, where $\tau_e$ is the time-scale of the environment, the relevant time separation is $\tau_{\rm eff}\equiv\min(T,\tau_e)$. 
As discussed in Appendix~\ref{app:finite-mass}, the two corrections are therefore controlled, up to factors of order unity, by
\begin{equation}
\epsilon_\varphi\sim \frac{|\mathbf q|^2\tau_{\rm eff}}{m_{\rm at}} \, ,
\qquad
\epsilon_\sigma \sim \left ( \frac{|\mathbf q| \sigma_p \tau_{\rm eff}}{m_\mathrm{at}}\right)^2 \, .
\end{equation}
Hence, the form of the influence action (see Eq.~\eqref{eq:h.p.}) entering the system's reduced density matrix at time $\mathcal{T}$ (see Eq.~\eqref{eq:rhos_step}) reads as
\begin{equation}
\left \langle S_{_\mathrm{IF}}\left [\overline{\boldsymbol{X}}_{\alpha,+} - \mi \tfrac{\boldsymbol{\delta}}{\boldsymbol{\delta j}_+},\overline{\boldsymbol{X}}_{\beta,-}+ \mi \tfrac{\boldsymbol{\delta}}{\boldsymbol{\delta j}_-}\right ] \right \rangle Z_0[\boldsymbol{j}_+,\boldsymbol{j}_-]\Big |_{\boldsymbol{j}_\pm = 0} = S_{_{\rm IF}}[\boldsymbol{\overline{x}}_{\alpha,+},\boldsymbol{\overline{x}}_{\beta,-}]  + \mathcal{O}(\epsilon_\varphi,\epsilon_\sigma ) \, .
\end{equation}
where $S_{_{\rm IF}}[\boldsymbol{\overline{x}}_{\alpha,+},\boldsymbol{\overline{x}}_{\beta,-}] $ is the influence action in the heavy-probe limit (see Eqs.~\eqref{eq:h.p.} and \eqref{eq:Sif_lim}).

For virialized dark matter, $v_0 \sim 10^{-3}$, $|\mathbf q|\lesssim m_\chi v_0$ and $\tau_e\simeq\tau_c=(m_\chi v_0^2)^{-1}$, giving
\begin{equation}
\epsilon_\varphi
\lesssim
\frac{m_\chi}{m_\mathrm{at}}
\min\!\left(1,\frac{T}{\tau_c}\right),
\qquad
\epsilon_\sigma
\lesssim
\left ( \frac{\sigma_p}{m_\mathrm{at} v_0} \right )^2 
\min\!\left(1,\frac{T^2}{\tau_c^2}\right).
\end{equation}
Thus, the heavy-probe approximation is controlled by $m_\chi\ll m_{\rm at}$ and $\sigma_p/ m_\mathrm{at} \ll v_0$, both of which are well satisfied in the applications considered here, where, e.g., $m_\chi \ll 1$~GeV and $\sigma_p \lesssim (100~\mu\mathrm{m})^{-1}$.

\section{Interferometer observables in many-atom systems}\label{sec:oEFT_multi}

Having established the map between MWI observables and the influence action for a single-atom system, we now construct the open EFT for a system of $N_\mathrm{at}$ atoms.
We distinguish between two classes of multi-atom MWIs, which differ in their initial state and, consequently, in the scaling of the observables with $N_\mathrm{at}$:
\begin{itemize}
  \item \textbf{Composite-particle interferometers (CPIs)}, which
    employ a \emph{mesoscopic bound object} (e.g.\ a silica or gold nanosphere) of total mass $M = N_\mathrm{at}\,m_\mathrm{at}$, in which all $N_\mathrm{at}$ constituents are approximately rigidly bound at lattice sites $\{\mathbf{r}_n\}$ relative to the center-of-mass. 
    The entire object is placed in a spatial superposition of its center-of-mass. 
    Proposed and existing CPI experiments employ optically trapped nanoparticles~\cite{Romero-Isart:2011prl,Bateman:2014natcomm,Neumeier:2024pnas,Roda-Llordes:2024prl}, magnetically levitated superconducting microspheres~\cite{Pino:2018erq}, and molecules~\cite{Nimmrichter:2011njp,Haslinger:2013natphys,Fein:2019natphys,Pedalino:2026nature}.
  \item \textbf{Atom interferometers (AIs)}, which employ a \emph{diffuse cloud} of $N_\mathrm{at}$ distinguishable\footnote{There also exist AIs employing an ensemble of $N_\mathrm{at}$ atoms in a BEC, e.g., Refs.~\cite{Aveline:2020kla,Frye_2021}, which we do not consider here.} and initially uncorrelated atoms of mass $m_\mathrm{at}$, each independently prepared in a spatial superposition. 
  This class encompasses light-pulse atom interferometers~\cite{Borde1989,Kasevich:1991zz,Riehle1991} and optically trapped atom interferometers~\cite{Xu2019,Panda2024}.
 \end{itemize}

Before we explore each class of experiments, a few general comments are in order.
The interaction Lagrangian is identical to the single-atom case (see~Eqs.~\eqref{eq:interaction_L} and \eqref{eq:nuclear-ff}). 
The multi-atom generalization enters exclusively through the action of the system operators in second-quantization on the many-body state.
Recalling that the quantized system operator appearing in the interaction Lagrangian is $\hat{O}_s(y) = \int d^3 r \varrho(\mathbf{r}) \sum_{\alpha} \hat \psi_\alpha^\dagger(\mathbf{y}-\mathbf{r})\,\hat \psi_\alpha(\mathbf{y}-\mathbf{r})$,\footnote{Here we assume that the object is composed of a single atomic species.}
 the action of the system operator on multiparticle states is given by
\begin{equation}\label{eq:action_Os_multi}
 \hat{O}_s(y) \prod_{n=1}^{N_\mathrm{at}}\hat \psi_\alpha^\dagger(\mathbf{x}+\mathbf{r}_n)\ket{0}_s = 
  \left [ \sum_{n=1}^{N_\mathrm{at}}\varrho (\mathbf{y}-\mathbf{x}-\mathbf{r}_n) \right ] 
    \prod_{n=1}^{N_\mathrm{at}}\hat \psi_\alpha^\dagger(\mathbf{x}+\mathbf{r}_n)\ket{0}_s \, ,
\end{equation}
such that $O_s \to \sum_{n=1}^{N_\mathrm{at}}\varrho (\mathbf{y}-\mathbf{x}-\mathbf{r}_n)$.

As shown in the following subsections, many-body effects enter entirely through the decoherence and phase form factors $F_\mathbb{D}(\mathbf{q})$ and $F_\mathbb{P}(\mathbf{q})$, which we collect in Table~\ref{tab:form-factors} and which all agree with previous results in the literature.
For the interferometer geometry introduced above, the leading-order decoherence and phase shift in the heavy-probe limit retain their single-atom structure, modified only by these form factors:
\begin{equation}
\boxed{
\begin{aligned}
  \mathbb{D}&= -\!\left(\tfrac{g}{\Lambda^D}\right)^{2}
  \int\!\dbar^4q |F_\mathbb{D}(\mathbf{q})|^2\;
  T^2 \mathrm{sinc}^2\!\left (\tfrac{q_0 T}{2}\right )
  \,
  \left [ 1-\cos(\mathbf{q}\cdot\boldsymbol{\Delta x})\right ]\,
  \mathrm{Im}\bigl[\widetilde{K}_{\mathrm{k}}(q)\bigr] + \mathcal{O}(g^3)  \, , \\
 \mathbb{P} &= \!\left(\tfrac{g}{\Lambda^D}\right)^{\!2}
  \!\int\!\dbar^4q 
  |F_\mathbb{P}(\mathbf{q})|^2 \sin \Bigl(\tfrac{\mathbf{q}\cdot\boldsymbol{\Delta x}}{2}\Bigr)\,
  \left[2T\deltabar(q_0)
    -4T^2\,\mathrm{sinc}^2\!\Bigl(\tfrac{q_0 T}{2}\Bigr)
    \sin^2\!\Bigl(\tfrac{\mathbf{q}\cdot\boldsymbol{\Delta x}}{4}\Bigr)
  \right]
  \mathrm{Im}\bigl[\widetilde{K}_{\mathrm{r}}(q)\bigr] + \mathcal{O}(g^3) \, .
\end{aligned}}\label{eq:int_obs}
\end{equation} 
Notably, for composite-particle interferometers (CPIs), the decoherence and phase shift form factors are identical. 
They both depend on the Debye-Waller form factor $e^{-2W(\mathbf q)}$, associated with the vibrational modes of a composite/bound mesoscopic object, and the form factor $|F_{\rm sphere}(\mathbf{q})|^2 \leq 1$, which depends on the (assumed to be spherical) geometry of the mesoscopic object. 
Hence, for $N_\mathrm{at}\gg1$, both phase shift and decoherence can inherit a large $N_\mathrm{at}^2$ enhancement in the limit of coherent interactions.\footnote{The atomic form factor is omitted here because the Debye-Waller form factor goes to zero for smaller values of $|\mathbf{q}|$.} 
For atom interferometers (AIs), instead, the decoherence and phase shift form factors differ. 
The decoherence and phase shift observables both depend on the atomic form factor $F_A(\mathbf{q})$, corresponding to the Helm form factor in the context of DM-nucleon interactions, while the phase shift observable is also sensitive to the total number of atoms in the sample via the cloud form factor $|F_{\rm cloud}(\mathbf{q})|^2 \leq 1$. 
Hence, in the limit of coherent interactions, only the phase shift can inherit a large $N_\mathrm{at}$ enhancement.

\begin{table*}[t!]
\begin{ruledtabular}
\begin{tabular}{lcc}
Form factor
 & CPI
 & AI \\
\colrule
$|F_\mathbb{D}(\mathbf q)|^2$
 & $e^{-2W(\mathbf q)} N_\mathrm{at}  \left[ 1+ (N_\mathrm{at}-1) \,|F_\mathrm{sphere}(\mathbf q)|^2 \right]$
 & $|F_A(\mathbf q)|^2$ \\[2mm]
$|F_\mathbb{P}(\mathbf q)|^2$
 & $e^{-2W(\mathbf q)} N_\mathrm{at} \left[ 1 + (N_\mathrm{at}-1) \,|F_\mathrm{sphere}(\mathbf q)|^2 \right]$
 & $|F_A(\mathbf q)|^2 \left[ 1 + (N_\mathrm{at}-1)\,|F_\mathrm{cloud}(\mathbf q)|^2 \right]$ \\[1mm]
\colrule\colrule
Observable & CPI & AI \\
\colrule
Number of counts
 & $\mathbb{D}_\mathrm{CPI},\ \mathbb{P}_\mathrm{CPI}$
 & $N_\mathrm{at}\,\mathbb{D}_\mathrm{AI},\ N_\mathrm{at}\,\mathbb{P}_\mathrm{AI}$ \\[2mm]
Shot-noise-limited sensitivity to $\mathbb{D}$, $\mathbb{P}$ \\
($N_\mathrm{exp}$ measurements)
 & $1/\sqrt{N_\mathrm{exp}}$
 & $1/\sqrt{N_\mathrm{at}N_\mathrm{exp}}$ \\
\end{tabular}
\end{ruledtabular}
\caption{
Many-body factors relevant to the results of this section, where $N_\mathrm{at}$ is the number of atoms in the target.
For composite-particle interferometers (CPIs), both the decoherence and phase shift form factors depend on the Debye-Waller form factor $e^{-2W(\mathbf q)}$ (see~Eq.~\eqref{eq:DW}) and $|F_{\rm sphere}(\mathbf{q})|^2$ (see~Eq.~\eqref{eq:sphere}), where the latter depends on the (assumed to be spherical) geometry of the mesoscopic object. 
For atom interferometers (AIs), instead, the decoherence and phase shift form factors differ: they both depend on the atomic form factor $F_A(\mathbf{q})$ (see Eq.~\eqref{eq:nuclear-ff}), but only the phase shift form factor is sensitive to the cloud form factor $|F_{\rm cloud}(\mathbf{q})|^2$~(see Eq.~\eqref{eq:AI_ff}). 
For both classes of MWI, phase and decoherence can be extracted from the expected number of counts in two output ports, corresponding to projective measurements in the qubit basis. 
Assuming shot-noise limited sensitivity, the error on $\mathbb{D}_\mathrm{AI}$ and $\mathbb{P}_\mathrm{AI}$ scales with $1/\sqrt{N_\mathrm{at}}$, while the error on $\mathbb{D}_\mathrm{CPI}$ and $\mathbb{P}_\mathrm{CPI}$ is $N_\mathrm{at}$ independent. 
For DM signals that are uncorrelated in time between successive measurements, both quantities receive a further $1/\sqrt{N_\mathrm{exp}}$ boost for $N_\mathrm{exp}$ independent measurements.
}
\label{tab:form-factors}
\end{table*}

\subsection{Composite-particle interferometers}\label{sec:oEFT_CPI}

In a composite-particle interferometer (CPI), the c.o.m. of the mesoscopic object is directly entangled with some measured degree of freedom, while the relative coordinates of the probe's constituents remain unobserved. 
It is therefore natural to write the worldline of the $n$-th atom on each Keldysh branch as
\begin{equation}
\boldsymbol X_{n,\pm}(t)
=
\boldsymbol{\mathcal X}_\pm(t)
+
\mathbf r_{n,\pm}(t),
\qquad 
\boldsymbol{\mathcal{X}}_{\pm}(t) = \frac{1}{N_\mathrm{at}}\sum_{n=1}^{N_\mathrm{at}} \boldsymbol{X}_{n,\pm}(t) \, ,
\qquad 
\sum_{n=1}^{N_\mathrm{at}}
m_\mathrm{at}\mathbf r_{n,\pm}(t)=0 \, ,
\label{eq:Xn_CPI_Jacobi}
\end{equation}
where $\boldsymbol{\mathcal{X}}_{\pm}$ is the worldline of the c.o.m., $\mathbf{r}_{n,\pm}$ is the coordinate of the $n$-th particle relative to the c.o.m., and $1\leq n \leq N_\mathrm{at}$. Hence, the CPI system currents in 3-momentum representation can be expressed as
\begin{equation}
\widetilde{\varrho}_{_\mathrm{CPI}}(\mathbf{q},t;\boldsymbol{\mathcal{X}}_\pm+\{\mathbf{r}_{n,\pm}\}) = \widetilde{\varrho}(\mathbf{q},t;\boldsymbol{\mathcal{X}}_\pm) \sum_{n=1}^{N_\mathrm{at}} e^{-\mi \mathbf{q}\cdot \mathbf{r}_{n,\pm}(t)} \, .
\end{equation}
Here, $\widetilde{\varrho}(\mathbf{q},t;\boldsymbol{\mathcal{X}}_\pm)$ are defined with respect to the time varying c.o.m. worldline, and are therefore structurally equivalent to the single-atom expressions under the replacement $\boldsymbol{X}_\pm\rightarrow \boldsymbol{\mathcal{X}}_\pm$ (see Eq.~\eqref{eq:rho_FT}). 

For translationally invariant binding interactions, which in a crystalline solid correspond to the lattice potential whose harmonic expansion yields the solid's phonon modes~\cite{kittel2005introduction}, this change of variables separates the free action into a c.o.m.\ sector and $N_\mathrm{at}-1$ internal Jacobi coordinates~\cite{greiner2012quantum}. 
As a result,  the c.o.m. and relative coordinates sectors decouple in the absence of system-environment interactions, and the external currents defining the left and right arms of the interferometer couple only to the c.o.m. coordinate. 
Therefore, after tracing over the relative coordinate sector, the 3-momentum system current bilinears entering the influence action read as
\begin{align}
\left \langle \widetilde{\varrho}_{_\mathrm{CPI}} (\mathbf{q},t;\boldsymbol{\mathcal{X}}_\pm+\{\mathbf{r}_{n,\pm}\}) \widetilde{\varrho}_{_\mathrm{CPI}} (-\mathbf{q},t';\boldsymbol{\mathcal{X}}_\pm+\{\mathbf{r}_{n,\pm}\}) \right \rangle_{\mathrm{rel}} & = \widetilde{\varrho} (\mathbf{q},t;\boldsymbol{\mathcal{X}}_\pm) \widetilde{\varrho} (-\mathbf{q},t';\boldsymbol{\mathcal{X}}_\pm) \mathcal{S}_{\mathrm{rel}}^{\pm\pm}(\mathbf{q},t-t') \, , \\
\left \langle \widetilde{\varrho}_{_\mathrm{CPI}} (\mathbf{q},t;\boldsymbol{\mathcal{X}}_\pm+\{\mathbf{r}_{n,\pm}\}) \widetilde{\varrho}_{_\mathrm{CPI}} (-\mathbf{q},t';\boldsymbol{\mathcal{X}}_\mp+\{\mathbf{r}_{n,\mp}\}) \right \rangle_{\mathrm{rel}} & = \widetilde{\varrho} (\mathbf{q},t;\boldsymbol{\mathcal{X}}_\pm) \widetilde{\varrho} (-\mathbf{q},t';\boldsymbol{\mathcal{X}}_\mp)\mathcal{S}_\mathrm{rel}^{\pm\mp}(\mathbf{q},t-t') \, ,
\end{align}
where the common relative-sector dependent quantity is the object's structure factor: 
\begin{equation}\label{eq:S_rel}
\begin{aligned}
\mathcal{S}_\mathrm{rel}^{\pm\pm}(\mathbf{q},t-t') = \left \langle \sum_{m,n=1}^{N_\mathrm{at}} e^{-\mi \mathbf{q} \cdot \left [\mathbf{r}_{n,\pm}(t)- \mathbf{r}_{m,\pm}(t') \right ]} \right \rangle_\mathrm{rel} \, , \, \,  
\mathcal{S}_\mathrm{rel}^{\pm\mp}(\mathbf{q},t-t') = \left \langle \sum_{m,n=1}^{N_\mathrm{at}} e^{-\mi \mathbf{q} \cdot \left [\mathbf{r}_{n,\pm}(t)- \mathbf{r}_{m,\mp}(t') \right ]} \right \rangle_\mathrm{rel} \, .
\end{aligned}
\end{equation}

To find a closed-form expression for the structure factor, we rewrite the relative coordinates as
\begin{equation}\label{eq:decomp}
\mathbf{r}_{n,\pm}(t) = \overline{\mathbf{r}}_n + \mathbf{u}_{n,\pm}(t) \, ,
\end{equation}
where $\overline{\mathbf{r}}_n$ are the static equilibrium positions, which are given by the minimum of the internal static binding potential, and $\mathbf{u}_{n,\pm}(t)$ are the time-dependent (and contour-dependent) fluctuations satisfying $\langle \mathbf{u}_{n,\pm}(t) \rangle_\mathrm{rel}=0$, as is the case for a solid in thermal equilibrium~\cite{Lin:2019uvt}. 

Hence, using Eq.~\eqref{eq:decomp}, Eq.~\eqref{eq:S_rel} may be expressed as 
\begin{equation}\label{eq:S_rel_f}
\begin{aligned}
\mathcal{S}^{\pm\pm}_\mathrm{rel}(\mathbf{q},t-t') &=  \sum_{m,n=1}^{N_\mathrm{at}} e^{-\mi \mathbf{q} \cdot \left (\overline{\mathbf{r}}_n- \overline{\mathbf{r}}_m \right )} \left \langle e^{-\mi \mathbf{q} \cdot (\mathbf{u}_{n,\pm}(t)-\mathbf{u}_{m,\pm}(t'))} \right \rangle_\mathrm{rel} \\
 & = \sum_{m,n=1}^{N_\mathrm{at}} e^{-\mi \mathbf{q} \cdot \left (\overline{\mathbf{r}}_n- \overline{\mathbf{r}}_m \right )} \left [ \left \langle e^{-\mi \mathbf{q} \cdot \mathbf{u}_n(t)} \right \rangle_\mathrm{rel}  \left \langle e^{\mi \mathbf{q} \cdot \mathbf{u}_m(t')} \right \rangle_\mathrm{rel} + \mathcal{O}(\text{phonon excitations}) \right ] \\
& = \underbrace{e^{-2W(\mathbf{q})}\sum_{m,n=1}^{N_\mathrm{at}} e^{-\mi \mathbf{q} \cdot \left (\overline{\mathbf{r}}_n- \overline{\mathbf{r}}_m \right )}}_{\mathcal{S}_\mathrm{rel}(\mathbf{q})}+ \mathcal{O}(\text{phonon excitations}) \, ,
\end{aligned}
 \end{equation}
 and similarly
\begin{equation}
\mathcal{S}^{\pm\mp}_\mathrm{rel}(\mathbf{q},t-t') = \mathcal{S}_\mathrm{rel}(\mathbf{q}) + \mathcal{O}(\text{phonon excitations}) \, ,
 \end{equation}
where the first term is the time-independent elastic contribution. 
The neglected terms describe inelastic processes, in which energy deposited by the environment excites phonons in the object~\cite{Griffin:2018bjn}. 

Here, $e^{-2W(\mathbf{q})}$ is the well-known Debye-Waller form factor~\cite{Squires_2012}, which quantifies the scattering rate suppression due to the mean relative spatial uncertainty of the atoms, whether from thermal or quantum fluctuations. 
For an isotropic and homogeneous solid at inverse temperature $\beta$ supporting phonon modes of energy $\omega_\mathrm{ph}$, 
\begin{equation}\label{eq:DW}
2W(\mathbf{q}) = \left \langle \left ( \mathbf{q}\cdot \mathbf{u}_n(t)\right )^2 \right \rangle_\mathrm{rel} = |\mathbf{q}|^2\sigma_u^2~,~\text{with}~\sigma_u^2 \simeq \frac{1}{\omega_\mathrm{ph} m_\mathrm{at}}\max \left ( 1, \frac{1}{\beta \omega_\mathrm{ph}}\right) \, ,
\end{equation}
where $\sigma_u^2$ characterizes the variance of the atoms' spatial fluctuations.
Therefore, for $|\mathbf{q}| \sigma_u \gg 1$, the Debye-Waller form factor goes to zero.  

The sum over the rigid positions of the atoms in the target, instead, depends on the geometry of the target, and can be expressed as
\begin{equation}
\sum_{m,n=1}^{N_\mathrm{at}} e^{-\mi \mathbf{q} \cdot \left (\overline{\mathbf{r}}_n- \overline{\mathbf{r}}_m \right )} = N_\mathrm{at} + \sum_{n\neq m} e^{-\mi\mathbf{q}\cdot(\overline{\mathbf{r}}_n-\overline{\mathbf{r}}_m)} \, ,
\end{equation} 
where the first and second terms account for incoherent and coherent effects, respectively. 
In the large $N_\mathrm{at}$ limit and assuming that the target is a uniform sphere of radius $r_c$ with normalized number density $n_\mathrm{sphere}(\mathbf{r}) = \frac{3}{4\pi r_c^3}\,\Theta(r_c-|\mathbf{r}|)$ the second term can be approximately rewritten as
\begin{equation}
\sum_{n\neq m} e^{-\mi\mathbf{q}\cdot(\overline{\mathbf{r}}_n-\overline{\mathbf{r}}_m)}\approx N_\mathrm{at}(N_\mathrm{at}-1) \!\left | \int\!\mathrm{d}^3r\;n_\mathrm{sphere}(\mathbf{r})\,e^{-\mi\mathbf{q}\cdot\mathbf{r}} \right |^2 = N_\mathrm{at}(N_\mathrm{at}-1) \,|F_\mathrm{sphere}(\mathbf{q})|^2 \, ,
\end{equation}
where the geometry is captured by the form factor
\begin{equation}\label{eq:sphere}
\boxed{
|F_\mathrm{sphere}(\mathbf{q})|^2= \left ( \frac{3\,j_1(|\mathbf{q}|r_c)}{|\mathbf{q}|r_c} \right )^2 
\, , } 
\end{equation}
with $j_1$ the spherical Bessel function of the first kind.

Neglecting phonon excitations, which require non-zero energy deposition, the classical-quantum bilinears appearing in the influence action hence take the simplified form
 \begin{align}
\left \langle \widetilde{\varrho}_{_\mathrm{CPI}}^\mathrm{q}(\mathbf{q},t;\boldsymbol{\mathcal{X}}_\pm+\{\mathbf{r}_{n,\pm}\}) \widetilde{\varrho}_{_\mathrm{CPI}}^\mathrm{cl} (-\mathbf{q},t';\boldsymbol{\mathcal{X}}_\pm+\{\mathbf{r}_{n,\pm}\}) \right \rangle_{\mathrm{rel}} & = \widetilde{\varrho}^\mathrm{q} (\mathbf{q},t;\boldsymbol{\mathcal{X}}_\pm) \widetilde{\varrho}^\mathrm{cl} (-\mathbf{q},t';\boldsymbol{\mathcal{X}}_\pm)\mathcal{S}_\mathrm{rel}(\mathbf{q})   \, , \\
\left \langle \widetilde{\varrho}_{_\mathrm{CPI}}^\mathrm{q} (\mathbf{q},t;\boldsymbol{\mathcal{X}}_\pm+\{\mathbf{r}_{n,\pm}\}) \widetilde{\varrho}_{_\mathrm{CPI}}^\mathrm{q} (-\mathbf{q},t';\boldsymbol{\mathcal{X}}_\pm+\{\mathbf{r}_{n,\pm}\}) \right \rangle_{\mathrm{rel}} & = \widetilde{\varrho}^\mathrm{q} (\mathbf{q},t;\boldsymbol{\mathcal{X}}_\pm) \widetilde{\varrho}^\mathrm{q} (-\mathbf{q},t';\boldsymbol{\mathcal{X}}_\pm)\mathcal{S}_\mathrm{rel}(\mathbf{q}) \, .
\end{align}
From this point, the heavy-probe derivation of the observables proceeds exactly as for the single-atom MWI of section~\ref{sec:toy}, by replacing the atom with the c.o.m. of the CPI, such that $m_\mathrm{at}\to M$. 
The c.o.m.\ wavepacket drops out, and the reduced density matrix takes the form of Eq.~\eqref{eq:rho_red_matrix}, with the influence functional evaluated on the classical c.o.m.\ trajectories $\boldsymbol{\overline{\mathcal{X}}}_\pm$. 
The composite initial state affects the result in two ways: the larger total mass $M = N_\mathrm{at} m_\mathrm{at}$ of the c.o.m. extends the validity of the heavy-probe approximation to larger momentum transfers, and therefore to larger DM masses, and the single-constituent form factor is replaced by its CPI counterpart. 
Both observables depend on the same form factor, $F_\mathbb{D}(\mathbf{q}) = F_\mathbb{P}(\mathbf{q}) = F_\mathrm{CPI}(\mathbf{q})$, given by the product of the geometric and elastic structure factors:
\begin{equation}
\boxed{
|F_{\mathrm{CPI}}(\mathbf{q})|^2 = e^{-2W(\mathbf q)} N_\mathrm{at} \left [ 1 + (N_\mathrm{at}-1) |F_{\rm sphere}(\mathbf{q})|^2 \right ] \, . } 
\end{equation}
Each factor has a simple interpretation: $F_\mathrm{sphere}$ describes the spatial arrangement of the atoms in the mesoscopic object, and $e^{-2W}$ suppresses the observables owing to fluctuations in the positions of the atoms in the solid.\footnote{Technically, this expression also depends on the atomic form factor $F_A(\mathbf{q})$, which depends on the charge distribution within each atom. However, this form factor is not relevant for the applications considered in this work.} 

The mapping between the observables $\mathbb{P}$ and $\mathbb{D}$ and the output port count is identical to the toy example considered earlier (see~Eq.~\eqref{eq:proj_meas}), e.g.,
\begin{equation}
\text{Expected number of counts in the `$-$' port} =\begin{cases}
\frac{1}{2}\mathbb{D}_\mathrm{CPI}  + \mathcal{O}(g^3) & \text{for $\theta = 0$,} \\
\frac{1}{2}\left (1-\mathbb{P}_\mathrm{CPI}  \right)  + \mathcal{O}(g^3) & \text{for $\theta = \pi/2$.}
\end{cases}
\end{equation}
Since the entire target either clicks in the left or right port, the shot-noise error on the count is Poissonian. 
Hence, by repeating the experiment $N_\mathrm{exp}$ times, the error on $\mathbb{D}_\mathrm{CPI}$ and $\mathbb{P}_\mathrm{CPI}$ scales with $1/\sqrt{N_\mathrm{exp}}$.
Furthermore, the single-atom limit is straightforwardly recovered for $N_\mathrm{at}=1$, where $F_\mathrm{sphere}=1$ identically, and these form factors agree with previous results in the literature~\cite{Riedel:2012ur,Riedel:2016acj,Du:2022ceh,Badurina:2024nge}. 
Notably, these form factors apply for both Markovian and non-Markovian dynamics, and are independent of the specific form of the DM operator and the DM occupation number.

Finally, note that our treatment neglected environment-induced phonon excitations which, if included, modify the form of the structure factor (see Eq.~\eqref{eq:S_rel_f}). We leave a study in this direction to future work.

\subsection{Atom interferometers}\label{sec:AI-CPI}

In an AI, $N_\mathrm{at}$ atoms are simultaneously interrogated, and each atom is independently prepared in a spatial superposition of states. 
This is qualitatively different from the CPI case: the atoms in the cloud are not bound into a single object whose c.o.m. is measured, but rather form a dilute ensemble of independently prepared probes.

Typically, the atoms are initially prepared in some trap and cooled to some temperature, such that the center of mass of each wavepacket is neither at rest nor located at the center of the trap. 
As a result, the atom cloud, which we assume to be dilute for simplicity, is expected to grow considerably during the interferometric sequence~\cite{Loriani:2018qej}.
However, it is possible to suppress atom cloud expansion during the interferometric sequence via matter-wave lensing techniques, such as delta-kick collimation, which effectively reduces the momentum spread of the cloud at the expense of a larger initial size~\cite{Chu:86,PhysRevLett.78.2088}.  
For simplicity, we therefore specialize to atomic fountains employing matter-wave lensing, which allow the atoms to freely propagate in spacetime during the interferometric sequence with negligible cloud expansion. 
Assuming that the c.o.m. of the $n$-th atom is located at $\overline{\mathbf{r}}_n$, in direct analogy to the equilibrium position of the atoms in a CPI (see~Eq.~\eqref{eq:decomp}), with $P(\{\overline{\mathbf{r}}_{n}\})$ the joint classical probability distribution over $\{\overline{\mathbf{r}}_n\}$,  the initial system density matrix can be expressed as
\begin{equation}\label{eq:state_AI}
  \hat{\rho}_{{s}}(t_i)
  = \!\left(\prod_{n=1}^{N_\mathrm{at}}\int \mathrm{d}^3 \overline{r}_{n}
    P\bigl(\{\overline{\mathbf{r}}_{n}\}\bigr) \right)
    \bigotimes_{n=1}^{N_\mathrm{at}}
    \ket{\Phi_{n}}\!\bra{\Phi_{n}} \,,
\end{equation}
where
\begin{equation}\label{eq:ket_AI}
\begin{aligned}
  \ket{\Phi_{n}}
      & = \frac{1}{\sqrt{2}}\int\!\mathrm{d}^3y\;
    \Phi(\mathbf{y})
    \bigl(\psi_L^\dagger(\mathbf{y}+\overline{\mathbf{r}}_{n})+\psi_R^\dagger(\mathbf{y}+\overline{\mathbf{r}}_{n})\bigr)\ket{0}_s 
   \end{aligned}
\end{equation}
is the single-atom state centered at $\overline{\mathbf{r}}_{n}$.\footnote{As for the single atom case, here $\Phi(\mathbf{y})$ is the atomic wavepacket centered at the origin (see~Eq.~\eqref{eq:initial_state}).}
For uncorrelated atoms drawn independently from a cloud of characteristic size $r_c$, the joint distribution factorizes,
\begin{equation}\label{eq:cloud_factorisation}
  P\bigl(\{\overline{\mathbf{r}}_{n}\}\bigr)
  = \prod_{n=1}^{N_\mathrm{at}} P(\overline{\mathbf{r}}_{n}) \,,
\end{equation}
where $P(\overline{\mathbf{r}}_{n})$ is the $n$-th atom's c.o.m. distribution, centered at the origin.\footnote{To capture the expansion of the cloud, it is necessary to consider the distribution $P(\overline{\mathbf{r}}_{n},\mathbf{p}_{0,n})$, where $\mathbf{p}_{0,n}$ is the initial momentum of the atoms, which yields a time-dependent atomic cloud form factor~\cite{Murgui:2025unt}.} 

As a result, the influence action depends on $N_\mathrm{at}$ pairs of worldlines $\{\boldsymbol{X}_{n,\pm}(t)\}$, such that the system currents take the form
\begin{equation}
\begin{aligned}\label{eq:AI_sys}
\varrho^\mathrm{cl}(x;\{\boldsymbol{X}_{n,\pm}+\overline{\mathbf{r}}_{n}\}) &  = \tfrac{1}{2} \sum_{n=1}^{N_\mathrm{at}}  \left [\varrho(\mathbf{x}-\boldsymbol{X}_{n,+}(t)-\overline{\mathbf{r}}_{n})+\varrho(\mathbf{x}-\boldsymbol{X}_{n,-}(t)-\overline{\mathbf{r}}_{n}) \right ] \, ,\\
\varrho^\mathrm{q}(x;\{\boldsymbol{X}_{n,\pm}+\overline{\mathbf{r}}_{n}\}) & = \sum_{n=1}^{N_\mathrm{at}}  \left [\varrho(\mathbf{x}-\boldsymbol{X}_{n,+}(t)-\overline{\mathbf{r}}_{n})-\varrho(\mathbf{x}-\boldsymbol{X}_{n,-}(t)-\overline{\mathbf{r}}_{n}) \right ] \, ,
\end{aligned}
\end{equation}
and the $(\{\alpha\},\{\beta\})$-component of the reduced density matrix for the $N_\mathrm{at}$-atom system at time $t_f$, where $\{\alpha\} = \{\alpha_1,\alpha_2,...,\alpha_{N_\mathrm{at}}\}$, reads as
\begin{equation}\label{eq:AI_PI}
\begin{aligned}
\bra{\{\alpha\}}\hat{\rho}_s(t_f)\ket{\{\beta\}}= \rho^{\{\alpha\},\{\beta\}}_{s}(t_f) & \equiv \frac{1}{2^{N_\mathrm{at}}} \prod_{n=1}^{N_\mathrm{at}} \left ( \int\!\!\mathrm{d}^3 x_{i,n,+}\int \mathrm{d}^3 x_{i,n,-} \Phi \left (\mathbf{x}_{i,n,+} \right ) \Phi^*\left (\mathbf{x}_{i,n,-} \right ) \right )\!\!\,  \\ 
& \qquad \qquad \times 
\prod_{n=1}^{N_\mathrm{at}} \left (\int\!\!\mathrm{d}^3 x_{f,n} \int_{\mathbf{x}_{i,n,+}}^{\mathbf{x}_{f,n}}  \!\!\!\!\mathcal{D}\boldsymbol{X}_{n,+}\!\!\int_{\mathbf{x}_{i,n,-}}^{\mathbf{x}_{f,n}}\!\!\mathcal{D}\boldsymbol{X}_{n,-} \right )
\\ 
& \qquad \qquad  \qquad \times \exp \left \{\mi \sum_{n=1}^{N_\mathrm{at}} S_s[\boldsymbol{X}_{n,+},\boldsymbol{J}_{\alpha_n,+}]-\mi \sum_{n=1}^{N_\mathrm{at}}S_s[\boldsymbol{X}_{n,-},\boldsymbol{J}_{\beta_n,-}] \right \}\, \\
& \qquad \qquad  \qquad \times 
\exp \left \{\mi S_{_\mathrm{IF}}[\{\boldsymbol{X}_{n,+}+\overline{\mathbf{r}}_{n}
\},\{\boldsymbol{X}_{n,-}+\overline{\mathbf{r}}_{n}
\}] \right \}  \, ,
\end{aligned}
\end{equation}
where, in direct analogy to Eq.~\eqref{eq:rho_PI_worldline}, $\mathbf{x}_{i,n,\pm}$ and $\mathbf{x}_{f,n}$ are the initial and final worldline configurations of the $n$-th atom, and $\boldsymbol{J}_{\alpha_n,\pm}$ is the external current acting on the $n$-th atom's worldline.
Notably, the free system part of Eq.~\eqref{eq:AI_PI} has the same structure as the single-atom path integral. 
The atom's initial position $\overline{\mathbf{r}}_{n}$, which differs from the single-atom expressions, only appears in the influence action.
Consequently, we can safely invoke the heavy-probe limit and the asymptotic time limit ${t_f=-t_i = \mathcal{T}\rightarrow \infty}$ as we justified in the single-atom case. 
In this case, exactly following the arguments of section~\ref{sec:toy}, the atomic wavepackets and the free system actions drop out of the reduced density matrix, and the atom worldlines reduce to $\boldsymbol{X}_{n,\pm}(t')\rightarrow \overline{\boldsymbol{x}}_{n,\pm}(t')$, where $\overline{\boldsymbol{x}}_{n,\pm}(t')$ is the particular solution of the classical worldlines driven by $\boldsymbol{J}_{\alpha_n,\pm}(t')$. 

Differently from CPIs, typical AI measurements rely on one-body observables, i.e., counting the number of atoms clicking in one of two ports in the experiment~\cite{abend2019atom}. 
The measurement is thus given by $\sum_{m=1}^{N_\mathrm{at}} \hat{P}^{(m)}_{\pm;\theta}$, where $\hat{P}^{(m)}_{\pm;\theta}$ corresponds to performing a projective measurement in the qubit basis on the $m$-th atom. 
Hence, the expectation value of the one-body observable is
\begin{equation}\label{eq:meas_AI}
\mathrm{Tr}\bigl[\hat{\rho}_\mathrm{s}(\mathcal{T}) \sum_{m=1}^{N_\mathrm{at}} \hat{P}^{(m)}_{\pm;\theta}\bigr] = N_\mathrm{at}\mathrm{Tr}\bigl[\hat{\rho}^{(m)}_\mathrm{s}(\mathcal{T})\hat{P}^{(m)}_{\pm;\theta}\bigr] \, , 
\end{equation}
where $\hat{\rho}^{(m)}_\mathrm{s}$ is the reduced density matrix of the $m$-th atom, obtained by tracing out the remaining $N_\mathrm{at}-1$ atoms. 
 As a result, we can focus on the reduced single-atom density matrix.

Crucially, the trace over $N_\mathrm{at}-1$ atoms impacts the structure of the influence functional. 
Operationally, this corresponds to setting $\alpha_{m'}= \beta_{m'}$ for all $m' \neq m$, such that the sum over the path labels of the remaining atoms includes $2^{N_\mathrm{at}-1}$ different terms, as shown in Refs.~\cite{Badurina:2024nge,Murgui:2025unt}. 
Equivalently, taking the trace corresponds to setting $\boldsymbol{J}_{\alpha_{m'}, +} = \boldsymbol{J}_{\beta_{m'}, -}$ for all $m' \neq m$ in the path integral, for $\alpha_{m'}\in\{L,R\}$, such that the heavy-probe limit of Eq.~\eqref{eq:AI_sys} reads as
\begin{equation}\label{eq:AI_sys_m}
\begin{aligned}
\varrho^\mathrm{q}(x';\{\overline{\boldsymbol{x}}_{n,\pm}+\overline{\mathbf{r}}_{n}
\}) & \to \varrho^\mathrm{q}(x';\overline{\boldsymbol{x}}_{m,\pm}+\overline{\mathbf{r}}_{m}
) \, , \\
 \varrho^\mathrm{cl}(x';\{\overline{\boldsymbol{x}}_{n,\pm}+\overline{\mathbf{r}}_{n}\}) & \to \varrho^\mathrm{cl}(x';\overline{\boldsymbol{x}}_{m,\pm}+\overline{\mathbf{r}}_{m}
) + \sum_{n\neq m} \varrho(x';\overline{\boldsymbol{x}}_{n}+\overline{\mathbf{r}}_{n}
) \, .
\end{aligned}
\end{equation}
Substituting the FT of Eq.~\eqref{eq:AI_sys_m} into the influence functional, three bilinear structures appear. 
Firstly, the $\widetilde{K}_{\mathrm{k}}$-dependent piece, which always gives rise to decoherence, involves only the contribution from the $m$-th atom, thereby yielding a result that is equivalent to the single-atom result. 
The $\widetilde{K}_{\mathrm{r,a}}$-dependent pieces, instead, depend both on a purely $m$-th atom-dependent contribution and on a cross term, which depends on a sum over $N_\mathrm{at}-1$ atoms. 
Using these results, and always working in the heavy-probe and asymptotic time regime, we find that the $m$-th atom's reduced density matrix is given by
\begin{equation}\label{eq:rho_AI}
\begin{aligned}
\lim_{\mathcal{T}\rightarrow \infty} \lim_{m_\mathrm{at}\rightarrow \infty}\hat{\rho}_{s}^{(m)}(\mathcal{T}) & = \frac{1}{2} \begin{pmatrix} e^{\mi \langle S_\mathrm{IF}[\overline{\boldsymbol{x}}_{L},\overline{\boldsymbol{x}}_{L}]\rangle_0 } & e^{\mi \langle  S_\mathrm{IF}[\overline{\boldsymbol{x}}_{L},\overline{\boldsymbol{x}}_{R}]\rangle_0 } \\ e^{\mi \langle S_\mathrm{IF}[\overline{\boldsymbol{x}}_{R},\overline{\boldsymbol{x}}_{L}] \rangle_0 } & e^{\mi \langle  S_\mathrm{IF}[\overline{\boldsymbol{x}}_{R},\overline{\boldsymbol{x}}_{R}]\rangle_0 } \end{pmatrix} \\
& \equiv \frac{1}{2} \begin{pmatrix} 1 & e^{-\mathbb{D}_\mathrm{AI} - \mi \mathbb{P}_\mathrm{AI}} \\ e^{-\mathbb{D}_\mathrm{AI} + \mi \mathbb{P}_\mathrm{AI}} & 1 \end{pmatrix} +\mathcal{O}(g^3)~,
\end{aligned}
\end{equation}
where $\langle \cdot \rangle_0$ is the weighted average over the statistical properties of the system, with decoherence and phase shift given by 
\begin{equation}\label{eq:D_AI}
  \mathbb{D}_\mathrm{AI} = -T^2\!\left(\tfrac{g}{\Lambda^D}\right)^{\!2}
  \int\!\dbar^4q\;
  \mathrm{sinc}^2\!\Bigl(\tfrac{q_0 T}{2}\Bigr)\,
  |F_A(\mathbf{q})|^2\,
  \bigl(1-\cos(\mathbf{q}\cdot\boldsymbol{\Delta x})\bigr)\,
  \mathrm{Im}\bigl[\widetilde{K}_{\mathrm{k}}(q)\bigr] \, ,
\end{equation}
and
\begin{equation}\label{eq:P_AI}
  \mathbb{P}_\mathrm{AI} = 
  T\!\left(\tfrac{g}{\Lambda^D}\right)^{\!2}
  \!\int\!\dbar^4q\;
  \sin\!\Bigl(\tfrac{\mathbf{q}\cdot\boldsymbol{\Delta x}}{2}\Bigr)\,
  |F_\mathrm{AI}(\mathbf{q})|^2
  \left[2\deltabar(q_0)
    -4T\,\mathrm{sinc}^2\!\Bigl(\tfrac{q_0 T}{2}\Bigr)
    \sin^2\!\Bigl(\tfrac{\mathbf{q}\cdot\boldsymbol{\Delta x}}{4}\Bigr)
  \right]
  \mathrm{Im}\bigl[\widetilde{K}_{\mathrm{r}}(q)\bigr] \, ,
\end{equation} 
respectively. 
Notably, while the decoherence observable only depends on the atomic form factor, the phase shift observable depends on the AI form factor 
\begin{equation}\label{eq:AI_ff}
\boxed{
  |F_\mathrm{AI}(\mathbf{q})|^2 \equiv |F_A(\mathbf{q})|^2 \left [ 1+(N_\mathrm{at}-1) |F_\mathrm{cloud}(\mathbf{q})|^2 \right ]
  \, , } 
\end{equation}
where $F_\mathrm{cloud}(\mathbf{q}) = \widetilde{P}(\mathbf{q})$ is the cloud form factor. 
For a Gaussian pdf with spread $r_c$, the modulus squared of the cloud form factor is given by 
\begin{equation}
\boxed{
|F_\mathrm{cloud}(\mathbf{q})|^2 = e^{-|\mathbf{q}|^2 r_c^2} 
\, . }
\end{equation}
By construction, $|F_\mathrm{cloud}(\mathbf{0})|^2 = 1$, and $|F_\mathrm{AI}(\mathbf{q})|^2 = |F_A(\mathbf{q})|^2$ for $N_\mathrm{at}=1$. 
Hence, only the phase shift can inherit an $N_\mathrm{at}$-enhancement, whereas the decoherence is always $N_\mathrm{at}$-independent.

As a result of the measurement protocol (see~Eq.~\eqref{eq:meas_AI}), which gives rise to the single-atom reduced density matrix (see Eq.~\eqref{eq:rho_AI}), the observable number of counts in either detection port is directly related to Eqs.~\eqref{eq:D_AI}-\eqref{eq:P_AI}, e.g.,
\begin{equation}
\text{Expected number of counts in the `$-$' port} = N_\mathrm{at} \times \begin{cases}
\frac{1}{2}\mathbb{D}_\mathrm{AI}  + \mathcal{O}(g^3) & \text{for $\theta = 0$,} \\
\frac{1}{2}\left (1-\mathbb{P}_\mathrm{AI} \right)  + \mathcal{O}(g^3) & \text{for $\theta = \pi/2$.}
\end{cases}
\end{equation}
Since each atom can now independently click in either port, in contrast with the CPI case, the shot-noise error on the count is Poissonian and scales with $1/\sqrt{N_\mathrm{at}}$ per shot. Hence, by repeating the experiment $N_\mathrm{exp}$ times, the error on $\mathbb{D}_\mathrm{AI}$ and $\mathbb{P}_\mathrm{AI}$ scales with $1/\sqrt{N_\mathrm{at}N_\mathrm{exp}}$.

The structure of the form factors, which agrees with Ref.~\cite{Badurina:2024nge} in the limit of a point-like cloud and with  Ref.~\cite{Murgui:2025unt} in the context of a static cloud, reveals a clear asymmetry between unitary and non-unitary effects. 
$\mathbb{D}_{\mathrm{AI}}$ is $N_\mathrm{at}$-independent, while $\mathbb{P}_\mathrm{AI}$ is at most linear in $N_\mathrm{at}$, as first shown in Ref.~\cite{Badurina:2024nge} in the context of coherent collisional decoherence.\footnote{Note that AIs are technically sensitive to a forward amplitude, or wake force~\cite{VanTilburg:2024xib,Day:2023mkb}, first identified in the context of elastic DM scattering in MWIs~\cite{Badurina:2024nge}. Although naively appearing as a $\mathcal{O}(g^2)$ phase shift effect, this effect appears in the observable at $\mathcal{O}(g^4)$.} 
Therefore, when the environment couples incoherently to individual atoms in the cloud, no correlations are generated among the atoms, and both the phase shift and the loss of contrast remain independent of $N_\mathrm{at}$.
Hence, in this regime the signal in an AI employing $N_\mathrm{at}$ atoms is equivalent, through $\mathcal{O}(g^2)$, to that of a single-atom interferometer. 
The sensitivity, however, is that of a single-atom interferometer repeated $N_\mathrm{at}$ times, since each atom provides an independent measurement (see Eq.~\eqref{eq:meas_AI}). 
In the coherent regime, by contrast, the phase acquires an $N_\mathrm{at}$-dependent contribution.
This arises because in this regime the environment mediates a force between the atoms in the cloud.

We conclude by highlighting that these expressions remain independent of the precise coupling between the environment and the system. 
They are also consistent with the general bounds of Ref.~\cite{Bogorad:2026ggm}.

\section{Spin-independent sub-GeV dark matter--nucleon scattering}\label{sec:DM-kernels}

We now turn our attention to computing the signatures of concrete DM models in MWIs.
In particular, we will focus on elastic spin-independent DM-nucleon interactions, where DM and nucleon fields are represented by $\chi$ and $N$, respectively. The two classes of models considered here include: (i) \textit{contact} interactions, i.e.,
\begin{equation}\label{eq:contact}
\mathcal{L} \supset - \frac{g}{\Lambda^2} \overline{N} N O_\chi \, ,
\end{equation}
and (ii) \textit{long-range} interactions mediated by a light hadrophilic real scalar $\phi$ of mass $m_\phi$, i.e.,
\begin{equation}\label{eq:long}
\mathcal{L} \supset -y_n \phi \overline{N} N  -y_\chi  \phi O_\chi \, ,
\end{equation}
where $y_n$ and $y_\chi$ are dimensionless coupling constants. In both cases the DM bilinears are given by
\begin{equation}\label{eq:DM_bilinears}
O_\chi = \begin{cases}
m_\chi \chi ^2 &   \text{for real scalar DM,}\\
\overline{\chi} \chi &  \text{for Dirac fermionic DM.} 
\end{cases}
\end{equation}
Clearly, $O_s\equiv \overline{N} N$ and $\{\chi,\phi\}$ is the environment field content. 

Equipped with these models, we can explicitly compute the entries of the environment matrix kernel $\widetilde{\mathsf{K}}$ (see the FT of Eqs.~\eqref{eq:retarded}-\eqref{eq:noise}), which directly enter the interferometer observables (see Eq.~\eqref{eq:int_obs}). 
Notably, for contact interactions the environment kernels can be understood in terms of in-in bubble diagrams involving only $\chi$ fields. 
In section~\ref{sec:contact} we show that, for single-component DM with $m_\chi \lesssim 10$~eV, a subset of these diagrams gives rise to Bose enhancements, which parametrically increase the decoherence signal while keeping the phase observable unchanged. 
In section~\ref{sec:long}, we show that, for long-range interactions, the same kernels correspond to \textit{dressed} in-in $\phi$ propagators, where the dressing is provided by the background of on-shell DM particles. 
Crucially, for sufficiently large DM-mediator coupling constants $y_\chi$ and small DM mass $m_\chi$, the DM environment, even in the case of a small DM subcomponent, introduces previously ignored corrections to the mediator's mass that can dramatically alter the phenomenology of these candidates. 
In particular, for sufficiently small mediator masses and hadrophilic interactions, the effective mediator mass in the DM medium becomes tachyonic, signalling a collective instability.

\subsection{Contact interactions}\label{sec:contact}

We first consider the case of contact interactions, distinguishing between fermionic and bosonic DM. 
In particular, we will show a simple, diagrammatic way of computing the environment-kernels using in-in bubble diagrams. 

Recall from section~\ref{sec:OQS_SK} that, before tracing out environmental degrees of freedom, the total action contains the interaction term 
\begin{equation}\label{eq:inin_int}
\begin{aligned}
S_\mathrm{int}[O_{s,+},O_{\chi,+}]-S_\mathrm{int}[O_{s,-},O_{\chi,-}] &\supset -\frac{g}{\Lambda^2} \left ( O_{s,+},O_{\chi,+} - O_{s,-},O_{\chi,-} \right ) \\
& = -\frac{g}{\Lambda^2} \left ( O_{s}^\mathrm{cl}O_{\chi}^\mathrm{q} + O_{s}^\mathrm{q}O_{\chi}^\mathrm{cl} \right ) \, ,
\end{aligned}
\end{equation}
where in the last line we used the Keldysh basis (see Eq.~\eqref{eq:cl/k}).
Writing out the environment currents explicitly in terms of $\chi$ fields in the cl-q basis then allows one to determine the environment kernels appearing in the phase and decoherence observables. 
Since, at $\mathcal{O}(g^2)$, $\mathbb{P}$ and $\mathbb{D}$ are controlled by $(\tfrac{g}{\Lambda^2})^2\,\widetilde{O}_s^\mathrm{cl}\widetilde{O}_s^\mathrm{q}\,\widetilde{K}_{r,a}$ and $(\tfrac{g}{\Lambda^2})^2\,\widetilde{O}_s^\mathrm{q}\widetilde{O}_s^\mathrm{q}\,\widetilde{K}_{k}$, respectively (see~Eqs.~\eqref{eq:int_obs}), the kernels, which were defined in terms of operators in Eqs.~\eqref{eq:retarded}-\eqref{eq:noise}, are diagrammatically obtained by joining appropriate combinations of two of the vertices above with $\chi$ in-in propagators (see~Eqs.~\eqref{eq:GreenBoson} and \eqref{eq:GreenFermion}). 
In detail, $\widetilde{K}_{r,a}$ is obtained from one $\widetilde{O}_s^\mathrm{cl}$ and one $\widetilde{O}_s^\mathrm{q}$ vertex, and $\widetilde{K}_{k}$ is obtained from two $\widetilde{O}_s^\mathrm{q}$ vertices (see Eq.~\eqref{eq:Seff_gaussian_momentum}), where in both cases the system currents are stripped off. 
The assignment of the internal lines is fixed by the $\mathrm{cl}/\mathrm{q}$ labels at each end, with each internal line corresponding to the appropriate in-in propagator. 
By sewing together $\chi$ fields using in-in propagators one obtains bubble diagrams.

\begin{itemize}

\item \textbf{Real scalar dark matter}

Using Eq.~\eqref{eq:cl/k} and the real scalar model in Eq.~\eqref{eq:DM_bilinears}, $O_\chi^\mathrm{cl,q}$ can be expressed in terms of $\chi^\mathrm{cl,q}$ fields such that Eq.~\eqref{eq:inin_int} reads as
\begin{equation}\label{eq:Feynman_scalar}
\begin{aligned}
S[O_s^\mathrm{cl}, O_s^\mathrm{q}, \chi^\mathrm{cl}, \chi^\mathrm{q}]  &\supset
-m_\chi \frac{g}{\Lambda^2}
\left(
2 O^\mathrm{cl}_s \chi^\mathrm{cl}\chi^\mathrm{q}
+
O^\mathrm{q}_s \chi^\mathrm{cl}\chi^\mathrm{cl}
+
\frac{1}{4}O^\mathrm{q}_s \chi^\mathrm{q}\chi^\mathrm{q}
\right)
\\[4pt]
&=
\begin{tikzpicture}[baseline=-0.5ex, scale=1.05]
  \draw[very thick, black, double, double distance=1pt]
    ({-sqrt(0.585^2+0.403^2)},0)
    node[left] {$O_s^\mathrm{cl}$} -- (0,0);
  \draw[very thick, black] (0,0) -- (0.585,0.403)
    node[right] {$\chi^\mathrm{cl}$};
  \draw[very thick, black, dotted] (0,0) -- (0.585,-0.403)
    node[right] {$\chi^\mathrm{q}$};
  \filldraw[black] (0,0.0) circle (1.6pt);
  \node[below=6pt] at (current bounding box.south)
    {$-2 m_\chi \frac{g}{\Lambda^2}$};
\end{tikzpicture}
\;+\;
\begin{tikzpicture}[baseline=-0.5ex, scale=1.05]
  \draw[very thick, black, dotted, double, double distance=1pt]
    ({-sqrt(0.585^2+0.403^2)},0)
    node[left] {$O_s^\mathrm{q}$} -- (0,0);
  \draw[very thick, black] (0,0) -- (0.585,0.403)
    node[right] {$\chi^\mathrm{cl}$};
  \draw[very thick, black] (0,0) -- (0.585,-0.403)
    node[right] {$\chi^\mathrm{cl}$};
  \filldraw[black] (0,0.0) circle (1.6pt);
  \node[below=6pt] at (current bounding box.south)
    {$-2 m_\chi \frac{g}{\Lambda^2}$};
\end{tikzpicture}
\;+\;
\begin{tikzpicture}[baseline=-0.5ex, scale=1.05]
  \draw[very thick, black, dotted, double, double distance=1pt]
    ({-sqrt(0.585^2+0.403^2)},0)
    node[left] {$O_s^\mathrm{q}$} -- (0,0);
  \draw[very thick, black, dotted] (0,0) -- (0.585,0.403)
    node[right] {$\chi^\mathrm{q}$};
  \draw[very thick, black, dotted] (0,0) -- (0.585,-0.403)
    node[right] {$\chi^\mathrm{q}$};
  \filldraw[black] (0,0.0) circle (1.6pt);
  \node[below=6pt] at (current bounding box.south)
    {$-\frac{1}{2}m_\chi \frac{g}{\Lambda^2}$};
\end{tikzpicture}
\, ,
\end{aligned}
\end{equation}
where in the second line we wrote down the in-in Feynman diagrams for each vertex and represented q- and cl-fields with dotted and solid lines, respectively. 
Single and double lines depict $\chi$ fields and $O_s$ currents, respectively.
At the level of the effective action (obtained after tracing out $\chi$), one can easily determine the FT of the environment kernels using the FT of the free-theory in-in $\chi$ propagators (cf. Eqs.~\eqref{eq:GreenBoson}): 
\begin{equation}
\begin{aligned}
\widetilde{G}^{^\chi}_\mathrm{a}(p) &= -\mi  \int \mathrm{d}^4 x \left\langle\!\!\left\langle \chi^\mathrm{q}(x)\chi^\mathrm{cl}(0) \right\rangle\!\!\right\rangle  e^{\mi p\cdot x} \, \\
& =  \dfrac{1}{(p_0 - \mi \epsilon)^2-E_{\mathbf{p}}^2}  = \mathrm{P}\left [ \frac{1}{p_0^2-E_\mathbf{p}^2}\right ] - \frac{\mi}{4E_\mathbf{p}}\left[\deltabar(p_0+E_\mathbf{p})-\deltabar(p_0-E_\mathbf{p})\right ]\, \\
& \equiv \begin{tikzpicture}[baseline=-0.5ex,scale=1.2]
  \draw[very thick,dotted] (0.24,0) -- (0.75,0);
  \draw[very thick] (0.75,0) -- (1.25,0);
\end{tikzpicture} , \\
\widetilde{G}^{^\chi}_\mathrm{r}(p) &=  -\mi  \int \mathrm{d}^4 x \left\langle\!\!\left\langle \chi^\mathrm{cl}(x)\chi^\mathrm{q}(0) \right\rangle\!\!\right\rangle  e^{\mi p\cdot x}  \,  \\
& = \dfrac{1}{(p_0 + \mi \epsilon)^2-E_{\mathbf{p}}^2}  = \mathrm{P}\left [ \frac{1}{p_0^2-E_\mathbf{p}^2}\right ] + \frac{\mi}{4E_\mathbf{p}}\left[\deltabar(p_0+E_\mathbf{p})-\deltabar(p_0-E_\mathbf{p})\right ] \, \\
& \equiv \begin{tikzpicture}[baseline=-0.5ex,scale=1.2]
  \draw[very thick] (0.24,0) -- (0.75,0);
  \draw[very thick,dotted] (0.75,0) -- (1.25,0);
\end{tikzpicture} \, , \\
\widetilde{G}^{^\chi}_\mathrm{k}(p) & = -\mi \int \mathrm{d}^4 x  \left\langle\!\!\left\langle \chi^\mathrm{cl}(x)\chi^\mathrm{cl}(0) \right\rangle\!\!\right\rangle e^{\mi p\cdot x} \\
 & = -\frac{\mi }{4E_{\mathbf p}}
\left[
(1+2 n_\chi(\mathbf p))\deltabar(p_0-E_{\mathbf p})
+
(1+2 n_\chi(-\mathbf p))\deltabar(p_0+E_{\mathbf p})
\right] \, \\
& \equiv  
\begin{tikzpicture}[baseline=-0.5ex,scale=1.2]
  \draw[very thick] (0.24,0) -- (1.25,0);
\end{tikzpicture} 
\, ,
\end{aligned}
\end{equation}
where $\mathrm{P}[\cdot]$ is the principal value, momentum flows from the left to the right,  $E_\mathbf{p} = \sqrt{m_\chi^2+|\mathbf{p}|^2}$, and the DM occupation per mode is $n_\chi(\mathbf{p}) \simeq \rho_\chi f_\chi(\mathbf{p})/m_\chi$. 
Since $\langle\!\langle \chi^\mathrm{q}\chi^\mathrm{q}\rangle\!\rangle = 0$ for bosonic fields, no bubble diagrams containing continuously dotted lines between vertices exist.
Since $S_\mathrm{IF} \supset (\tfrac{g}{\Lambda^2})^2\,\widetilde{O}_s^\mathrm{cl}\widetilde{O}_s^\mathrm{q}\,\widetilde{K}_\mathrm{r}$ (see~Eqs.~\eqref{eq:Seff_gaussian_momentum}), $\widetilde{K}_\mathrm{r}$ is consequently obtained by sewing together the $\chi$ legs from the first and second vertex in Eq.~\eqref{eq:Feynman_scalar}.  
Similarly, because $S_\mathrm{IF} \supset (\tfrac{g}{\Lambda^2})^2\,\widetilde{O}_s^\mathrm{q}\widetilde{O}_s^\mathrm{q}\,\widetilde{K}_\mathrm{k}$, $\widetilde{K}_\mathrm{k}$ is given by contractions of $\chi$ from the last two vertices in Eq.~\eqref{eq:Feynman_scalar}.
Hence, given the interaction term in Eq.~\eqref{eq:Feynman_scalar} and the in-in free theory $\chi$ propagators, the only non-vanishing one-loop and system-source-stripped diagrams contributing to the observables  (see~Eq.~\eqref{eq:int_obs}) are 
\begin{equation}\label{eq:bubble_scalar}
\begin{aligned}
\left (\tfrac{g}{\Lambda^2} \right )^2 \widetilde{K}_\mathrm{r}(q)&=  \mi \, \, \begin{tikzpicture}[baseline=-0.5ex, scale=1.0]
  \draw[very thick] (0,0.0) arc (180:0:0.80);
  \draw[very thick,dotted] (1.60,0) arc (0:-90:0.80);
  \draw[very thick] (0.80,-0.8) arc (-90:-180:0.80);
  \filldraw[black] (0,0.0) circle (1.6pt);
  \filldraw[black] (1.6,0.0) circle (1.6pt);
\end{tikzpicture} 
= \mi \left ( 2 m_\chi \tfrac{g}{\Lambda^2} \right )^2 \!\! \int \dbar^4 p \,  \widetilde{G}^\chi_\mathrm{k}(p)\widetilde{G}^\chi_\mathrm{r}(q-p)
 \equiv
 \widetilde{\Pi}^{\chi}_{\mathrm{r}}(q)  \, ,  \\
\left ( \tfrac{g}{\Lambda^2} \right )^2 \widetilde{K}_\mathrm{k}(q) &  = \mi \, \, \begin{tikzpicture}[baseline=-0.5ex, scale=1.0]
  \draw[very thick] (0,0) arc (180:0:0.80);
  \draw[very thick] (1.60,0) arc (0:-180:0.80);
  \filldraw[black] (0,0.0) circle (1.6pt);
  \filldraw[black] (1.6,0.0) circle (1.6pt);
\end{tikzpicture} 
+\mi \, \, \begin{tikzpicture}[baseline=-0.5ex, scale=1.0]
  \draw[very thick,dotted] (1.60,0) arc (0:90:0.80);
  \draw[very thick] (0.80,0.8) arc (90:180:0.80);
  \draw[very thick,dotted] (1.60,0) arc (0:-90:0.80);
  \draw[very thick] (0.80,-0.8) arc (-90:-180:0.80);
  \filldraw[black] (0,0.0) circle (1.6pt);
  \filldraw[black] (1.6,0.0) circle (1.6pt);
  \end{tikzpicture} 
  +\mi \, \, \begin{tikzpicture}[baseline=-0.5ex, scale=1.0]
  \draw[very thick] (1.60,0) arc (0:90:0.80);
  \draw[very thick,dotted] (0.80,0.8) arc (90:180:0.80);
  \draw[very thick] (1.60,0) arc (0:-90:0.80);
  \draw[very thick,dotted] (0.80,-0.8) arc (-90:-180:0.80);
  \filldraw[black] (0,0.0) circle (1.6pt);
  \filldraw[black] (1.6,0.0) circle (1.6pt);
  \end{tikzpicture} 
\\
&= \tfrac{\mi}{2} \left ( 2 m_\chi \tfrac{g}{\Lambda^2} \right )^2 \!\!  \int \dbar^4 p  \left [ \widetilde{G}^\chi_\mathrm{k}(p)\widetilde{G}^\chi_\mathrm{k}(q-p) + \tfrac{1}{4} \widetilde{G}^\chi_\mathrm{r}(p)\widetilde{G}^\chi_\mathrm{r}(q-p) + \tfrac{1}{4} \widetilde{G}^\chi_\mathrm{a}(p)\widetilde{G}^\chi_\mathrm{a}(q-p)  \right ] 
\\
& \equiv 
\widetilde{\Pi}^{\chi}_{\mathrm{k}}(q) \, , 
\end{aligned}
\end{equation}
where in the last diagrams we applied the symmetry factor of $1/2$. Note that these can also be computed using the operator definition of the noise kernels (see the FT of Eqs.~\eqref{eq:retarded}-\eqref{eq:noise}). 

\item \textbf{Dirac fermion dark matter} 

In the case of fermionic dark matter, we rewrite the fermions in the Larkin-Ovchinnikov field basis (see Eq.~\eqref{eq:Larkin}) such that, from Eqs.~\eqref{eq:contact} and \eqref{eq:DM_bilinears} the closed theory action will contain the terms
\begin{equation}\label{eq:Feynman_fermion}
\begin{aligned}
& S[O_s^\mathrm{cl}, O_s^\mathrm{q}, \chi^\mathrm{cl},  \chi^\mathrm{q}, \overline{\chi}^\mathrm{cl},\overline{\chi}^\mathrm{q}]  \supset
-\frac{g}{\Lambda^2}
\left(
O^\mathrm{cl}_s \overline{\chi}^\mathrm{cl}\chi^\mathrm{cl}
+
O^\mathrm{cl}_s \overline{\chi}^\mathrm{q}\chi^\mathrm{q}
+
\tfrac{1}{2}O^\mathrm{q}_s \overline{\chi}^\mathrm{cl}\chi^\mathrm{q}
+
\tfrac{1}{2}O^\mathrm{q}_s \overline{\chi}^\mathrm{q}\chi^\mathrm{cl}
\right)
\\[4pt]
& \qquad \qquad  =
\begin{tikzpicture}[baseline=-0.5ex, scale=1.05]
  \draw[very thick, black, double, double distance=1pt]
    ({-sqrt(0.585^2+0.403^2)},0)
    node[left] {$O_s^\mathrm{cl}$} -- (0,0);
  \draw[very thick, black] (0,0) -- (0.585,0.403)
    node[right] {$\overline{\chi}^\mathrm{cl}$};
  \draw[very thick, black] (0,0) -- (0.585,-0.403)
    node[right] {$\chi^\mathrm{cl}$};
  \midarrow{0.2925}{0.2015}{214.6}
  \midarrow{0.2925}{-0.2015}{-34.6}
  \filldraw[black] (0,0.0) circle (1.6pt);
  \node[below=6pt] at (current bounding box.south)
    {$-\tfrac{g}{\Lambda^2}$};
\end{tikzpicture}
\;+\;
\begin{tikzpicture}[baseline=-0.5ex, scale=1.05]
  \draw[very thick, black, double, double distance=1pt]
    ({-sqrt(0.585^2+0.403^2)},0)
    node[left] {$O_s^\mathrm{cl}$} -- (0,0);
  \draw[very thick, black, dotted] (0,0) -- (0.585,0.403)
    node[right] {$\overline{\chi}^\mathrm{q}$};
  \draw[very thick, black, dotted] (0,0) -- (0.585,-0.403)
    node[right] {$\chi^\mathrm{q}$};
  \midarrow{0.2925}{0.2015}{214.6}
  \midarrow{0.2925}{-0.2015}{-34.6}
  \filldraw[black] (0,0.0) circle (1.6pt);
  \node[below=6pt] at (current bounding box.south)
    {$-\tfrac{g}{\Lambda^2}$};
\end{tikzpicture}
\;+\;
\begin{tikzpicture}[baseline=-0.5ex, scale=1.05]
  \draw[very thick, black, dotted, double, double distance=1pt]
    ({-sqrt(0.585^2+0.403^2)},0)
    node[left] {$O_s^\mathrm{q}$} -- (0,0);
  \draw[very thick, black] (0,0) -- (0.585,0.403)
    node[right] {$\overline{\chi}^\mathrm{cl}$};
  \draw[very thick, black, dotted] (0,0) -- (0.585,-0.403)
    node[right] {$\chi^\mathrm{q}$};
  \midarrow{0.2925}{0.2015}{214.6}
  \midarrow{0.2925}{-0.2015}{-34.6}
  \filldraw[black] (0,0.0) circle (1.6pt);
  \node[below=6pt] at (current bounding box.south)
    {$-\tfrac{1}{2}\tfrac{g}{\Lambda^2}$};
\end{tikzpicture}
\;+\;
\begin{tikzpicture}[baseline=-0.5ex, scale=1.05]
  \draw[very thick, black, dotted, double, double distance=1pt]
    ({-sqrt(0.585^2+0.403^2)},0)
    node[left] {$O_s^\mathrm{q}$} -- (0,0);
  \draw[very thick, black, dotted] (0,0) -- (0.585,0.403)
    node[right] {$\overline{\chi}^\mathrm{q}$};
  \draw[very thick, black] (0,0) -- (0.585,-0.403)
    node[right] {$\chi^\mathrm{cl}$};
  \midarrow{0.2925}{0.2015}{214.6}
  \midarrow{0.2925}{-0.2015}{-34.6}
  \filldraw[black] (0,0.0) circle (1.6pt);
  \node[below=6pt] at (current bounding box.south)
    {$-\tfrac{1}{2}\tfrac{g}{\Lambda^2}$};
\end{tikzpicture}
\, ,
\end{aligned}
\end{equation}
where single lines denote $\chi$ particles and antiparticles, with arrows pointing toward and away from vertices for particles and antiparticles, respectively.
Keeping the spinor indices $A,B$ explicit, the free theory fermionic propagators (see~Eq.~\eqref{eq:GreenFermion}) are given by 
\begin{equation}
\begin{aligned}
[\widetilde{G}^{^\chi}_\mathrm{a}(p)]_{_{AB}} &= -\mi  \int d^4 x \Big\langle\!\!\Big\langle \chi^\mathrm{q}(x) \overline{\chi}^\mathrm{q}(0)  \Big\rangle\!\!\Big\rangle_{_{AB}} e^{\mi p \cdot x} \\
&=  \frac{(\slashed{p}+m_\chi)_{_{AB}}}{(p_0 - \mi\epsilon)^2-E_\mathbf{p}^2} = (\slashed{p}+m_\chi)_{_{AB}}\left \{ \mathrm{P}\left [\frac{1}{p_0^2-E_{\mathbf{p}}^2} \right] - \frac{\mi}{4E_\mathbf{p}}\left[\deltabar(p_0+E_\mathbf{p})-\deltabar(p_0-E_\mathbf{p})\right]\right \}  \\
& = \begin{tikzpicture}[baseline=-0.5ex,scale=1.2]
  \draw[very thick,dotted] (0.24,0) -- (0.75,0);
  \draw[very thick,dotted] (0.75,0) -- (1.25,0);
  \rarrowhead{0.75}{0}
  \end{tikzpicture} \, \, ,
  \end{aligned}
\end{equation}
\begin{equation}
\begin{aligned}
[\widetilde{G}^{^\chi}_\mathrm{r}(p)]_{_{AB}} &= -\mi  \int d^4 x \left\langle\!\!\left\langle \chi^\mathrm{cl}(x) \overline{\chi}^\mathrm{cl}(0)  \right\rangle\!\!\right\rangle_{_{AB}} e^{\mi p \cdot x} \\ 
& = \frac{(\slashed{p}+m_\chi)_{_{AB}}}{(p_0 + \mi\epsilon)^2-E_\mathbf{p}^2} = (\slashed{p}+m_\chi)_{_{AB}}\left \{ \mathrm{P}\left [\frac{1}{p_0^2-E_{\mathbf{p}}^2} \right] + \frac{\mi}{4E_\mathbf{p}}\left[\deltabar(p_0+E_\mathbf{p})-\deltabar(p_0-E_\mathbf{p})\right]\right \}  \\
& = \begin{tikzpicture}[baseline=-0.5ex,scale=1.2]
  \draw[very thick] (0.24,0) -- (0.75,0);
  \draw[very thick] (0.75,0) -- (1.25,0);
  \rarrowhead{0.75}{0}
\end{tikzpicture} \, \, , 
\end{aligned}
\end{equation}
\begin{equation}
\begin{aligned}
[\widetilde{G}^{^\chi}_\mathrm{k}(p)]_{_{AB}} &= -\mi  \int d^4 x \left\langle\!\!\left\langle \chi^\mathrm{cl}(x) \overline{\chi}^\mathrm{q}(0)  \right\rangle\!\!\right\rangle_{_{AB}} e^{\mi p \cdot x} \\  
& = -\frac{\mi}{2E_\mathbf{p}} (\slashed{p}+m_{_\chi})_{_{AB}} \left [ \deltabar(p_0-E_\mathbf{p}) (1-n_{_\chi}(\mathbf{p})) + \deltabar(p_0+E_\mathbf{p}) \right ] \, \\
& = \begin{tikzpicture}[baseline=-0.5ex,scale=1.2]
  \draw[very thick] (0.24,0) -- (0.75,0);
  \draw[very thick,dotted] (0.75,0) -- (1.25,0);
  \rarrowhead{0.75}{0}
\end{tikzpicture}
\, ,
\end{aligned}
\end{equation}
where $n_\chi(\mathbf{p})$ is the \textit{spin-summed} occupation number per three-momentum $\mathbf{p}$, and we assumed that the environment has no antiparticles in the initial state. 
Note that only the two-point function $\langle\!\langle \chi^\mathrm{q}\overline{\chi}^\mathrm{cl}\rangle\rangle$ vanishes for fermions.
Thus, with our diagrammatic conventions, any diagram containing a propagator whose arrow points from a dotted segment to a solid segment vanishes.
Since $S_\mathrm{IF} \supset (\tfrac{g}{\Lambda^2})^2\,\widetilde{O}_s^\mathrm{q}\widetilde{O}_s^\mathrm{cl}\,\widetilde{K}_\mathrm{r}$ and $S_\mathrm{IF} \supset (\tfrac{g}{\Lambda^2})^2\,\widetilde{O}_s^\mathrm{q}\widetilde{O}_s^\mathrm{q}\,\widetilde{K}_\mathrm{k}$ (see~Eqs.~\eqref{eq:Seff_gaussian_momentum}), for the interactions in Eq.~\eqref{eq:Feynman_fermion}, the non-vanishing diagrams contributing to the observables (see~Eq.~\eqref{eq:int_obs}) read as
\begin{equation}\label{eq:bubble_fermion}
\begin{aligned}
\left ( \tfrac{g}{\Lambda^2} \right )^2 \widetilde{K}_\mathrm{r}(q)&=  -\mi \, \, \begin{tikzpicture}[baseline=-0.5ex, scale=1.0]
  \draw[very thick,dotted] (0,0.0) arc (180:90:0.80);
  \draw[very thick] (0.80,0.8) arc (90:-0:0.80);
  \larrowhead{0.80}{0.80}
  \draw[very thick] (1.60,0.0) arc (0:-90:0.80);
  \draw[very thick] (0.80,-0.8) arc (-90:-180:0.80);
  \rarrowhead{0.80}{-0.80}
  \filldraw[black] (0,0) circle (1.6pt);
  \filldraw[black] (1.6,0) circle (1.6pt);
\end{tikzpicture}
- \mi \, 
\begin{tikzpicture}[baseline=-0.5ex, scale=1.0]
  \draw[very thick,dotted] (0,0.0) arc (180:90:0.80);
  \draw[very thick,dotted] (0.80,0.8) arc (90:-0:0.80);
  \larrowhead{0.80}{0.80}
  \draw[very thick,dotted] (1.60,0.0) arc (0:-90:0.80);
  \draw[very thick] (0.80,-0.8) arc (-90:-180:0.80);
  \rarrowhead{0.80}{-0.80}
\filldraw[black] (0,0) circle (1.6pt);
  \filldraw[black] (1.6,0) circle (1.6pt);
\end{tikzpicture}  \\
&= -\tfrac{\mi}{2} \, \left (\tfrac{g}{\Lambda^2} \right )^2 \int \dbar^4 p \,\mathrm{tr} \big [\widetilde{G}^{^\chi}_{\mathrm{k}}(p) \widetilde{G}^{^\chi}_{\mathrm{r}}(p+q) + \widetilde{G}^{^\chi}_{\mathrm{a}}(p)\widetilde{G}^{^\chi}_{\mathrm{k}}(p+q) \big ]
 \equiv 
 \widetilde{\Pi}^{\chi}_{\mathrm{r}}(q)  \, ,  \\
\left ( \tfrac{g}{\Lambda^2} \right )^2 \widetilde{K}_\mathrm{k}(q) &  = -\mi \,  \, \begin{tikzpicture}[baseline=-0.5ex, scale=1.0]
  \draw[very thick,dotted] (0,0.0) arc (180:90:0.80);
  \draw[very thick] (0.80,0.8) arc (90:-0:0.80);
  \larrowhead{0.80}{0.80}
  \draw[very thick,dotted] (1.60,0.0) arc (0:-90:0.80);
  \draw[very thick] (0.80,-0.8) arc (-90:-180:0.80);
  \rarrowhead{0.80}{-0.80}
\filldraw[black] (0,0) circle (1.6pt);
  \filldraw[black] (1.6,0) circle (1.6pt);
\end{tikzpicture}
-\mi \,  \,
\begin{tikzpicture}[baseline=-0.5ex, scale=1.0]
  \draw[very thick] (0,0.0) arc (180:90:0.80);
  \draw[very thick] (0.80,0.8) arc (90:-0:0.80);
  \larrowhead{0.80}{0.80}
  \draw[very thick,dotted] (1.60,0.0) arc (0:-90:0.80);
  \draw[very thick,dotted] (0.80,-0.8) arc (-90:-180:0.80);
  \rarrowhead{0.80}{-0.80}
\filldraw[black] (0,0) circle (1.6pt);
  \filldraw[black] (1.6,0) circle (1.6pt);
\end{tikzpicture}
- \mi \,  \,
\begin{tikzpicture}[baseline=-0.5ex, scale=1.0]
  \draw[very thick,dotted] (0,0.0) arc (180:90:0.80);
  \draw[very thick,dotted] (0.80,0.8) arc (90:-0:0.80);
  \larrowhead{0.80}{0.80}
  \draw[very thick] (1.60,0.0) arc (0:-90:0.80);
  \draw[very thick] (0.80,-0.8) arc (-90:-180:0.80);
  \rarrowhead{0.80}{-0.80}
\filldraw[black] (0,0) circle (1.6pt);
  \filldraw[black] (1.6,0) circle (1.6pt);
\end{tikzpicture}  \\
& = -\tfrac{\mi}{4} \left ( \tfrac{g}{\Lambda^2} \right )^2 \! \! \int \dbar^4 p \, \mathrm{tr} \! \left [ \widetilde{G}^\chi_\mathrm{k}(p)\widetilde{G}^\chi_\mathrm{k}(p+q)
+ \widetilde{G}^\chi_\mathrm{r}(p)\widetilde{G}^\chi_\mathrm{a}(p+q)
+ \widetilde{G}^\chi_\mathrm{a}(p)\widetilde{G}^\chi_\mathrm{r}(p+q) \right ]
\\
& \equiv 
\widetilde{\Pi}^{\chi}_{\mathrm{k}}(q) \, ,
\end{aligned}
\end{equation}
where $\mathrm{tr}[\cdot]$ is the trace over spinor indices and the overall minus sign comes from the fermion loop.
\end{itemize}

Equipped with the form of the environment kernels, we now invoke the Markovian limit, corresponding to the regime in which the time scale of the environment is much shorter than the system's time scale. 
In our application, the DM coherence time $\tau_c$ sets the environmental timescale, while the interferometric/interrogation time $T$ sets the system's timescale.  
For DM masses $m_\chi \gtrsim 10^{-7}$~eV, $\tau_c \ll T \sim \mathcal{O}(1~\mathrm{s})$, such that the environment kernels can be approximated as being effectively local in time, i.e., $\mathsf{K}(x-x')\propto\delta(t-t')$. 
Formally, this is equivalent to taking the limit $T\rightarrow\infty$, which implies the distributional identity
\begin{equation}\label{eq:identity}
T^2 \mathrm{sinc}^2\left ( \frac{q_0 T}{2}\right ) \rightarrow T \deltabar(q_0) \, .
\end{equation} 
Under this identification, and after performing the $q_0$ integral, Eq.~\eqref{eq:int_obs} reads as
\begin{equation}
\begin{aligned}
  \mathbb{D}&= -T \int\!\dbar^3 q |F_\mathbb{D}(\mathbf{q})|^2
  \,
  \left [ 1-\cos(\mathbf{q}\cdot\boldsymbol{\Delta x})\right ]\,
  \mathrm{Im}\bigl[\widetilde{\Pi}^\chi_{\mathrm{k}}(0,\mathbf{q})\bigr] + \mathcal{O}(g^3)\, , \\
 \mathbb{P} &= T \!\int\!\dbar^3 q 
  |F_\mathbb{P}(\mathbf{q})|^2 \sin \left (\mathbf{q}\cdot\boldsymbol{\Delta x}\right )\,
  \mathrm{Im}\bigl[\widetilde{\Pi}^\chi_{\mathrm{r}}(0,\mathbf{q})\bigr] + \mathcal{O}(g^3) \, .
\end{aligned}
\end{equation} 
Notably, the emergence of the energy-conserving delta function dictates that open-system dynamics requires no energy deposition from the environment. 

Since DM in the laboratory is non-relativistic, the imaginary part of the environment kernels (see Eqs.~\eqref{eq:bubble_scalar} and \eqref{eq:bubble_fermion} for scalar and Dirac fermion DM, respectively) can be approximated by 
\begin{equation}\label{eq:imaginary_Pi}
\begin{aligned}
\mathrm{Im}\left [ \widetilde{\Pi}^\chi_\mathrm{r}(0,\mathbf{q})\right ] & \approx \tfrac{1}{2} \left(\tfrac{g}{\Lambda^2}\right)^{2}   \int \dbar^3p \, \left [n_\chi(\mathbf{p})-n_\chi(\mathbf{p-q}) \right ] \deltabar(E_\mathbf{p}-E_\mathbf{p-q}) \, , \\
\mathrm{Im}\left [ \widetilde{\Pi}^\chi_\mathrm{k}(0,\mathbf{q}) \right ] & \approx -\tfrac{1}{2} \left(\tfrac{g}{\Lambda^2}\right)^{2} \int \dbar^3p \Big [ n_\chi(\mathbf{p}) S_\chi(\mathbf{p-q}) + n_\chi(\mathbf{p-q}) S_\chi(\mathbf{p})\Big ] \deltabar(E_\mathbf{p}-E_\mathbf{p-q}) \, ,
\end{aligned}
\end{equation}
 where we introduced the (spin-statistics dependent) DM statistical factor
\begin{equation}
\boxed{
\begin{aligned}
S_\chi(\mathbf{p-q}) &= \begin{cases}  1 + n_\chi(\mathbf{p-q}) \quad \text{for bosonic DM,} \\
1 - \tfrac{1}{2} n_\chi(\mathbf{p-q}) \quad \text{for fermionic DM.}
\end{cases}
\end{aligned}
}
\end{equation}  
Hence, the decoherence and phase shift observables may be expressed as
\begin{empheq}[box=\fbox]{align}
\mathbb{D}&=T\!\left(\tfrac{g}{\Lambda^2}\right)^{\!2}\!\!\int\! \dbar^3q \, |F_\mathbb{D}(\mathbf{q})|^2 [1-\cos(\mathbf{q}\!\cdot\!\boldsymbol{\Delta x})] g_{_\mathbb{D}}(\mathbf{q})\,,\label{eq:D_scatt}\\
\mathbb{P}&=T\!\left(\tfrac{g}{\Lambda^2}\right)^{\!2}\!\!\int\! \dbar^3q \, |F_\mathbb{P}(\mathbf{q})|^2 \sin(\mathbf{q}\!\cdot\!\boldsymbol{\Delta x}) g_{_\mathbb{P}}(\mathbf{q})\,,\label{eq:P_scatt}
\end{empheq}
with $F_\mathbb{D,P}$ the detector-specific decoherence/phase shift form factor (see~Table~\ref{tab:form-factors}), with the atomic form factor $F_A(\mathbf{q}) \propto A$ replaced by the nuclear or Helm form factor~\cite{PhysRev.104.1466}, where $A$ is the atomic mass number, and 
\begin{empheq}[box=\fbox]{align}
g_{_\mathbb{D}}(\mathbf{q}) &= \int\! \dbar^3p \, n_\chi(\mathbf{p}) S_\chi(\mathbf{p-q}) \deltabar(E_\mathbf{p}\!-\!E_{\mathbf{p}-\mathbf{q}}) \, , \label{eq:g_d_def}  \\
g_{_\mathbb{P}}(\mathbf{q}) &= \int\! \dbar^3p \, n_\chi(\mathbf{p}) \deltabar(E_\mathbf{p}\!-\!E_{\mathbf{p}-\mathbf{q}}) \, . \label{eq:g_p_def}
\end{empheq}
Introducing the vector $\hat{\mathbf{q}}=\mathbf{q}/|\mathbf{q}|$, the latter function can be expressed in terms of the DM velocity distribution assuming the SHM (see Eq.~\eqref{eq:SHM}) as~\cite{Trickle:2019nya}
\begin{equation}\label{eq:g_p}
\begin{aligned}
g_{_\mathbb{P}}(\mathbf{q}) &= \frac{\rho_\chi}{m_\chi} \frac{1}{ |\mathbf{q}|} \underbrace{\int d^3 v f_\chi(\mathbf{v})\deltabar(\mathbf{v}\cdot\hat{\mathbf{q}}-|\mathbf{q}|/2m_\chi)}_{g(\mathbf{q})} \\
& = \frac{\rho_\chi}{m_\chi} \frac{2\pi^2v_0^2}{N_0 |\mathbf{q}|} \left [ \exp \left ( - v_\mathrm{min}(\mathbf{q})^2/v_0^2 \right) -\exp \left (-v_\mathrm{esc}^2/v_0^2 \right ) \right  ] \, ,
\end{aligned}
\end{equation}
with 
\begin{equation}\label{eq:vmin}
v_\mathrm{min}(\mathbf{q}) = \min \left ( \boldsymbol{v}_\oplus\cdot\hat{\mathbf{q}}+|\mathbf{q}|/2m_\chi , v_\mathrm{esc} \right) \, .
\end{equation}

\begin{figure}[t]
\includegraphics[width=0.99\linewidth]{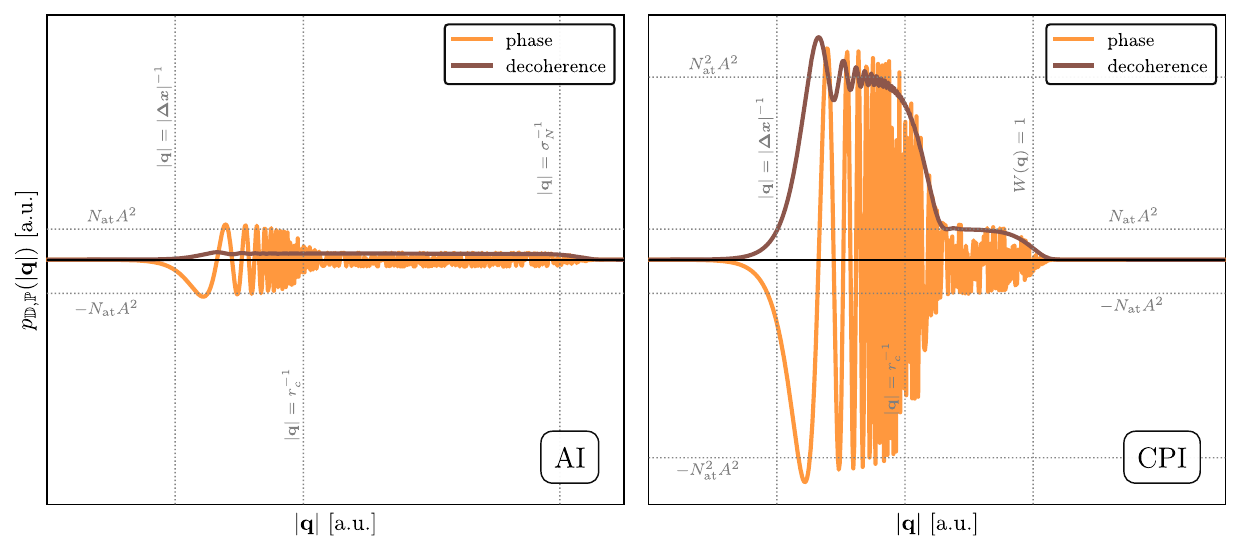}
\caption{Detector-dependent normalized DM effect (see Eq.~\eqref{eq:norm_DM_eff}) as a function of the (magnitude of the) three-momentum transfer $|\mathbf{q}|$ in arbitrary units (a.u.). The decoherence and phase shift-dependent pieces are shown with purple and orange solid lines, respectively. The atom interferometer and composite-particle interferometer curves are shown in the left and right panel, respectively.}\label{fig:response_function}
\end{figure}

The three boxed expressions capture some of the central results of this section. 
For fixed DM occupation number, decoherence saturates for momentum transfers $|\mathbf q|\gtrsim |\boldsymbol{\Delta x}|^{-1}$, while the phase shift remains nonzero exclusively in anisotropic environments and becomes a fast oscillating function for $|\mathbf q|\gg |\boldsymbol{\Delta x}|^{-1}$~\cite{PhysRevA.42.38,Hornberger:2003umw}. 
We can observe this behavior in Fig.~\ref{fig:response_function}, where we plot the experiment-dependent ``normalized DM effect''\footnote{These expressions assume $\boldsymbol{\Delta x} \parallel \mathbf{v}_\oplus $ and are what remains of the integrands on the right-hand sides of Eqs.~\eqref{eq:D_scatt}-\eqref{eq:P_scatt} once: (i) the factor $g_{_\mathbf{D}}(\mathbf{q})$ is stripped off and only the cross term depending on both $\boldsymbol{v}_\oplus$ and $\mathbf{q}$ in $g_{_\mathbf{P}}(\mathbf{q})$ is kept, and (ii) the solid angle integral is performed.}~\cite{Du:2022ceh}:
\begin{equation}\label{eq:norm_DM_eff}
\begin{aligned}
p_{_\mathbb{D}}(|\mathbf{q}|) &= \frac{1}{2} \int \mathrm{d}\cos \theta \, |F_\mathbb{D}(\mathbf{q})|^2 \left [1-\cos(|\mathbf{q}||\boldsymbol{\Delta x}|\cos \theta)\right ]\, \\
& = |F_\mathbb{D}(\mathbf{q})|^2 \left (1-\mathrm{sinc}(|\mathbf{q}||\boldsymbol{\Delta x}|) \right ) \, , \\
p_{_\mathbb{P}}(|\mathbf{q}|)  & = \frac{1}{2} \int \mathrm{d}\cos \theta  \, |F_\mathbb{P}(\mathbf{q})|^2 \sin(|\mathbf{q}||\boldsymbol{\Delta x}|\cos \theta) e^{-|\mathbf{v}_\oplus| |\mathbf{q}|\cos \theta/m_\chi v_0^2} \, \\
& =|F_{\mathbb P}(\mathbf q)|^2 \frac{\cos (|\mathbf q||\boldsymbol{\Delta x}|)\,\sinh(\nu_\oplus |\mathbf q||\boldsymbol{\Delta x}|)
-\nu_\oplus \sin (|\mathbf q||\boldsymbol{\Delta x}|) \,\cosh(\nu_\oplus |\mathbf q||\boldsymbol{\Delta x}|)}{|\mathbf q||\boldsymbol{\Delta x}|(1+\nu_\oplus^2)}\, ,
\end{aligned}
\end{equation} 
where $\mathrm{sinc}(x)=\sin(x)/x$, $\nu_\oplus = |\boldsymbol{v}_\oplus|/m_\chi v_0^2 |\boldsymbol{\Delta x}|$, and we used the fact that all form factors considered here only depend on $|\mathbf{q}|$. 
From Eq.~\eqref{eq:norm_DM_eff}, $p_{_\mathbb{D}}\propto (|\mathbf q||\boldsymbol{\Delta x}|)^2$ and $p_{_\mathbb{P}}\propto (|\mathbf q||\boldsymbol{\Delta x}|)^2 \nu_\oplus \simeq |\mathbf q|^2|\boldsymbol{\Delta x}|/m_\chi v_0 $ in the limit $|\mathbf q|\ll |\boldsymbol{\Delta x}|^{-1}$, and  $p_{_\mathbb{P}}$ vanishes for $|\boldsymbol{v}_\oplus| = 0$.

In light of the many-body and interferometer-specific form factors, and recalling that $\sigma_N < r_c < |\boldsymbol{\Delta x}|$, where $\sigma_N$ is the size of the nucleus, Fig.~\ref{fig:response_function} also shows how the initial state of the interferometer affects the observables as a function of $|\mathbf{q}|$. 
This can be seen in three kinematic regimes. 
For $|\mathbf{q}| \lesssim 1/r_c$, the DM cannot resolve the internal structure of either the composite object or the atomic cloud, so all atoms and nucleons interact in phase with the environment; hence, $|F_{\mathrm{CPI}}(\mathbf{q})|^2 \propto A^2 N_\mathrm{at}^2$, such that both CPI observables scale with $A^2 N_\mathrm{at}^2$, and $|F_{\mathrm{AI}}(\mathbf{q})|^2 \propto A^2 N_\mathrm{at}$, which only affects the AI phase shift (see Table~\ref{tab:form-factors}). 
For $|\mathbf{q}| \gtrsim 1/r_c$, the DM resolves the bulk of the object and the structure of the cloud. 
In AIs, as long as $|\mathbf{q}| \lesssim 1/\sigma_N$, the DM does not resolve the internal nuclear distribution of each atom, and $\mathbb{D}_\mathrm{AI}$ and $\mathbb{P}_\mathrm{AI}$ reduce to the ($N_\mathrm{at}$-independent) single-atom result. 
When $|\mathbf{q}| \gtrsim 1/\sigma_N$, both AI observables vanish in light of the Helm form factor. 
In CPIs, instead, as long as $|\mathbf{q}| \lesssim 1/\sigma_u$, the DM does not resolve the individual lattice sites, and $|F_{\mathrm{CPI}}(\mathbf{q})|^2 \propto N_\mathrm{at}^2A^2/(|\mathbf{q}|r_c)^4$ (cf. Eq.~\eqref{eq:sphere}). 
The DM resolves the position of the individual atoms for $|\mathbf{q}| \gtrsim 1/\sigma_u$, for which the CPI observables are exponentially suppressed by the Debye-Waller form factor $e^{-2W}$. 
For typical materials, $\omega_\mathrm{ph} \gtrsim 10$~meV~\cite{Trickle:2019nya}, such that, well below room temperature, the signal is parametrically suppressed for $|\mathbf{q}| \gtrsim \sqrt{\omega_\mathrm{ph}m_\mathrm{at}} \simeq 30~\mathrm{keV} \sqrt{m_\mathrm{at}/100~\mathrm{GeV}}$ (see Eq.~\eqref{eq:DW}).

\begin{figure}[t]
\includegraphics[width=0.55\linewidth]{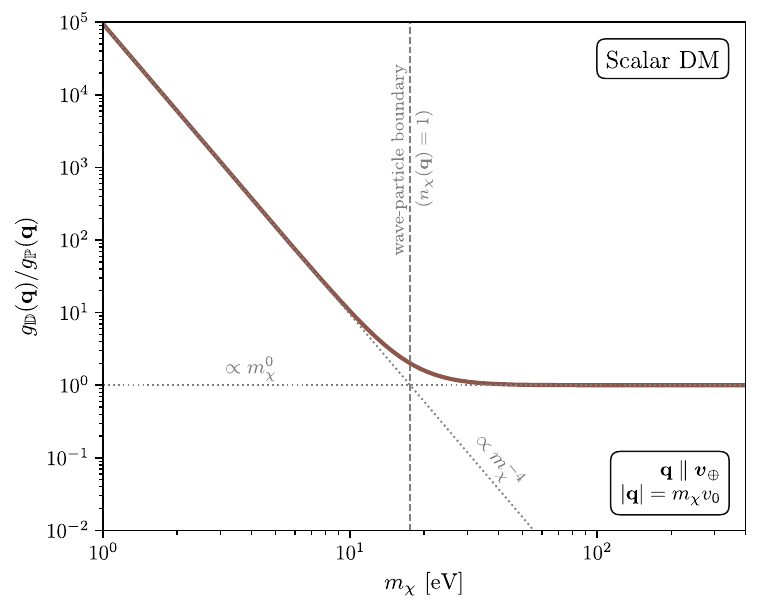}
\caption{Ratio $g_{_\mathbb{D}}(\mathbf{q})/g_{_\mathbb{P}}(\mathbf{q})$ as a function of $m_\chi$ for scalar DM for $|\mathbf{q}| = m_\chi v_0$ with $\mathbf{q} \parallel \boldsymbol{v}_\oplus$. For masses below the wave-particle boundary (i.e., $n_\chi(\mathbf{q}) \gtrsim 1$) the decoherence observable is enhanced relative to the phase observable due to Bose statistics.}\label{fig:Pauli_Bose}
\end{figure}

Through the DM statistical factor $S_\chi(\mathbf p-\mathbf q)$, $\mathbb D$ automatically acquires a non-thermal Bose-enhancement factor $n_\chi(\mathbf{p}) [1+n_\chi(\mathbf p-\mathbf q)]$ for bosonic DM, and a Pauli blocking factor $n_\chi(\mathbf{p})[1-\tfrac12 n_\chi(\mathbf p-\mathbf q)]$ for fermionic DM, with the spin-summed occupation bounded by $n_\chi(\mathbf p)\leq2$. 
Notably, $\mathbb P$ remains linear in the occupation number and is therefore insensitive to Bose enhancement or Pauli blocking. 
We show this in Fig.~\ref{fig:Pauli_Bose}, where we plot the ratio $g_{_\mathbb{D}}(\mathbf{q})/g_{_\mathbb{P}}(\mathbf{q})$ for scalar DM as a function of $m_\chi$, with $|\mathbf{q}| = m_\chi v_0$.  
The wave-particle boundary is defined at $n_\chi(\mathbf q) = 1$.
For masses above the wave-particle boundary, the occupation number is small, DM is therefore dilute and $g_\mathbb{D}(\mathbf{q})/g_\mathbb{P}(\mathbf{q}) \approx 1$. 
In this limit, our expressions reduce to those of Refs.~\cite{Riedel:2012ur,Riedel:2016acj,Du:2022ceh}. 
For masses below the boundary, the occupation number per mode exceeds unity and $g_\mathbb{D}(\mathbf{q})/g_\mathbb{P}(\mathbf{q}) \sim \rho_\chi/m_\chi^{4}v_0^{3} \simeq \lambda_c^3 \rho_\chi /m_\chi \gg 1$, where $\lambda_c$ is the DM coherence length/de Broglie length.
 
\subsubsection{Projected sensitivity}

To understand the phenomenological consequences of these previously overlooked finite density effects, we study the reach of the AI and CPI benchmarks (see Table~\ref{tab:benchmarks}) to single-component sub-MeV DM, i.e., for ${\rho_\chi = 0.4~\mathrm{GeV}/\mathrm{cm}^3}$. 
To evade warm DM constraints, which forbid thermally-produced DM lighter than $\mathcal{O}(1)$~keV~\cite{Irsic:2023equ,Gilman:2026uvq}, we assume a non-thermal production mechanism. 
The simplest model providing the necessary ingredients is $\mathbb{Z}_2$-symmetric Higgs portal DM with the relic abundance generated via a non-thermal production mechanism, e.g., the misalignment mechanism. 
The matching from the UV completion to the effective operator of interest at atomic scales is given by
\begin{equation}\label{eq:HiggPortal}
\mathcal{L} \supset -g_H \chi^2 H^\dagger H \rightarrow \underbrace{\frac{2}{9} \frac{g_H m_N}{m_h^2 m_\chi}}_{-g/\Lambda^2} m_\chi \chi^2 \overline{N}N \, ,
\end{equation}
where $g_H<0$ is a dimensionless coupling constant, $H$ is the Higgs doublet, $m_h$ the Higgs mass and $m_N$ the nucleon mass~\cite{Graham:2015ifn}.\footnote{Note that, for our choice of conventions and for fixed $g_H$, $g/\Lambda^2$ decreases with the DM mass.} 

\begin{figure}[t]
\includegraphics[width=0.55\linewidth]{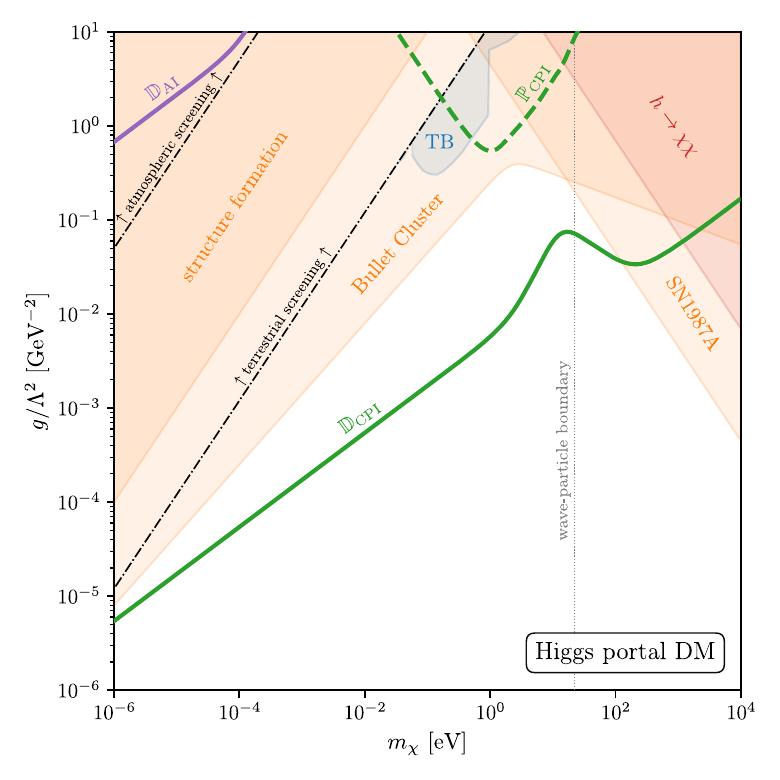}
\caption{
Projected constraints on the dimensionful coupling strength $g/\Lambda^2$ as a function of the DM mass $m_\chi$ for single-component quadratically-coupled scalar DM assuming the Higgs portal DM UV completion (see Eq.~\eqref{eq:HiggPortal}) for the CPI (green) and AI (purple) benchmarks (see Table~\ref{tab:benchmarks}). 
The projected reach from the phase shift ($\mathbb{P}$) and decoherence ($\mathbb{D}$) observables is shown with dashed and solid lines, respectively. 
In light blue we show existing laboratory constraints on background-enhanced fifth forces from torsion balance experiments (TB) and others~\cite{VanTilburg:2024xib,Matsumoto:2025rcz}; 
in orange we show the regions of parameter space excluded by structure formation~\cite{Arvanitaki:2014faa}, the Bullet Cluster~\cite{Randall:2008ppe,VanTilburg:2024xib}, and SN1987A and stellar cooling~\cite{Raffelt:1990yz,Olive:2007aj}. 
In red we show bounds from invisible Higgs decay at the LHC~\cite{ATLAS:2023tkt}. 
Note that for repulsive interactions the DM is shielded by the laboratory environment~\cite{VanTilburg:2024xib,Day:2023mkb}, so the projected interferometer constraints are only valid below the dash-dotted black lines for a terrestrial experiment.}\label{fig:contact_reach}
\end{figure}

The hierarchy between the phase-shift and decoherence channels is visible directly in Fig.~\ref{fig:contact_reach}, which shows the projected reach of the AI and CPI benchmarks in Table~\ref{tab:benchmarks} to single-component bosonic DM. 
The main qualitative features are: (i) the oscillation close to the wave-particle boundary, which appears only in the decoherence observable; (ii) the different low- and high-mass scaling of the two observables; and (iii) the different roles played by coherent atom-number enhancement in CPIs and AIs. These features follow from the interplay between the experimental length scales, the DM coherence properties, and the CPI/AI form factors, as summarized below.

\begin{table}[t]
\setlength{\tabcolsep}{10pt}
\begin{tabular}{cccccc}
\hline
\hline
Experiment type & Atomic species    & $N_\mathrm{at}$  & $\Delta x$ [m] & $r_c$ [m] & $T$ [s] \\
\hline
CPI &                    Nb-93  & $2\times 10^{11}$       & $2.5\times 10^{-4}$     & $10^{-6}$  & $0.5$ \\
AI &                      Rb-87 & $10^{8}$    & $5$ &            $10^{-3}$ 
  & $1$   \\
\hline
\hline
\end{tabular}
\caption{%
  Benchmark experimental parameters assumed for the two interferometer
  configurations considered in this work. For both examples, we assume shot-noise limited sensitivity and a total measurement campaign $T_\mathrm{int} = 10^8$~s, such that the dataset consists of $N_\mathrm{exp}=T_\mathrm{int}/T$ data points. These parameters have been adapted from existing proposals/detector concepts~\cite{Pino:2018erq}.}\label{tab:benchmarks}
\end{table}

\paragraph{Phase shifts.} The sensitivity from the $\mathcal{O}(g^2)$ phase shift can be understood by studying two distinct regimes. 
\begin{itemize}
\item $m_\chi \ll (|\boldsymbol{\Delta x}|v_0)^{-1}$.---In this mass range $|\mathbf{q}|\ll|\boldsymbol{\Delta x}|^{-1}$, so $p_\mathbb{P} \propto (|\mathbf q||\boldsymbol{\Delta x}|)^2/m_\chi v_0 \Delta x$ (see Eq.~\eqref{eq:norm_DM_eff}) and DM cannot resolve the superposition. Since $r_c \ll |\boldsymbol{\Delta x}|$, DM cannot resolve the atoms in the target either, and $|F_\mathrm{CPI}|^2\sim A^2 N_\mathrm{at}^2$ and $|F_\mathrm{AI}|^2\sim A^2 N_\mathrm{at}$. Consequently, the rate is dominated by $|\mathbf{q}| \sim m_\chi v_0$, for which  $g_\mathbb{P}\sim \rho_{_\mathrm{DM}}/m_\chi |\mathbf{q}|v_0$ (see Eqs.~\eqref{eq:g_p} and \eqref{eq:vmin}), and ${\mathbb{P}_\mathrm{CPI} \sim T \left ( \tfrac{g}{\Lambda^2} \right )^2 N_\mathrm{at}^2A^2 \rho_{_\mathrm{DM} } (m_\chi v_0)^2 |\boldsymbol{\Delta x}|}$ and ${\mathbb{P}_\mathrm{AI} \sim T \left ( \tfrac{g}{\Lambda^2} \right )^2 N_\mathrm{at} A^2 \rho_{_\mathrm{DM} } (m_\chi v_0)^2 |\boldsymbol{\Delta x}|}$. Assuming shot noise sensitivity, which implies $1/\sqrt{N_\mathrm{at}}$ and 1 radian resolution per shot for AIs and CPIs, respectively, and $N_\mathrm{exp}$ runs, the sensitivity scales as 
\begin{equation}
\frac{g}{\Lambda^2}\gtrsim \left (A  m_\chi v_0 \right )^{-1}\left ( \rho_{_\mathrm{DM}} |\boldsymbol{\Delta x}| T \right )^{-1/2} N_\mathrm{exp} ^{-1/4} 
\begin{cases}
N_\mathrm{at} ^{-1}& \text{for CPIs,} \\
N_\mathrm{at}^{-3/4} & \text{for AIs.} 
\end{cases}
\end{equation}
\item $m_\chi \gg (|\boldsymbol{\Delta x}|v_0)^{-1}$.---Since $p_\mathbb{P}$ is a fast oscillating function for $|\mathbf{q}|\gg|\boldsymbol{\Delta x}|^{-1}$  (see Fig.~\ref{fig:response_function}), the rate is dominated by $|\mathbf{q}| \sim  |\boldsymbol{\Delta x}|^{-1}\ll m_\chi v_0$, for which $p_\mathbb{P}\sim |F_\mathbb{P}|^2$. As a result, the DM cannot resolve the individual atoms in the target, i.e., $|F_\mathrm{CPI}|^2\sim N_\mathrm{at}^2A^2$ and  $|F_\mathrm{AI}|^2\propto N_\mathrm{at} A^2$, and $g_\mathbb{P}\sim\rho_{_\mathrm{DM}}/m_\chi^2 v_0^2$. Hence,  $\mathbb{P}_\mathrm{CPI} \sim T \left ( \tfrac{g}{\Lambda^2} \right )^2 N_\mathrm{at}^2A^2 \rho_{_\mathrm{DM} } (m_\chi v_0)^{-2} |\boldsymbol{\Delta x}|^{-3}$ and ${\mathbb{P}_\mathrm{AI} \sim T \left ( \tfrac{g}{\Lambda^2} \right )^2 N_\mathrm{at} A^2 \rho_{_\mathrm{DM} } (m_\chi v_0)^{-2} |\boldsymbol{\Delta x}|^{-3}}$, and 
\begin{equation}
\frac{g}{\Lambda^2}\gtrsim |\boldsymbol{\Delta x}|^{3/2} \, m_\chi v_0 \, A^{-1}\left ( \rho_{_\mathrm{DM}}T  \right )^{-1/2}N_\mathrm{exp}^{-1/4} \begin{cases}
N_\mathrm{at} ^{-1}& \text{for CPIs,} \\
N_\mathrm{at}^{-3/4} & \text{for AIs.} 
\end{cases}
\end{equation}
\end{itemize}

The scalings above explain why the phase-shift reach is optimized near the transition between the two regimes, $m_\chi \sim 10^3/|\boldsymbol{\Delta x}|$, for which 
\begin{equation}
\frac{g}{\Lambda^2} \Big |_\mathrm{optimal} \sim |\boldsymbol{\Delta x}|^{1/2} \, A^{-1}\left ( \rho_{_\mathrm{DM}}T  \right )^{-1/2}N_\mathrm{exp}^{-1/4} \begin{cases}
N_\mathrm{at} ^{-1}& \text{for CPIs,} \\
N_\mathrm{at}^{-3/4} & \text{for AIs.} 
\end{cases}
\end{equation}
For the parameters displayed in Table~\ref{tab:benchmarks}, this gives $g/\Lambda^2 |_\mathrm{optimal} \sim 10^{7}~\mathrm{GeV}^{-2}$ at $m_\chi \sim 4\times 10^{-5}$~eV, and $g/\Lambda^2 |_\mathrm{optimal} \sim 10^{-1}~\mathrm{GeV}^{-2}$ at $m_\chi \sim 1$~eV for the AI and CPI benchmarks, respectively, both of which are in agreement with Fig.~\ref{fig:contact_reach}. 
Thus, even with the coherent atom-number enhancement and shot-noise-limited readout, the phase-shift channel in the AI benchmark does not reach unexplored parameter space once existing astrophysical and laboratory constraints are included (see Appendix~\ref{app:Higgs} for a discussion). 
By contrast, the CPI benchmark, aided by its much larger effective coherent target and maximal separation $|\boldsymbol{\Delta x}|\lesssim250~\mu{\rm m}$, may still place meaningful constraints for optimized parameters.
 
\paragraph{Decoherence.} The decoherence observable displays much richer phenomenology.
This is because the observable is sensitive both to the wave-particle transition and to whether the DM resolves the superposition or the target constituents. 
The Bose enhancement in the wavelike regime changes the parametric scaling of the two observables with the DM mass and is therefore responsible for the pronounced low-mass improvement of the decoherence curves in Fig.~\ref{fig:contact_reach}. 
In light of the choice of $|\boldsymbol{\Delta x}|$ and $r_c$ for the two benchmarks, we distinguish the common CPI/AI regimes from the detector-specific high-mass regimes below. 
\begin{itemize}
\item CPI and AI
\begin{itemize}
\item $m_\chi \ll (|\boldsymbol{\Delta x}| v_0)^{-1}$.---In light of the choice of $|\boldsymbol{\Delta x}|$, this limit corresponds to $m_\chi \ll 1$~eV, which is deep in the wave-like regime for both the CPI and AI benchmarks. As a result, $g_\mathbb{D}\approx g_\mathbb{P} \, \rho_{_\mathrm{DM}}/m_\chi^{4}v_0^3$. In this limit, however, the DM can resolve neither the superposition nor the individual atoms within the target, such that $p_\mathbb{D}\propto |\mathbf{q}|^2|\boldsymbol{\Delta x}|^2$, and $|F_\mathrm{CPI}|^2 \propto N_\mathrm{at}^2 A^2$. Note that the AI observable in this case only inherits an $A^2$ enhancement in light of its form factor. The rate is then dominated by $|\mathbf{q}|\sim m_\chi v_0$. As a result, $g_\mathbb{D} \sim \rho_{_\mathrm{DM}}^2/m_\chi^5 v_0^4 |\mathbf{q}|$, $\mathbb{D}_\mathrm{CPI} \sim T \left ( \tfrac{g}{\Lambda^2} \right )^2 N_\mathrm{at}^2A^2 \rho_{_\mathrm{DM} }^2 m_\chi^{-1} |\boldsymbol{\Delta x}|^2$ and $\mathbb{D}_\mathrm{AI} \sim T \left ( \tfrac{g}{\Lambda^2} \right )^2 A^2 \rho_{_\mathrm{DM} }^2 m_\chi^{-1} |\boldsymbol{\Delta x}|^2$, and 
\begin{equation}
\frac{g}{\Lambda^2}\gtrsim m_\chi^{1/2} T^{-1/2} \left (A \rho_{_\mathrm{DM}}|\boldsymbol{\Delta x}| \right )^{-1} N_\mathrm{exp}^{-1/4}  
\begin{cases}
N_\mathrm{at}^{-1} & \text{for CPIs,} \\
N_\mathrm{at}^{-1/4} & \text{for AIs.} 
\end{cases}
\end{equation}
\item $(|\boldsymbol{\Delta x}| v_0)^{-1} \lesssim m_\chi \lesssim 10$~eV.---In this regime, DM is still wavelike, such that $g_\mathbb{D}\approx g_\mathbb{P} \, \rho_{_\mathrm{DM}}/m_\chi^{4}v_0^3$, and DM can resolve the superpositions while not resolving the individual atoms in the target. Thus, the rate is dominated by $|\mathbf{q}|\sim m_\chi v_0$, $|F_\mathrm{CPI}|^2\propto N_\mathrm{at}^2 A^2$, $p_\mathbb{D}$ is independent of $\mathbf{q}$ and $\boldsymbol{\Delta x}$, and $g_\mathbb{D} \sim \rho_{_\mathrm{DM}}^2/m_\chi^5 v_0^4 |\mathbf{q}|$. As a result, ${\mathbb{D}_\mathrm{CPI} \propto T \left ( \tfrac{g}{\Lambda^2} \right )^2 N_\mathrm{at}^2A^2 \rho_{_\mathrm{DM} }^2 m_\chi^{-3}v_0^{-2}}$, ${\mathbb{D}_\mathrm{AI} \propto T \left ( \tfrac{g}{\Lambda^2} \right )^2 A^2 \rho_{_\mathrm{DM} }^2 m_\chi^{-3}v_0^{-2}}$ and the reach is parametrically
\begin{equation}
\frac{g}{\Lambda^2}\gtrsim m_\chi^{3/2} v_0 T^{-1/2} (A \rho_{_\mathrm{DM}})^{-1}  N_\mathrm{exp} ^{-1/4} \begin{cases}
N_\mathrm{at}^{-1} & \text{for CPIs,} \\
N_\mathrm{at}^{-1/4} & \text{for AIs.}
\end{cases}
\end{equation}
\end{itemize}
\item CPI-only
\begin{itemize}
\item $10~\mathrm{eV} \lesssim m_\chi  \lesssim (r_c \, v_0)^{-1}$.---In this mass range, DM can be modeled as a ballistic particle, such that $g_\mathbb{D}\sim \rho_{_\mathrm{DM}}/m_\chi |\mathbf{q}|v_0$. As in the previous regime, DM can resolve the superpositions while not resolving the individual atoms in the target, such that the rate is again dominated by $|\mathbf{q}| = m_\chi v_0$. Hence, $|F_\mathrm{CPI}|^2\propto N_\mathrm{at}^2 A^2$, and $p_\mathbb{D}$ is independent of $\mathbf{q}$ and $\boldsymbol{\Delta x}$. Hence, $\mathbb{D}_\mathrm{CPI} \propto T \left ( \tfrac{g}{\Lambda^2} \right )^2 N_\mathrm{at}^2A^2 \rho_{_\mathrm{DM} } m_\chi v_0$ and
 \begin{equation}
\frac{g}{\Lambda^2}\gtrsim \left ( \rho_{_\mathrm{DM}} m_\chi v_0 T  \right )^{-1/2}  (A N_\mathrm{at})^{-1} N_\mathrm{exp}^{-1/4} \, .
\end{equation}
\item $m_\chi  \gg (r_c \, v_0)^{-1}$.---In this regime, the rate can inherit a coherent enhancement only for $|\mathbf{q}|\lesssim r_c^{-1}$. Because $p_\mathbb{D}$ is independent of $\mathbf{q}$ and $\boldsymbol{\Delta x}$ in this range, and $g_\mathbb{D}\sim \rho_{_\mathrm{DM}}/m_\chi |\mathbf{q}|v_0$, $\mathbb{D}_\mathrm{CPI} \propto T \left ( \tfrac{g}{\Lambda^2} \right )^2 N_\mathrm{at}^2A^2 \rho_{_\mathrm{DM} } (m_\chi v_0)^{-1}r_c^{-2}$,  and
\begin{equation}
\frac{g}{\Lambda^2}\gtrsim r_c \left (m_\chi v_0 \right)^{1/2} \left (\rho_{_\mathrm{DM}} T \right )^{-1/2} (A N_\mathrm{at})^{-1}  N_\mathrm{exp}^{-1/4} \, .
\end{equation}
\end{itemize}
\item AI-only
\begin{itemize}
\item $ m_\chi \gg 10 $~eV.---In this regime, DM is particle-like, such that $g_\mathbb{D}\sim \rho_{_\mathrm{DM}}/m_\chi |\mathbf{q}|v_0$. Because the observable is insensitive to the total number of atoms in the target, the rate is dominated by $|\mathbf{q}| = m_\chi v_0$, for which $p_\mathbb{D}$ is independent of $\mathbf{q}$ and $\boldsymbol{\Delta x}$. Hence, $\mathbb{D}_\mathrm{AI} \propto T \left ( \tfrac{g}{\Lambda^2} \right )^2 A^2 \rho_{_\mathrm{DM} } m_\chi v_0$ and
 \begin{equation}
\frac{g}{\Lambda^2}\gtrsim \left ( \rho_{_\mathrm{DM}} m_\chi v_0 T  \right )^{-1/2}  A^{-1} (N_\mathrm{exp}N_\mathrm{at})^{-1/4} \, .
\end{equation}
\end{itemize}
\end{itemize}

The decoherence scalings derived above reproduce the structure of the curves in Fig.~\ref{fig:contact_reach} and clarify the nontrivial hierarchy between the two observables. 
Importantly, below the wave--particle boundary, Bose enhancement makes decoherence parametrically more sensitive than phase shifts to DM scattering signatures. 
This also applies to the AI benchmark, despite the large $N_\mathrm{at}$ enhancement that boosts the phase shift signal while keeping the decoherence observable unchanged.

Finally, Fig.~\ref{fig:contact_reach} shows that the Bose-enhanced decoherence reach is not equally useful for the two detector classes once existing constraints and shielding effects are included. For the repulsive interactions assumed here, only the CPI benchmark probes previously unexplored regions of parameter space in which the Earth and atmosphere do not screen the DM as seen by a terrestrial experiment~\cite{VanTilburg:2024xib,Day:2023mkb} (see the discussion in Appendix~\ref{app:DM_mod}).\footnote{The laboratory would \textit{not} shield the DM for \textit{attractive} DM-nucleon interactions. In this case, however, regions of parameter space above the black dot-dashed line in Fig.~\ref{fig:contact_reach} would be subject to DM reflection~\cite{VanTilburg:2024xib}. We leave a study of these non-perturbative effects to future work.} Hence, the CPI benchmark provides the best opportunity for probing this model across the wave-particle boundary. 

Finally, note that the operator considered here can produce a warm DM subcomponent via freeze-in and thermalize with the SM plasma in the early universe. Therefore, for sufficiently large reheat temperatures, large regions of parameter space are excluded. To evade such bounds, setting a much smaller reheat temperature is consistent with the plot. We therefore leave a detailed study that depends on the specific cosmology to future work.

\subsection{Long-range interactions}\label{sec:long}

We now consider DM-nucleon interactions mediated by a light hadrophilic scalar $\phi$ with mass $m_\phi$ (see~Eq.~\eqref{eq:long}).
In particular, we shall consider the case where the mediator mass $m_\phi$ is small relative to the typical three-momentum transferred by the DM in typical interactions. 
Setting $O_s = \overline{N}N$, the closed-theory action in the Keldysh basis contains the terms
\begin{equation}\label{eq:closed_long}
\begin{aligned}
S[O_s^\mathrm{cl},O_s^\mathrm{q},\phi^{\mathrm{cl}},\phi^{\mathrm{q}},O_\chi^\mathrm{cl},O_\chi^\mathrm{q}] \supset  - y_n \phi^\mathrm{cl}\,O^\mathrm{q}_s - y_n \phi^\mathrm{q}\,O^\mathrm{cl}_s -  y_\chi \phi^\mathrm{cl}\,O^\mathrm{q}_\chi - y_\chi \phi^\mathrm{q}\,O^\mathrm{cl}_\chi  \, ,
\end{aligned}
\end{equation}
where $y_\chi$ and $y_n$ are dimensionless couplings, and $O_\chi$ are the DM bilinears (see~Eq.~\eqref{eq:DM_bilinears}).

As before, we can obtain the effective action by tracing out environmental degrees of freedom. 
In this case, however, the environment contains \textit{both} $\chi$ and $\phi$, and the environment's initial state factorizes into the mediator and the DM sectors: $\hat{\rho}_e = \hat{\rho}_\phi \otimes \hat{\rho}_\chi$, with $\hat{\rho}_\phi = \ket{0}\bra{0}$ and $\hat{\rho}_\chi$ the (by assumption Gaussian) DM density matrix. 
In light of Eq.~\eqref{eq:closed_long}, it immediately follows that the influence action will depend on the retarded, advanced and Keldysh propagators of \textit{both} $\chi$ and $\phi$. 
However, differently from the case of contact interactions, the perturbative expansion will be organized with respect to $y_n$ and $y_\chi$. 

Because of fifth force constraints from tests of the inverse square law~\cite{Chen:2014oda} and from stellar cooling~\cite{Hardy:2016kme,Bottaro:2023gep,Hardy:2024gwy,Fiorillo:2025zzx}, we will be considering a light mediator with very small couplings to nucleons, i.e.,  $y_n \ll 1$ for $m_\phi \ll 1$~GeV. 
In particular we will consider that $\chi$ is either a small DM subcomponent or saturates the local DM energy density.
In the former case, we assume that $\chi$ makes up 5\% of the DM in order to evade current $N_\mathrm{eff}$ and thermalization constraints~\cite{Knapen:2017xzo}. 
We also fix $y_\chi = 10^{-3}$, which ensures perturbativity, and choose mediator masses satisfying $m_\phi > y_\chi^2 m_\chi/(4\pi)$ to avoid two-body Yukawa bound state formation~\cite{Tulin:2013teo,Wise:2014jva,Petraki:2016cnz}.
Furthermore, assuming that this DM subcomponent decoupled from the thermal bath above the QCD phase transition and ensuring that its typical velocity be less than the escape velocity of the Milky Way, we will focus on DM masses $m_\chi \gtrsim 10$~eV~\cite{Du:2022ceh}.
In the latter case, to be consistent with self-interacting DM (SIDM) constraints~\cite{Tulin:2013teo}, we choose $y_\chi^2/(4\pi) = 10^{-10} (m_\chi/\mathrm{MeV})^{3/2}$~\cite{Knapen:2017xzo}. 
Furthermore, to be consistent with warm DM constraints (assuming a thermal production mechanism), we focus on DM masses larger than $10$~keV.

As a direct consequence of these assumptions and constraints, we compute the effective action through $\mathcal{O}(y_n^2)$, while resumming the one-loop diagrams of $\chi$, thereby obtaining expressions that are valid to higher order in $y_\chi$. 
Because the influence action is computed by tracing over all environmental degrees of freedom, the entries of the environment kernel matrix $\widetilde{\mathsf{K}}$ correspond to the \textit{dressed} $\phi$ propagators $\widetilde{D}^\phi_\mathrm{r,a,k}$, where the dressing is provided by the DM environment.
As we shall explicitly see below, the dressing is the result of the finite density of DM particles, which implies that $\phi$  exchanges and fluctuations are in-medium.
Given the structure of the interactions, the dressed $\phi$ propagators can be computed at one-loop, where $\chi$ particles run in the loop.
Given the free theory $\phi$ propagators
\begin{equation}\label{eq:undressed_phi}
\begin{aligned}
\widetilde{G}^{\phi}_\mathrm{a}(p)
&=
\begin{tikzpicture}[baseline=-0.5ex,scale=1.2]
  \draw[very thick,dotted,gray] (0.24,0) -- (0.75,0);
  \draw[very thick,gray] (0.75,0) -- (1.25,0);
\end{tikzpicture}
=
\dfrac{1}{(p_0-\mi\epsilon)^2-E_{\mathbf p}^2}
\, ,
\\
\widetilde{G}^{\phi}_\mathrm{r}(p)
&=
\begin{tikzpicture}[baseline=-0.5ex,scale=1.2]
  \draw[very thick,gray] (0.24,0) -- (0.75,0);
  \draw[very thick,dotted,gray] (0.75,0) -- (1.25,0);
\end{tikzpicture}
=
\dfrac{1}{(p_0+\mi\epsilon)^2-E_{\mathbf p}^2}
\, ,
\\
\widetilde{G}^{\phi}_\mathrm{k}(p)
&=
\begin{tikzpicture}[baseline=-0.5ex,scale=1.2]
  \draw[very thick,gray] (0.24,0) -- (1.25,0);
\end{tikzpicture}
=
-\frac{\mi}{4E_{\mathbf p}}
\left[
\deltabar(p_0-E_{\mathbf p})
+
\deltabar(p_0+E_{\mathbf p})
\right]
\, ,
\end{aligned}
\end{equation}
with $E_\mathbf{p} = \sqrt{|\mathbf{p}|^2+m_\phi^2}$, the in-in dressed $\phi$ propagators at one-loop can be computed via the Dyson series. 
Mapping $ \tfrac{g}{\Lambda^2} O_s^\mathrm{cl,q} O_\chi^\mathrm{q,cl}$ in Eq.~\eqref{eq:inin_int} to  $y_\chi \phi^\mathrm{cl,q}O_\chi^\mathrm{q,cl}$ in Eq.~\eqref{eq:closed_long}, it immediately follows that the dressing can be computed using Eq.~\eqref{eq:bubble_scalar} for real scalar DM and Eq.~\eqref{eq:bubble_fermion} for fermionic DM. Employing the relation $\widetilde{\Pi}_{\mathrm a}^{\chi}(q) = \left [\widetilde{\Pi}_{\mathrm r}^{\chi}(q) \right ]^\dagger$ and the un-dressed $\phi$ propagators (see Eq.~\eqref{eq:undressed_phi}), we find 
\begin{equation}
\begin{aligned}
\widetilde{D}_\mathrm{r}^\phi(q)
&=
\begin{tikzpicture}[baseline=-0.5ex,scale=1.2]
  \draw[very thick,gray] (0.24,0) -- (0.75,0);
  \draw[very thick,dotted,gray] (0.75,0) -- (1.25,0);
\end{tikzpicture}
+
\begin{tikzpicture}[baseline=-0.5ex,scale=1.0]
  \draw[very thick,gray]
    (-1.20,0) -- (-0.60,0);
  \draw[very thick,dotted,gray]
    (-0.60,0) -- (0,0);
  \draw[very thick,gray]
    (0.62,0) -- (1.22,0);
  \draw[very thick,dotted,gray]
    (1.22,0) -- (1.82,0);
  \draw[very thick,fill=white]
    (0,-0.31) rectangle (0.62,0.31);
  \node at (0.31,0)
    {\scriptsize $\widetilde{\Pi}_{\mathrm r}^{\chi}$};
\end{tikzpicture}
+
\begin{tikzpicture}[baseline=-0.5ex,scale=1.0]
  \draw[very thick,gray]
    (-1.20,0) -- (-0.60,0);
  \draw[very thick,dotted,gray]
    (-0.60,0) -- (0,0);
  \draw[very thick,gray]
    (0.62,0) -- (1.22,0);
  \draw[very thick,dotted,gray]
    (1.22,0) -- (1.82,0);
  \draw[very thick,gray]
    (2.44,0) -- (3.04,0);
  \draw[very thick,dotted,gray]
    (3.04,0) -- (3.64,0);
  \draw[very thick,fill=white]
    (0,-0.31) rectangle (0.62,0.31);
  \node at (0.31,0)
    {\scriptsize $\widetilde{\Pi}_{\mathrm r}^{\chi}$};
  \draw[very thick,fill=white]
    (1.82,-0.31) rectangle (2.44,0.31);
  \node at (2.13,0)
    {\scriptsize $\widetilde{\Pi}_{\mathrm r}^{\chi}$};
\end{tikzpicture}
+\cdots
\\
&=
\widetilde{G}^{\phi}_{\mathrm r}(q)
+
\widetilde{G}^{\phi}_{\mathrm r}(q)\,
\widetilde{\Pi}^{\chi}_{\mathrm r}(q)\,
\widetilde{D}^{\phi}_{\mathrm r}(q)
\\
&=
\frac{1}{
(q_0+\mi\epsilon)^2
-E_{\mathbf q}^2
-\widetilde{\Pi}^{\chi}_{\mathrm r}(q)
}
\, ,
\end{aligned}
\end{equation}
\begin{equation}
\begin{aligned}
\widetilde{D}_\mathrm{k}^\phi(q)
&=
\begin{tikzpicture}[baseline=-0.5ex,scale=1.2]
  \draw[very thick,gray] (0.24,0) -- (0.75,0);
  \draw[very thick,gray] (0.75,0) -- (1.25,0);
\end{tikzpicture}
+
\begin{tikzpicture}[baseline=-0.5ex,scale=1.0]
  \draw[very thick,gray]
    (-1.20,0) -- (-0.60,0);
  \draw[very thick,dotted,gray]
    (-0.60,0) -- (0,0);
  \draw[very thick,gray]
    (0.62,0) -- (1.22,0);
  \draw[very thick,gray]
    (1.22,0) -- (1.82,0);
  \draw[very thick,fill=white]
    (0,-0.31) rectangle (0.62,0.31);
  \node at (0.31,0)
    {\scriptsize $\widetilde{\Pi}_{\mathrm r}^{\chi}$};
\end{tikzpicture}
+
\begin{tikzpicture}[baseline=-0.5ex,scale=1.0]
  \draw[very thick,gray]
    (-1.20,0) -- (-0.60,0);
  \draw[very thick,gray]
    (-0.60,0) -- (0,0);
  \draw[very thick,dotted,gray]
    (0.62,0) -- (1.22,0);
  \draw[very thick,gray]
    (1.22,0) -- (1.82,0);
  \draw[very thick,fill=white]
    (0,-0.31) rectangle (0.62,0.31);
  \node at (0.31,0)
    {\scriptsize $\widetilde{\Pi}_{\mathrm a}^{\chi}$};
\end{tikzpicture}
+
\begin{tikzpicture}[baseline=-0.5ex,scale=1.0]
  \draw[very thick,gray]
    (-1.20,0) -- (-0.60,0);
  \draw[very thick,dotted,gray]
    (-0.60,0) -- (0,0);
  \draw[very thick,dotted,gray]
    (0.62,0) -- (1.22,0);
  \draw[very thick,gray]
    (1.22,0) -- (1.82,0);
  \draw[very thick,fill=white]
    (0,-0.31) rectangle (0.62,0.31);
  \node at (0.31,0)
    {\scriptsize $\widetilde{\Pi}_{\mathrm k}^{\chi}$};
\end{tikzpicture}
+\cdots
\\
&=
\widetilde{D}^{\phi}_{\mathrm r}(q) \left [
\widetilde{G}^{\phi}_{\mathrm r}(q)^{-1} \widetilde{G}^{\phi}_{\mathrm k}(q) \widetilde{G}^{\phi}_{\mathrm a}(q)^{-1}
+
\widetilde{\Pi}^{\chi}_{\mathrm k}(q) 
\right ] \widetilde{D}^{\phi}_{\mathrm a}(q) \, ,
\end{aligned}
\end{equation}
where $\widetilde{D}^{\phi}_{\mathrm a}(q)= [\widetilde{D}^{\phi}_{\mathrm r}(q) ]^*$ ~\cite{Jauho_1994,Kamenev:2009jj}.

Because of the DM mass range of interest, $\tau_c \ll T$ and the Markovian approximation is valid. Hence, we can invoke the distributional identity Eq.~\eqref{eq:identity}, such that the MWI observables in Eq.~\eqref{eq:int_obs} read as 
\begin{align}
\mathbb{D}&=-T y_n^2 \int\! \dbar^3q \, |F_\mathbb{D}(\mathbf{q})|^2 [1-\cos(\mathbf{q}\!\cdot\!\boldsymbol{\Delta x})] \mathrm{Im} \left [ \widetilde{D}^\phi_\mathrm{k} \left ( 0, \mathbf{q} \right ) \right ] + \mathcal{O}(y_n^3)  \,,\label{eq:D_scatt_long}\\
\mathbb{P}&= T y_n^2 \int\!\dbar^3q \,|F_\mathbb{P}(\mathbf{q})|^2
\sin(\mathbf{q}\!\cdot\!\boldsymbol{\Delta x})\,\mathrm{Im} \left [ \widetilde{D}^\phi_\mathrm{r} \left ( 0, \mathbf{q} \right ) \right ] + \mathcal{O}(y_n^3)  \,. 
\label{eq:P_scatt_long} 
\end{align}
Because $\widetilde{G}^\phi_\mathrm{k}(0,\mathbf{q}) = 0$ for a massive mediator (see Eq.~\eqref{eq:undressed_phi}), the imaginary parts of the dressed $\phi$ propagators are given by 
\begin{equation}\label{eq:im_D}
\begin{aligned}
\mathrm{Im} \left [ \widetilde{D}^\phi_\mathrm{r} \left ( 0, \mathbf{q} \right ) \right ] &= \frac{\mathrm{Im} \left [ \widetilde{\Pi}^\chi_\mathrm{r} \left ( 0, \mathbf{q} \right ) \right ]}{\left ( |\mathbf{q}|^2+m_{\phi}^2 + \mathrm{Re} \left [ \widetilde{\Pi}^\chi_\mathrm{r} \left ( 0, \mathbf{q} \right ) \right ] \right )^2 + \mathrm{Im} \left [ \widetilde{\Pi}^\chi_\mathrm{r} \left ( 0, \mathbf{q} \right ) \right ]^2} \, , \\
\mathrm{Im} \left [ \widetilde{D}^\phi_\mathrm{k} \left ( 0, \mathbf{q} \right ) \right ] &= \frac{\mathrm{Im} \left [ \widetilde{\Pi}^\chi_\mathrm{k} \left ( 0, \mathbf{q} \right ) \right ]}{\left ( |\mathbf{q}|^2+m_{\phi}^2+\mathrm{Re} \left [ \widetilde{\Pi}^\chi_\mathrm{r} \left ( 0, \mathbf{q} \right ) \right ]\right )^2 + \mathrm{Im} \left [ \widetilde{\Pi}^\chi_\mathrm{r} \left ( 0, \mathbf{q} \right ) \right ]^2} \, ,
\end{aligned}
\end{equation}
with $\mathrm{Im} \left [ \widetilde{\Pi}^\chi_\mathrm{r,k} \left ( 0, \mathbf{q} \right ) \right ]$ computed in Eq.~\eqref{eq:imaginary_Pi} under the replacement $g/\Lambda^2\to y_\chi$ and 
\begin{equation}
\mathrm{Re} \left [ \widetilde{\Pi}^\chi_\mathrm{r} \left ( 0, \mathbf{q} \right ) \right ] \approx y_\chi^2 \int \dbar^3 p \, \mathrm{P}\left [ \frac{n_\chi(\mathbf{p})-n_\chi(\mathbf{p-q})}{E_\mathbf{p}-E_\mathbf{p-q}}\right ] \, ,
\end{equation}
where $\mathrm{P}$ is the principal value.\footnote{Note that $\mathrm{Re} \left [ \widetilde{\Pi}^\chi_\mathrm{r} \left ( 0, \mathbf{q} \right ) \right ]$ also contains a contribution that is independent of the DM number density. This contribution, however, can be absorbed by the mediator mass.}

In the heavy-mediator regime, the in-medium $\phi$ propagators reduce to
\begin{equation}
y_n^2\, \mathrm{Im}\!\left[\widetilde{D}^\phi_{\mathrm{r,k}}(0,\mathbf{q})\right] 
\simeq
\frac{y_n^2}{m_\phi^4}\,
\mathrm{Im}\!\left[\widetilde{\Pi}^\chi_{\mathrm{r,k}}(0,\mathbf{q})\right] \, , \qquad \text{for } m_\phi \gg |\mathbf{q}| \, .
\end{equation}
This reproduces the contact-interaction result under the mapping $g\to -y_\chi y_n$ and $\Lambda^2\to m_\phi^2$ (see Eq.~\eqref{eq:imaginary_Pi}), as expected from standard EFT arguments. 
Indeed, in the NR limit and for $m_\phi \gg |\mathbf{q}|$, one can integrate out the heavy mediator, obtaining the effective interaction\footnote{Integrating out $\phi$ also gives rise to nucleon self-interactions: $\mathcal{L}\supset \tfrac{y_n^2}{2 m_\phi^2}(\overline{N}N)^2$. This term, however, does not contribute to the MWI observable for the geometries assumed here.}
\begin{equation}
\mathcal{L}\supset
\frac{y_\chi y_n}{m_\phi^2}\,\overline{N}N\,O_\chi ,
\qquad [O_\chi]=3 .
\end{equation}

In the light-mediator regime, the dressed $\phi$ propagators reveal previously overlooked effects. 
Since $\mathrm{Re}\,[\widetilde{\Pi}^\chi_{\rm r}(0,\mathbf{q})]$ is bounded from below by $\lim_{|\mathbf{q}|\to 0}\mathrm{Re}\,[\widetilde{\Pi}^\chi_{\rm r}(0,\mathbf{q})]\leq 0$, the mass of $\phi$ in-medium can become tachyonic. 
Using the definition of the in-medium mass squared of $\phi$, \begin{equation}\label{eq:meffsq}
\begin{aligned}
m_{\phi,\mathrm{eff}}^2 &\equiv m_\phi^2 + \lim_{|\mathbf q|\to0} \mathrm{Re}\!\left[ \widetilde{\Pi}^\chi_{\rm r}(0,\mathbf q) \right]  \simeq m_\phi^2 - 2 y_\chi^2\frac{\rho_\chi}{m_\chi^2 v_0^2}  \, ,
\end{aligned}
\end{equation}
tachyonic instabilities occur for
\begin{equation}
m_{\phi,\mathrm{eff}}^2 < 0 \implies m_\phi \lesssim y_\chi \frac{\sqrt{2\rho_\chi}}{m_\chi v_0} \simeq 5\times 10^{-4}~\mathrm{eV} \left ( \frac{y_\chi}{1} \right ) \left ( \frac{1~\mathrm{keV}}{m_\chi} \right ) \left ( \frac{10^{-3}}{v_0} \right ) \left ( \frac{\rho_\chi/\rho_{_\mathrm{DM}}}{0.05} \right )^{1/2} \, .
\end{equation}
Hence, even a small DM subcomponent with large $y_\chi$ and small $m_\chi$ can drive $m_{\phi,\mathrm{eff}}^2$ negative.
Because interactions are attractive, in this regime DM can clump, thereby invalidating the plane-wave approximation and the homogeneity assumption used throughout.
We therefore do not consider regions of parameter space in which $m_{\phi,\mathrm{eff}}^2<0$.

Using the property $\mathrm{Im} [ \widetilde{\Pi}^\chi_\mathrm{r} \left ( 0, -\mathbf{q} \right )  ]=-\mathrm{Im}  [ \widetilde{\Pi}^\chi_\mathrm{r} \left ( 0, \mathbf{q} \right ) ]$ and $\mathrm{Re} [ \widetilde{\Pi}^\chi_\mathrm{r} \left ( 0, -\mathbf{q} \right ) ] = \mathrm{Re} [ \widetilde{\Pi}^\chi_\mathrm{r} \left ( 0, \mathbf{q} \right )  ]$, Eqs.~\eqref{eq:D_scatt_long} and \eqref{eq:P_scatt_long} can be written analogously to the contact interactions case:
\begin{empheq}[box=\fbox]{align}
\mathbb{D}&=T y_n^2 y_\chi^2 \!\int\! \dbar^3q \, |F_\mathbb{D}(\mathbf{q})|^2 [1-\cos(\mathbf{q}\!\cdot\!\boldsymbol{\Delta x})] \frac{1}{( |\mathbf{q}|^2+m_{\phi}^2+\mathrm{Re} [ \widetilde{\Pi}^\chi_\mathrm{r} \left ( 0, \mathbf{q} \right )  ] )^2 + \mathrm{Im} [ \widetilde{\Pi}^\chi_\mathrm{r} \left ( 0, \mathbf{q} \right ) ] ^2}g_{_\mathbb{D}}(\mathbf{q}) \,,\label{eq:D_scatt_long_final}\\
\mathbb{P}&=T y_n^2 y_\chi^2 \!\int\! \dbar^3q \, |F_\mathbb{P}(\mathbf{q})|^2 \sin(\mathbf{q}\!\cdot\!\boldsymbol{\Delta x}) \frac{1}{( |\mathbf{q}|^2+m_{\phi}^2+\mathrm{Re} [ \widetilde{\Pi}^\chi_\mathrm{r} \left ( 0, \mathbf{q} \right )  ] )^2 + \mathrm{Im} [ \widetilde{\Pi}^\chi_\mathrm{r} \left ( 0, \mathbf{q} \right ) ] ^2} g_{_\mathbb{P}}(\mathbf{q})\,,\label{eq:P_scatt_long_final}
\end{empheq}
with $g_{_\mathbb{D}}$ and $g_{_\mathbb{P}}$ defined in Eqs.~\eqref{eq:g_d_def} and \eqref{eq:g_p_def}, respectively.   
In the dilute DM limit, which is valid in the mass range considered in this section, $g_{_\mathbb{P}} \approx g_{_\mathbb{D}}$. 
Hence, finite density effects simply modify the medium propagator appearing in the rate. 
Rather than depending on $|\widetilde{G}^\phi_\mathrm{r}(0,\mathbf{q})|^2 = (|\mathbf{q}|^2+m_\phi^2)^{-2}$, which was used in Refs.~\cite{Riedel:2016acj,Du:2022ceh,Murgui:2025unt,Murgui:2026mlr}, the rate depends on $|\widetilde{D}^\phi_\mathrm{r}(0,\mathbf{q})|^2 = [(|\mathbf{q}|^2+m_\phi^2 + \mathrm{Re} [ \widetilde{\Pi}^\chi_\mathrm{r} \left ( 0, \mathbf{q} \right )  ] )^2 + \mathrm{Im} [ \widetilde{\Pi}^\chi_\mathrm{r} \left ( 0, \mathbf{q} \right ) ] ^2]^{-1}$. 
Notably, when $m_{\phi,\mathrm{eff}}^2$ remains positive, this finite-density correction can increase the range of the force mediated by $\phi$ by decreasing the mediator's in-medium mass. 
In the high-$|\mathbf{q}|$ regime, both propagators scale with $|\mathbf{q}|^{-4}$, so finite-density effects are negligible. 
In the small-$|\mathbf{q}|$ limit, instead, $|\widetilde{G}^\phi_\mathrm{r}(0,\mathbf{q})|^2$ and $|\widetilde{D}^\phi_\mathrm{r}(0,\mathbf{q})|^2$ asymptote to $m_\phi^{-4}$ and $\sim m_{\phi,\mathrm{eff}}^{-4}$.
Therefore, in the small momentum transfer regime, finite density corrections can parametrically boost the sensitivity of MWIs.

\subsubsection{Projected sensitivity}

\begin{figure}[t]
\includegraphics[width=0.99\linewidth]{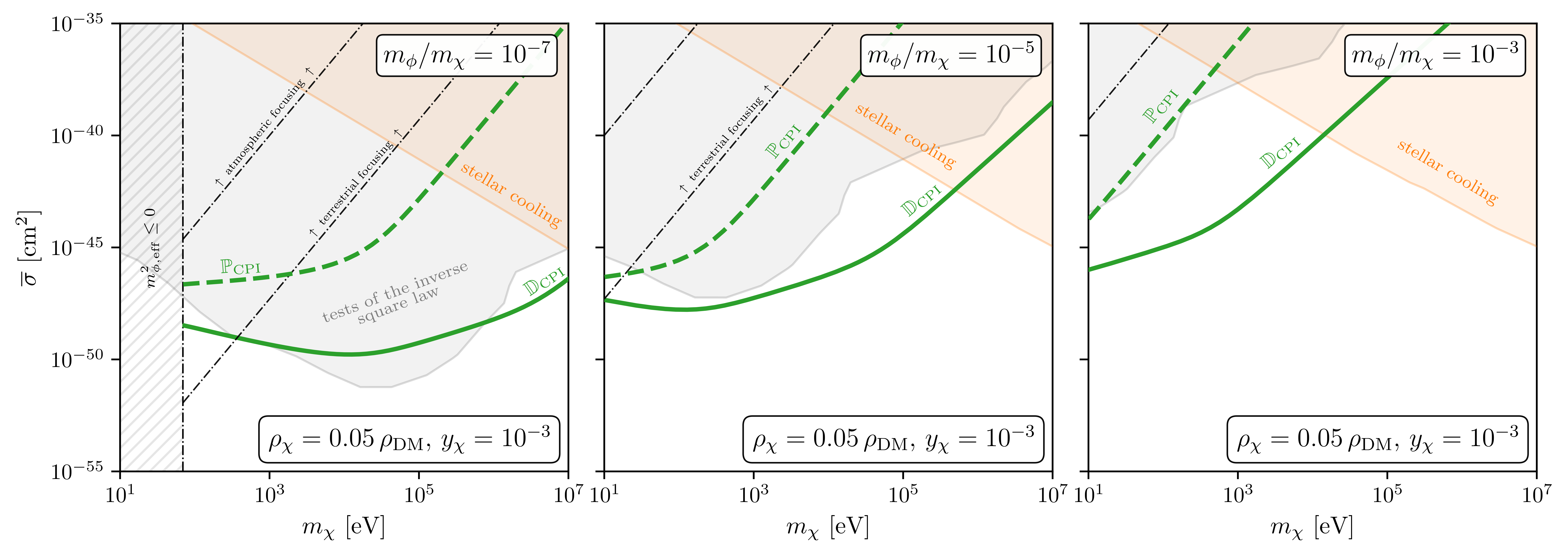}
\caption{Projected sensitivity of the CPI (green) benchmark to spin-independent elastic DM-nucleon scattering via the exchange of a light hadrophilic mediator (see Eqs.~\eqref{eq:long} and \eqref{eq:DM_bilinears}), assuming a $5\%$ DM subcomponent. We plot the reference cross section $\overline{\sigma}$ (see Eq.~\eqref{eq:ref_cs}) as a function of the DM mass $m_\chi$ for $y_\chi=10^{-3}$.
We fix the ratio $R_{\phi,\chi}=m_\phi/m_\chi$ to be equal to either $10^{-7}$ (left panel), $10^{-5}$ (central panel) or $10^{-3}$ (right panel).
The projected reach from the phase shift ($\mathbb{P}_\mathrm{CPI}$) and decoherence ($\mathbb{D}_\mathrm{CPI}$) observables is shown with dashed and solid lines, respectively. 
Regions of model space where the mediator mass becomes tachyonic, i.e., $m_{\phi,\mathrm{eff}}^2\leq 0$, are hatched.
In grey we show existing laboratory constraints from tests of the inverse square law~\cite{Chen:2014oda}, while in orange we show regions of parameter space excluded by neutron star and stellar cooling~\cite{Hardy:2016kme,Bottaro:2023gep,Hardy:2024gwy,Fiorillo:2025zzx}. 
Above the black dot-dashed curves, refraction by the Earth or atmosphere can induce order-one fractional changes in the DM speed and density, i.e., focusing. 
}
\label{fig:sigma_reach}
\end{figure}

\begin{figure}[t]
\includegraphics[width=0.99\linewidth]{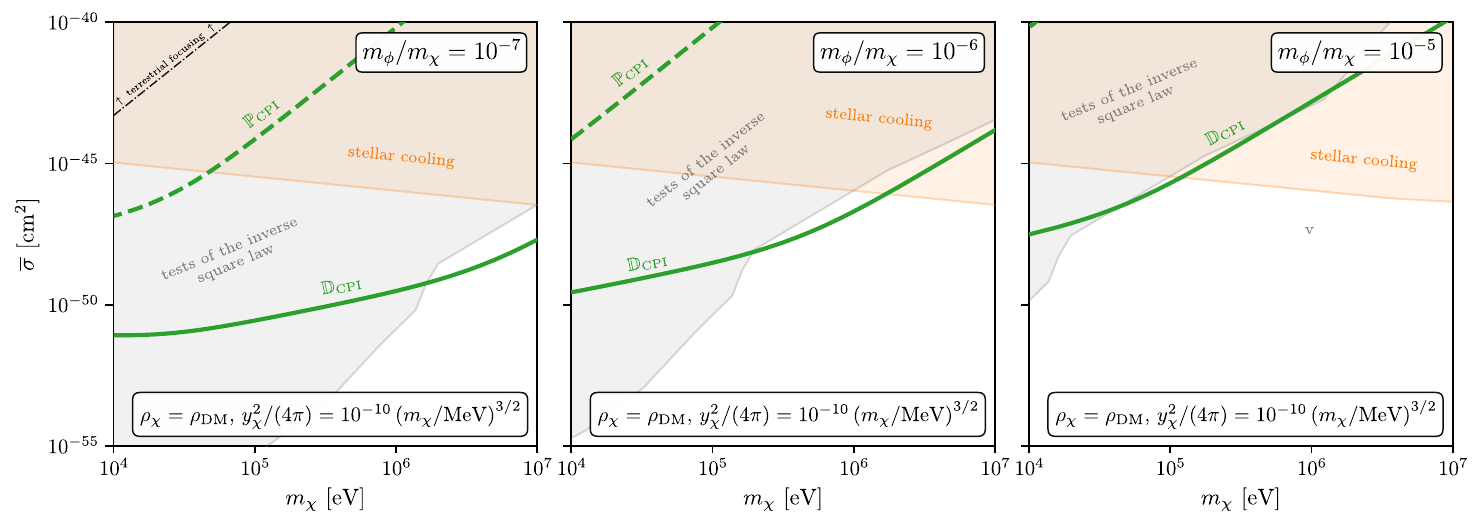}
\caption{Projected sensitivity of the CPI (green) benchmark to spin-independent elastic DM-nucleon scattering via the exchange of a light hadrophilic mediator (see Eqs.~\eqref{eq:long} and \eqref{eq:DM_bilinears}), assuming that $\chi$ is the totality of DM. To be consistent with SIDM constraints, we plot the reference cross section $\overline{\sigma}$ (see Eq.~\eqref{eq:ref_cs}) as a function of the DM mass $m_\chi$ for $y_\chi^2/(4\pi)=10^{-10} (m_\chi/\mathrm{MeV})^{3/2}$~\cite{Knapen:2017xzo}.
We fix the ratio $R_{\phi,\chi}=m_\phi/m_\chi$ to be equal to either $10^{-7}$ (left panel), $10^{-6}$ (central panel) or $10^{-5}$ (right panel).
All other plotting conventions and shaded regions are identical to those in Fig.~\ref{fig:sigma_reach}.
}
\label{fig:sigma_reach_tot}
\end{figure}

Equipped with Eqs.~\eqref{eq:D_scatt_long_final} and \eqref{eq:P_scatt_long_final}, in Figs.~\ref{fig:sigma_reach} and \ref{fig:sigma_reach_tot} we show the projected reach of the benchmarks to elastic spin-independent DM-nucleon scattering via the exchange of a light hadrophilic scalar.
We plot the projected sensitivity to the reference cross section $\overline{\sigma}$,
\begin{equation}\label{eq:ref_cs}
\overline{\sigma} = \frac{y_\chi^2y_n^2}{4\pi} \frac{m_\chi^2}{(m_\chi^2v_0^2+m_\phi^2)^2} \, ,
\end{equation}
for fixed ratios $R_{\phi,\chi}\equiv m_\phi/m_\chi$. 
For all ratios and DM masses considered here, $m_\phi^{-1}$ is much smaller than the size of the shield housing the experiment, such that laboratory effects can be neglected.
Furthermore, for the range of reference cross sections shown, DM thermalization and isotropization by the atmosphere or by the local laboratory environment are negligible~\cite{Riedel:2016acj}.
Using Eq.~\eqref{eq:ref_cs} and in the dilute DM limit, Eqs.~\eqref{eq:D_scatt_long_final} and \eqref{eq:P_scatt_long_final} can be rewritten as
\begin{equation}\label{eq:P_scatt_long_final_ref}
\begin{aligned}
\mathbb{D}&=4 \pi T \overline{\sigma} \tfrac{\rho_\chi}{m_\chi^3} \int\! \tfrac{\dbar^3q}{|\mathbf{q}|} \, |F_\mathbb{D}(\mathbf{q})|^2 |\mathcal{F}_\mathrm{\phi,\mathrm{eff}}(\mathbf{q})|^2 [1-\cos(\mathbf{q}\!\cdot\!\boldsymbol{\Delta x})] g(\mathbf{q}) + \mathcal{O}(y_n^3)  , \\
\mathbb{P}&=4 \pi T \overline{\sigma} \tfrac{\rho_\chi}{m_\chi^3} \int\! \tfrac{\dbar^3q}{|\mathbf{q}|}  \, |F_\mathbb{P}(\mathbf{q})|^2 |\mathcal{F}_\mathrm{\phi,\mathrm{eff}}(\mathbf{q})|^2 \sin(\mathbf{q}\!\cdot\!\boldsymbol{\Delta x}) g(\mathbf{q}) + \mathcal{O}(y_n^3) \,,
\end{aligned}
\end{equation}
with $\mathcal{F}_\mathrm{\phi,\mathrm{eff}}(\mathbf{q})$ the \textit{dressed} DM mediator form factor:
\begin{equation}
|\mathcal{F}_\mathrm{\phi,\mathrm{eff}}(\mathbf{q})|^2 = \frac{(m_\chi^2v_0^2+m_\phi^2)^2}{(|\mathbf{q}|^2+m_\phi^2 + \mathrm{Re} [ \widetilde{\Pi}^\chi_\mathrm{r} \left ( 0, \mathbf{q} \right )  ] )^2 + \mathrm{Im} [ \widetilde{\Pi}^\chi_\mathrm{r} \left ( 0, \mathbf{q} \right ) ] ^2} \, .
\end{equation}

The scaling of the CPI and AI curves can be understood by studying different kinematic regimes. To facilitate this analysis, we define $R_{\mathrm{eff},\phi}\equiv m_{\phi,\mathrm{eff}}/m_{\phi}$. 
Note that, since $R_{\phi,\chi} \lesssim v_0$, all curves satisfy the condition $m_\phi < m_\chi v_0 $, such that $|\mathbf{q}|$ can always be set equal to $m_\phi$. 
\paragraph{Phase shift.}
We begin by studying the sensitivity arising from the phase shift observable.
\begin{itemize}
\item $m_\chi \ll (R_{\phi,\chi}|\boldsymbol{\Delta x}|)^{-1}$.---This regime is equivalent to $m_\phi \ll |\boldsymbol{\Delta x}|^{-1}$. 
The rate is dominated by $|\mathbf{q}|\sim m_{\phi,\mathrm{eff}}$, such that $p_\mathbb{P} \propto |\mathbf{q}|^2 |\boldsymbol{\Delta x}|/m_\chi v_0$ and $|\mathcal{F}_{\phi,\mathrm{eff}}|^2 \sim (m_\chi v_0/m_{\phi,\mathrm{eff}})^4$.
Since $r_c \ll |\boldsymbol{\Delta x}|$, coherent enhancements are present: $|F_\mathrm{CPI}|^2\propto N_\mathrm{at}^2A^2$ and $|F_\mathrm{AI}|^2\propto N_\mathrm{at}A^2$. 
Hence, $\mathbb{P}_\mathrm{CPI}\sim T \overline{\sigma} \rho_\chi A^2 N_\mathrm{at}^2 |\boldsymbol{\Delta x}| v_0^2$ and $\mathbb{P}_\mathrm{AI}\sim T \overline{\sigma} \rho_\chi A^2 N_\mathrm{at} |\boldsymbol{\Delta x}| v_0^2$, and the shot noise limited reach after $N_\mathrm{exp}$ runs is parametrically 
\begin{equation}
\overline{\sigma} \gtrsim N_\mathrm{exp}^{-1/2}(\rho_\chi T |\boldsymbol{\Delta x}|)^{-1}(v_0 A)^{-2} \begin{cases}
N_\mathrm{at}^{-2} & \text{for CPIs,} \\
N_\mathrm{at}^{-3/2}& \text{for AIs.} 
\end{cases}
\end{equation}
Note that the scaling here is independent of $m_\chi$ and improves for large arm separations. 
Also, for $m_{\phi,\mathrm{eff}}^2 >0$, this regime is insensitive to finite density effects.
\item $m_\chi \gg (R_{\phi,\chi}|\boldsymbol{\Delta x}|)^{-1}$.---In this regime, $m_\phi \gg |\boldsymbol{\Delta x}|^{-1}$, so that the phase factor oscillates rapidly for large momentum transfer. 
Provided that $m_{\phi,\mathrm{eff}}\gg |\boldsymbol{\Delta x}|^{-1}$, the rate is dominated by $|\mathbf{q}|\sim |\boldsymbol{\Delta x}|^{-1}$. 
At this momentum transfer, $|\mathcal{F}_{\phi,\mathrm{eff}}|^2 \sim (v_0/R_{\phi,\chi}R_{\mathrm{eff},\phi})^4$ and $p_\mathbb{P} \propto (m_\chi v_0 |\boldsymbol{\Delta x}|)^{-1}$. 
Since $r_c \ll |\boldsymbol{\Delta x}|$,  coherent enhancements are still kinematically allowed, such that $|F_\mathrm{CPI}|^2\propto N_\mathrm{at}^2A^2$ and $|F_\mathrm{AI}|^2\propto N_\mathrm{at}A^2$, and $\mathbb{P}_\mathrm{CPI} \sim T \overline{\sigma} \rho_\chi A^2 N_\mathrm{at}^2 v_0^2 (R_{\phi,\chi}R_{\mathrm{eff},\phi}m_\chi)^{-4} |\boldsymbol{\Delta x}|^{-3}$ and $\mathbb{P}_\mathrm{AI} \sim T \overline{\sigma} \rho_\chi A^2 N_\mathrm{at} v_0^2 (R_{\phi,\chi}R_{\mathrm{eff},\phi}m_\chi)^{-4} |\boldsymbol{\Delta x}|^{-3}$. 
The sensitivity to the reference cross section is therefore 
\begin{equation}
\overline{\sigma} \gtrsim (R_{\phi,\chi}R_{\mathrm{eff},\phi}m_\chi)^{4} |\boldsymbol{\Delta x}|^{3} N_\mathrm{exp}^{-1/2} (\rho_\chi T)^{-1}(v_0 A)^{-2} \begin{cases}
N_\mathrm{at}^{-2} & \text{for CPIs,} \\
N_\mathrm{at}^{-3/2}& \text{for AIs.} 
\end{cases}
\end{equation}
In this regime, the reach deteriorates as $m_\chi^4$ and dramatically improves for light mediators via $R_{\phi,\chi}^4$. 
Furthermore, in this regime we also note that the sensitivity can be parametrically enhanced when $R_{\mathrm{eff},\phi}\ll1$.
\end{itemize}

In light of these scalings, and in the limit where finite density effects are negligible, i.e., $R_{\mathrm{eff},\phi}\approx1$, both CPIs and AIs have peak sensitivity for $m_\chi \lesssim (R_{\phi,\chi}|\boldsymbol{\Delta x}|)^{-1}$. 
This corresponds to $m_\chi \lesssim (4\times 10^{-8}/R_{\phi,\chi})~\mathrm{eV}$ and $m_\chi \lesssim (10^{-3}/R_{\phi,\chi})~\mathrm{eV}$ for the AI and CPI benchmarks, respectively. 
Note that, for our choice of parameters, the AI benchmark's phase could only probe reference cross sections well above $10^{-35}~\mathrm{cm}^2$, so its projected reach lies outside the range shown in Figs.~\ref{fig:sigma_reach} and \ref{fig:sigma_reach_tot}.

\paragraph{Decoherence.} We now study the sensitivity from the decoherence observable. Because decoherence in AIs is insensitive to the atom cloud structure, we study the AI and CPI cases separately where appropriate.
\begin{itemize}
\item CPI and AI
\begin{itemize}
\item
$m_\chi \ll (R_{\phi,\chi}|\boldsymbol{\Delta x}|)^{-1}$.---In this regime, $m_\phi \ll |\boldsymbol{\Delta x}|^{-1}$ and the rate is dominated by $|\mathbf{q}|\sim m_{\phi,\mathrm{eff}}$, as for the phase shift case. Differently from the phase shift case, $p_\mathbb{D} \propto (|\mathbf{q}||\boldsymbol{\Delta x}|)^{2}$, and only the CPI observable receives a large $N_\mathrm{at}$ enhancement. Consequently, 
$\mathbb{D}_\mathrm{CPI}\sim T \overline{\sigma} \rho_\chi A^2 N_\mathrm{at}^2 m_\chi |\boldsymbol{\Delta x}|^2 v_0^3$ and $\mathbb{D}_\mathrm{AI} \sim T \overline{\sigma} \rho_\chi A^2 m_\chi |\boldsymbol{\Delta x}|^2 v_0^3$, such that the shot-noise limited sensitivity is parametrically
\begin{equation}
\overline{\sigma} \gtrsim N_\mathrm{exp}^{-1/2}(\rho_\chi T m_\chi)^{-1} (A|\boldsymbol{\Delta x}|)^{-2}v_0^{-3} \begin{cases}
N_\mathrm{at}^{-2} & \text{for CPIs,} \\
N_\mathrm{at}^{-1/2}& \text{for AIs.} 
\end{cases}
\end{equation}
As for the phase shift, in this regime finite density effects are not relevant.
\end{itemize}
\item AI only
\begin{itemize}
\item $m_\chi \gg (R_{\phi,\chi} |\boldsymbol{\Delta x}|)^{-1}$.---In this regime, $m_\phi  \gg |\boldsymbol{\Delta x}|^{-1}$. 
Because $p_\mathbb{D}\sim |F_\mathbb{A}|^2$ for $|\mathbf{q}|\gtrsim |\boldsymbol{\Delta x}|^{-1}$, the rate is dominated by $ |\mathbf{q}|\sim m_{\phi,\mathrm{eff}}$. 
Since we truncate the plot at $m_\chi =10$~MeV, the observable is not suppressed by the nuclear form factor. Hence, $\mathbb{D}_\mathrm{AI}\sim T \overline{\sigma} \rho_\chi A^2 v_0^3 m_\chi^{-1} (R_{\phi,\chi}R_{\mathrm{eff},\phi})^{-2}$, and 
\begin{equation}
\overline{\sigma} \gtrsim (R_{\phi,\chi}R_{\mathrm{eff},\phi})^{2} m_\chi (N_\mathrm{exp}N_\mathrm{at})^{-1/2}(\rho_\chi T)^{-1} A^{-2}v_0^{-3} 
\end{equation}
Hence, the sensitivity deteriorates as $m_\chi$ but can be parametrically enhanced when $m_{\phi,\mathrm{eff}}^2\to 0$.
\end{itemize}
\item CPI only
\begin{itemize}
\item $(R_{\phi,\chi}|\boldsymbol{\Delta x}|)^{-1} \lesssim m_\chi \lesssim (R_{\phi,\chi}r_c)^{-1}$.---In this regime, $|\boldsymbol{\Delta x}|^{-1} \lesssim m_\phi \lesssim r_c^{-1}$. Because $p_\mathbb{D}\sim |F_\mathbb{D}|^2$ for $|\mathbf{q}|\gtrsim |\boldsymbol{\Delta x}|^{-1}$, and if $|\boldsymbol{\Delta x}|^{-1} \lesssim m_{\phi,\mathrm{eff}} \lesssim r_c^{-1}$, the rate is dominated by $ |\mathbf{q}|\sim m_{\phi,\mathrm{eff}}$, and inherits coherent enhancements. 
Hence, $\mathbb{D}_\mathrm{CPI}\sim T \overline{\sigma} \rho_\chi A^2 N_\mathrm{at}^2 v_0^3 m_\chi^{-1} (R_{\phi,\chi}R_{\mathrm{eff},\phi})^{-2}$ and 
\begin{equation}
\overline{\sigma} \gtrsim (R_{\phi,\chi}R_{\mathrm{eff},\phi})^{2} m_\chi N_\mathrm{exp}^{-1/2}(\rho_\chi T)^{-1} (AN_\mathrm{at})^{-2}v_0^{-3} \, .
\end{equation}
In this regime, the sensitivity also deteriorates with $m_\chi$ and is susceptible to finite density enhancements.
\item $m_\chi \gtrsim (R_{\phi,\chi}r_c)^{-1}$.---This is the last regime that we shall consider and which is relevant to Fig.~\ref{fig:sigma_reach}.
 Here, $m_\phi \gg r_c^{-1}$. 
 To ensure coherent enhancements, $|\mathbf{q}|\lesssim r_c^{-1}$, so the rate is dominated by $|\mathbf{q}|\sim r_c^{-1}$ provided that $m_{\phi,\mathrm{eff}} \gg r_c^{-1}$.  Hence, $\mathbb{D}_\mathrm{CPI}\sim T \overline{\sigma} \rho_\chi A^2 N_\mathrm{at}^2 v_0^3 m_\chi^{-3} (R_{\phi,\chi}R_{\mathrm{eff},\phi})^{-4}r_c^{-2}$, and 
\begin{equation}
\overline{\sigma} \gtrsim (R_{\phi,\chi}R_{\mathrm{eff},\phi})^{4} m_\chi^3 r_c^2 N_\mathrm{exp}^{-1/2}(\rho_\chi T)^{-1} (N_\mathrm{at}A)^{-2}v_0^{-3} \, .
\end{equation}
\end{itemize}
\end{itemize}

In the limit where finite density effects can be neglected, our results agree exactly with Ref.~\cite{Du:2022ceh} up to differences in the choice of experimental parameters. Because $|\boldsymbol{\Delta x}| > r_c$ in both benchmarks considered here, the peak sensitivity from this observable occurs at $m_\chi \sim (R_{\phi,\chi} |\boldsymbol{\Delta x}|)^{-1}$, corresponding to $m_\chi \sim (4\times 10^{-8}/R_{\phi,\chi})~\mathrm{eV}$ and $m_\chi \sim (10^{-3}/R_{\phi,\chi})~\mathrm{eV}$ for the AI and CPI benchmarks, respectively. 
As with the phase observable, for our choice of parameters the AI benchmark is only sensitive to reference cross sections via decoherence above $10^{-35}~\mathrm{cm}^2$, so its projected reach lies outside the range shown in Figs.~\ref{fig:sigma_reach} and \ref{fig:sigma_reach_tot}.

Notably, in light of constraints from stellar cooling and tests of the inverse square law, we can see from  Fig.~\ref{fig:sigma_reach} that a 5\% DM subcomponent with mass between $\sim10$~eV and $\sim10$~MeV could be detectable for both large and small mediator-DM couplings by the CPI benchmark. 
In light of the onset of the tachyonic instability and the lack of $N_\mathrm{at}^2$ enhancements, AIs are \textit{not} expected to probe any unexplored region of parameter space that is not subject to DM condensation. 
Furthermore, tachyonic instabilities occur over large regions of parameter space.
Finally, note that in large regions of parameter space, DM refraction caused by the Earth and atmosphere can induce order-one changes in the DM speed, and therefore in the DM density, at terrestrial experiments (see Appendix~\ref{app:DM_mod} for a derivation). 
This is the non-gravitational analogue of gravitational focusing of DM~\cite{Griest:1987vc,Kim:2021yyo}, and has previously been explored in the context of very light scalar mediators~\cite{Davoudiasl:2020ypv}.\footnote{This is a mean-field effect and is distinct from terrestrial accumulation of DM due to large spin-independent DM-nucleon cross sections. Notably, this effect is important for much larger cross sections and DM masses than those shown in Figs.~\ref{fig:sigma_reach} and \ref{fig:sigma_reach_tot}~\cite{Mack:2007xj,McKeen:2022poo,Leane:2022hkk}.}
Remarkably, once SIDM constraints are taken into account, the CPI benchmark could probe scenarios in which $\chi$ makes up all of the DM, as shown in Fig.~\ref{fig:sigma_reach_tot}. 
Indeed, the $\mathbb{D}_\mathrm{CPI}$ observable reaches well below current stellar-cooling bounds and into unexplored regions of parameter space.

\section{Summary \& conclusions}\label{sec:discussion}

In this work we presented an \textit{open} EFT for computing DM-induced effects in MWIs. By making use of the SK formalism, we extended the single-atom and contact interactions case considered in Ref.~\cite{Badurina:2026owr} to also encompass long-range interactions and realistic MWI experiments employing $N_\mathrm{at}\gg 1$ atoms, e.g., AIs and CPIs.

Firstly, by mapping the open EFT from the field level to system worldlines, we rigorously showed the precise sense in which MWIs are sensitive to DM signatures in the absence of DM-induced corrections to the system worldlines. 
This is equivalent to the observation that DM can induce changes in the MWI reduced density matrix in the absence of energy deposition.
Notably, we showed that this approximation is valid across the DM mass range considered in this work.

Secondly, by extending our analysis to many-atom systems, we recovered multi-atom form factors previously identified in the literature in the context of elastic spin-independent DM-nucleon scattering~\cite{Riedel:2012ur,Riedel:2016acj,Du:2022ceh,Badurina:2024nge,Murgui:2026mlr}.
Notably, these form factors we derived are independent of the DM-MWI coupling and are valid also in the non-Markovian regime.
For CPI experiments, we find that all observables can receive at leading order an $N_\mathrm{at}^2$-enhancement. 
For AIs, at leading order the decoherence and phase form factors differ: phase shifts can receive an $N_\mathrm{at}$-enhancement but decoherence remains strictly $N_\mathrm{at}$ independent. 
Nevertheless, for a shot-noise limited experiment, employing $N_\mathrm{at}$ atoms can increase the sensitivity to DM-SM couplings via Poisson statistics. 
This can be understood intuitively by mapping the phase and decoherence to the number of atom counts in either measurement port.

Most importantly, our results also provide a new phenomenological perspective on DM searches with MWIs. 
For contact interactions, Bose enhancements parametrically increase the sensitivity of CPIs to scalar DM candidates with masses well below $\mathcal{O}(10~\mathrm{eV})$. 
Taking our benchmarks at face value, this enhancement may allow proposed and ambitious CPIs to probe otherwise unexplored regions of Higgs-portal DM parameter space.
Furthermore, in the context of attractive long-range interactions, as is the case for DM-nucleon interactions via a light hadrophilic scalar mediator, the DM background can give rise to a \textit{negative} correction to the mediator mass squared. 
For sufficiently small DM masses and/or large mediator-DM coupling strengths, this correction can give rise to a tachyonic instability in relevant regions of DM parameter space. 
Provided that $m^2_{\phi,\mathrm{eff}} \geq 0$, the in-medium correction to the mediator's mass could in principle increase the interaction range, thereby enhancing the CPI and AI reach to sub-MeV DM. 
We find, however, this effect to be negligible for the DM and MWI benchmarks considered here.

This framework also opens up new avenues for exploring the DM landscape. 
By neglecting finite DM energy depositions, we automatically excluded contributions to the CPI observable from phonon excitations.
Because phonon excitations in solids and other materials are expected to provide promising reach to DM with masses $m_\chi \gtrsim \text{MeV}$ (see, e.g., Ref.~\cite{Kahn:2021ttr} for a review), we will explore these signatures in future work.
Our formalism also clarifies the necessary steps required to compute the reach to heavier DM candidates, where the heavy-probe approximation breaks down. 
Furthermore, a phenomenological survey of the model-dependent and model-independent constraints at the under-explored wave-particle boundary is warranted. 

Finally, our formalism can also be used to go beyond the leading order contribution to the decoherence and phase shift observables, and to study ordinary environmental backgrounds in MWI experiments, such as gas collisions, blackbody radiation, vibration noise, and laser phase noise. 
Indeed, having identified the relationship between the effective action and MWI observables, and having established how to construct environment kernels using in-in Feynman rules, one can systematically compute corrections to the phase and decoherence for any MWI to arbitrary order in perturbation theory. 
We leave this more technical direction to future work.

\begin{acknowledgments}
We thank Itay Bloch, Federico Cima, Sebastian A. R. Ellis, Angelo Esposito, Roni Harnik, Junwu Huang, Christopher McCabe, Michele Papucci, Pablo Quilez, Jesse Shelfhout, Allic Sivaramakrishnan, and Jordan Wilson-Gerow for discussions. 
L.B. and K.Z. are supported by the U.S. Department of Energy, Office of Science, Office of High Energy Physics under Award Number DE-SC0011632, and by the Walter Burke Institute for Theoretical Physics. 
K.Z. is also supported by a Simons Investigator award, and by the Heising-Simons Foundation ``Observational Signatures of Quantum Gravity'' collaboration grant. 
This work was performed in part at the Aspen Center for Physics, which is supported by National Science Foundation grant PHY-2210452. 
L.B. also thanks the Mainz Institute for Theoretical Physics (MITP) of the Cluster of Excellence PRISMA+ (Project ID 390831469), and the CERN Theory group for their hospitality and support during the completion of this work. 
L.B. would also like to thank the participants of the TH-Institute ``Quantum sensing for fundamental physics" for many stimulating discussions.
GPT-6 Astra and Claude Fable 5.1 were used under human supervision to produce all figures and to proof-read the paper. 
We take full responsibility for the accuracy of all aspects of this work.
\end{acknowledgments}

\clearpage

\appendix

\section{Single-atom reduced density matrix at finite mass}\label{app:finite-mass}

In this appendix, we show how to derive the expressions for the phase shift and decoherence observable at finite atomic mass $m_\mathrm{at}$. 
Although we focus on a single-atom MWI, these results also apply to multi-atom systems.

In section~\ref{sec:toy}, we showed that, after expanding around the classical trajectories, setting $t_f=-t_i=\mathcal{T}$ and assuming closed and symmetric interferometric sequences, the reduced density matrix of the system can be expressed as 
\begin{equation}\label{app:eq:rho_red}
\begin{aligned}
\rho_s^{\alpha \beta}(\mathcal{T}) & = \tfrac{1}{2} \left ( \tfrac{m_\mathrm{at}}{4\pi\mathcal{T}} \right )^3 \int\!\!\mathrm{d}^3 x_{i,+}\int \mathrm{d}^3 x_{i,-} \!\!\,  \Phi(\mathbf{x}_{i,+})\Phi^*(\mathbf{x}_{i,-})
\\ & \qquad \times
\int\!\!\mathrm{d}^3 x_f \, e^{\tfrac{\mi m_\mathrm{at}}{4\mathcal{T}} \left (|\mathbf{x}_f-\mathbf{x}_{i,+}|^{2}-|\mathbf{x}_f-\mathbf{x}_{i,-}|^{2}\right)} e^{\mi S_{_\mathrm{IF}}\left [\overline{\boldsymbol{X}}_{\alpha,+} - \mi \tfrac{\boldsymbol{\delta}}{\boldsymbol{\delta j}_+},\overline{\boldsymbol{X}}_{\beta,-}+ \mi \tfrac{\boldsymbol{\delta}}{\boldsymbol{\delta j}_-}\right ]}  Z_0[\boldsymbol{j}_+,\boldsymbol{j}_-]\Big |_{\boldsymbol{j}_\pm = 0} \\
& \equiv \tfrac{1}{2} +  \tfrac{\mi}{2} \left \langle S_{_\mathrm{IF}}\left [\overline{\boldsymbol{X}}_{\alpha,+} - \mi \tfrac{\boldsymbol{\delta}}{\boldsymbol{\delta j}_+},\overline{\boldsymbol{X}}_{\beta,-}+ \mi \tfrac{\boldsymbol{\delta}}{\boldsymbol{\delta j}_-}\right ] \right \rangle Z_0[\boldsymbol{j}_+,\boldsymbol{j}_-]\Big |_{\boldsymbol{j}_\pm = 0} +\mathcal{O}(g^3) \, ,
\end{aligned}
\end{equation}
(see~Eq.~\eqref{eq:rhos_step} and the surrounding discussion).
Introducing the vector
\begin{equation}\label{app:eq:W}
\begin{aligned}
\vec{\widetilde{\mathcal{W}}}(\mathbf{q},t;\boldsymbol{X}_{\pm})^\mathsf{T} &=   \left (\widetilde{\mathcal{W}}^\mathrm{cl}(\mathbf{q},t;\boldsymbol{X}_{\pm}) \quad \widetilde{\mathcal{W}}^\mathrm{q}(\mathbf{q},t;\boldsymbol{X}_{\pm}) \right ) \, ,
\end{aligned}
\end{equation}
which depends on the worldline histories via Eq.~\eqref{eq:W}, and using Eq.~\eqref{eq:Sif}, the second term in last line of Eq.~\eqref{app:eq:rho_red} can be rewritten as
\begin{equation}\label{app:eq:Sif_q}
\begin{aligned}
& \text{2\textsuperscript{nd} term  of Eq.~\eqref{app:eq:rho_red}} = 
-\tfrac{\mi}{4}\!\left(\tfrac{g}{\Lambda^D}\right)^{\!2}\!\! \int_{-\mathcal{T}}^{\mathcal{T}} \mathrm{d} t \!\!  \int_{-\mathcal{T}}^{\mathcal{T}} \mathrm{d} t' \!\!  \int \dbar^3 q
|F_A(\mathbf{q})|^2 \\
&\quad \quad \quad \times 
\left \langle \vec{\widetilde{W}}\left (-\mathbf{q},t;\overline{\boldsymbol{X}}_{\pm}\mp \mi \tfrac{\boldsymbol{\delta}}{\boldsymbol{\delta j}_\pm}\right)^{\mathsf T}
\,\widetilde{\mathsf{K}}(\mathbf{q},t-t')\,
\vec{\widetilde{W}}\left(\mathbf{q},t';\overline{\boldsymbol{X}}_{\pm}\mp \mi \tfrac{\boldsymbol{\delta}}{\boldsymbol{\delta j}_\pm} \right) \right \rangle Z_0[\boldsymbol{j}_+,\boldsymbol{j}_-]\Big |_{\boldsymbol{j}_\pm = 0} \, .
\end{aligned}
\end{equation}

In light of Eqs.~\eqref{app:eq:W} and \eqref{eq:W}, we can factorise the basic worldline-dependent objects entering Eq.~\eqref{app:eq:Sif_q} into classical path dependent and worldline fluctuation dependent pieces: for $m,n\in\{+,-\}$, the relevant object is
\begin{equation}\label{app:eq:average1}
\text{2\textsuperscript{nd} line of Eq.~\eqref{app:eq:Sif_q}} \supset 
\left\langle
e^{-\mi\mathbf{q}\cdot\overline{\boldsymbol{X}}_{m}(t)}
e^{+\mi\mathbf{q}\cdot\overline{\boldsymbol{X}}_{n}(t')}
\right\rangle \underbrace{e^{-m\mathbf{q}\cdot\tfrac{\boldsymbol{\delta}}{\boldsymbol{\delta j}_m(t)}}e^{n\mathbf{q}\cdot\tfrac{\boldsymbol{\delta}}{\boldsymbol{\delta j}_n(t')}} Z_0[\boldsymbol{j}_+,\boldsymbol{j}_-]\Big |_{\boldsymbol{j}_\pm = 0}}_{e^{\mi \theta_{mn}(t,t')}} \, ,
\end{equation}
where $\theta_{mn}(t,t')$ captures the correction from worldline fluctuations. 
Using the definition of the generating functional for worldline fluctuations (see Eq.~\eqref{eq:gen_func}), the phases $\theta_{mn}(t,t')$ can be expressed in terms of Dirichlet propagators (cf. Eq.~\eqref{eq:Dirichlet_prop}) as
\begin{equation}\label{app:eq:theta}
\begin{aligned}
\theta_{++}(t,t') = - \theta_{--}(t,t')
&=
\tfrac{|\mathbf{q}|^2}{2}
\left[
G_{\rm d}(t,t)
+
G_{\rm d}(t',t')
-
2G_{\rm d}(t,t')
\right],
\\
\theta_{+-}(t,t')
= -\theta_{-+}(t,t') &= 
\tfrac{|\mathbf{q}|^2}{2}
\left[
G_{\rm d}(t,t)
-
G_{\rm d}(t',t')
\right] \, .
\end{aligned}
\end{equation}
Using Eq.~\eqref{eq:Xbar}, the classical (unperturbed) worldline-dependent contribution in Eq.~\eqref{app:eq:average1} reads as
\begin{equation}\label{app:eq:average}
\begin{aligned}
\left\langle
e^{-\mi\mathbf{q}\cdot\overline{\boldsymbol{X}}_{m}(t)}
e^{+\mi\mathbf{q}\cdot\overline{\boldsymbol{X}}_{n}(t')}
\right\rangle &= \left\langle
e^{-\mi\mathbf{q}\cdot\left[\mathbf{x}_{i,m}\left(\tfrac{\mathcal{T}-t}{2\mathcal{T}}\right) - \mathbf{x}_{i,n}\left(\tfrac{\mathcal{T}-t'}{2\mathcal{T}}\right) + \mathbf{x}_f \left ( \tfrac{t-t'}{2\mathcal{T}} \right )+\overline{\boldsymbol{x}}_{m}(t)-\overline{\boldsymbol{x}}_{n}(t')\right]}
\right\rangle \\
& =  e^{-\mi\mathbf{q}\cdot\left(\overline{\boldsymbol{x}}_{m}(t)-\overline{\boldsymbol{x}}_{n}(t')\right)} \\
& \quad \times \int\!\!\mathrm{d}^3 x_{i,+}\int \mathrm{d}^3 x_{i,-} \Phi(\mathbf{x}_{i,+})\Phi^*(\mathbf{x}_{i,-}) \delta^{(3)}\!\left(
\mathbf{x}_{i,-}-\mathbf{x}_{i,+}
+\tfrac{t'-t}{m_\mathrm{at}}\mathbf{q}
\right) \\
& \quad  \times e^{-\mi\mathbf{q}\cdot\left[\mathbf{x}_{i,m}\left(\tfrac{\mathcal{T}-t}{2\mathcal{T}}\right) - \mathbf{x}_{i,n}\left(\tfrac{\mathcal{T}-t'}{2\mathcal{T}}\right) \right]}e^{\tfrac{\mi m_\mathrm{at}}{4\mathcal{T}}(|\mathbf{x}_{i,+}|^2-|\mathbf{x}_{i,-}|^2)} \\
& = e^{-\mi\mathbf{q}\cdot\left(\overline{\boldsymbol{x}}_{m}(t)-\overline{\boldsymbol{x}}_{n}(t')\right)} e^{\mi \mathcal{\phi}_{mn}(t,t')} \underbrace{\int\!\!\mathrm{d}^3 x_{i,+} \Phi(\mathbf{x}_{i,+})\Phi^*\left (\mathbf{x}_{i,+}+\tfrac{t-t'}{m_\mathrm{at}}\mathbf{q} \right)}_{F_W(\mathbf{q},t-t')} \, ,
\end{aligned}
\end{equation} 
where, in going to the second line, the integral over $\mathbf{x}_f$ produces the delta function, and in the final step we introduced the phase
\begin{equation}\label{app:eq:phi}
\phi_{mn}(t,t')
=
\tfrac{|\mathbf{q}|^2(t-t')}{2m_\mathrm{at}\mathcal T}
\left[
\nu_n(\mathcal T-t')
-
\nu_m(\mathcal T-t)
\right]
-
\tfrac{|\mathbf{q}|^2(t-t')^2}{4m_\mathrm{at}\mathcal T},
\, \,  \text{with} \, \, 
\nu_+=0, \, 
\nu_-=1 \, ,
\end{equation}
and the time-dependent wavepacket form factor $F_W(\mathbf{q},t-t')$.
Using Eqs.~\eqref{app:eq:theta}--\eqref{app:eq:phi}, and defining 
\begin{equation}
\varphi_{mn}(t-t') \equiv \phi_{mn}(t,t')+\theta_{mn}(t,t') \, , 
\end{equation}
which parametrizes the eikonal phase due to environment-induced recoils, Eq.~\eqref{app:eq:average1} reads as 
\begin{equation}\label{app:eq:averaged_final}
\begin{aligned}
\left\langle
e^{-\mi\mathbf{q}\cdot\overline{\boldsymbol{X}}_{m}(t)}
e^{+\mi\mathbf{q}\cdot\overline{\boldsymbol{X}}_{n}(t')}
\right\rangle e^{\mi \theta_{mn}(t,t')} &= e^{-\mi\mathbf{q}\cdot\left(\overline{\boldsymbol{x}}_{m}(t)-\overline{\boldsymbol{x}}_{n}(t')\right)} e^{\mi\varphi_{mn}(t-t')}  F_W(\mathbf{q},t-t') \, ,
\end{aligned}
\end{equation}
where
\begin{equation}\label{app:eq:varphi}
\begin{aligned}
\varphi_{++}(t-t') &= -\varphi_{--}(t-t') = -\tfrac{|\mathbf{q}|^2}{2m_\mathrm{at}}|t-t'| \, ,\\
\varphi_{+-}(t-t') &= -\varphi_{-+}(t-t') = \tfrac{|\mathbf{q}|^2}{2m_\mathrm{at}}(t-t') \, , 
\end{aligned}
\end{equation}
and
\begin{equation}\label{app:eq:FW}
F_W(\mathbf{q},t-t') = e^{-|\mathbf{q}|^2(t-t')^2/8m_\mathrm{at}^2\sigma^2} \, ,
\end{equation}
for Gaussian wavepackets with width $\sigma$.

As is evident from Eqs.~\eqref{app:eq:averaged_final}--\eqref{app:eq:FW}, finite mass corrections enter the observable via $\varphi_{mn}(t-t')$ and $F_W(\mathbf{q},t-t')$.
Since the system currents have support for $t,t' \in [-T_s/2,T_s/2]$, where $T_s$ is the interferometer time, and $\widetilde{\mathsf{K}}(\mathbf{q};t-t')$ in Eq.~\eqref{app:eq:Sif_q} has support for $|t-t'| \lesssim \tau_e$, where $\tau_e$ is the time-scale of the environment (e.g., the DM coherence time), these corrections are parametrically governed by $|\mathbf{q}|^2 \tau_\mathrm{eff}/m_\mathrm{at}$ and $|\mathbf{q}|^2 \tau_\mathrm{eff}^2/m_\mathrm{at}^2\sigma^2 \sim |\mathbf{q}|^2 \tau_\mathrm{eff}^2\sigma_p^2/m_\mathrm{at}^2$, where $\tau_\mathrm{eff} = \min(T_s,\tau_e)$ and $\sigma_p \sim 1/\sigma$.  
Equivalently, these corrections arise from considering the finite energy deposited by the environment in typical interactions, as heuristically explained in section~\ref{sec:toy}. 

The heavy-probe limit can then be easily taken at fixed $\mathcal{T}$ and $\sigma$. 
In this limit, the finite mass phase $\varphi_{mn}$ vanishes and $F_W(\mathbf{q},t-t') \rightarrow 1$, such that 
\begin{equation}
\lim_{m_\mathrm{at}\rightarrow \infty}\left\langle
e^{-\mi\mathbf{q}\cdot\overline{\boldsymbol{X}}_{m}(t)}
e^{+\mi\mathbf{q}\cdot\overline{\boldsymbol{X}}_{n}(t')}
\right\rangle e^{\mi \theta_{mn}(t,t')} = e^{-\mi \mathbf{q}\cdot(\overline{\boldsymbol{x}}_m(t)-\overline{\boldsymbol{x}}_n(t'))} \, ,
\end{equation}
and the driven classical histories remain encoded in the influence action via the particular solution $\overline{\boldsymbol{x}}_\pm$ (see Eq.~\eqref{eq:Xbar}).
Taking the asymptotic limit $\mathcal{T}\rightarrow \infty$, Eq.~\eqref{app:eq:Sif_q} simplifies to Eq.~\eqref{eq:Sif_lim}.

\section{Derivation of the system currents and geometry dependence}\label{app:sys_operators}

In this appendix, we explicitly compute the Fourier transform of the system charge current 
\begin{equation}\label{app:eq:FT_sys}
\int\!\mathrm{d}^4x\;
  \varrho(\mathbf{x} -\overline{\boldsymbol{x}}_{{\alpha}}(t))\,e^{\mi q\cdot x}
 =  F_A(\mathbf{q}) \widetilde{\mathcal{W}}(q;\overline{\boldsymbol{x}}_{{\alpha}}) \, ,
\end{equation}
where $F_A(\mathbf{q})$ is the atomic charge form factor, $\overline{\boldsymbol{x}}_{{\alpha}}(t)$ is the time-dependent classical worldline of the $\alpha \in \{L,R\}$ path. 
We show, in particular, how the explicit form of $\widetilde{\mathcal{W}}(q;\overline{\boldsymbol{x}}_{\alpha})$ in Eq.~\eqref{app:eq:FT_sys} depends on the details of the interferometric sequence.

We begin by considering the split and hold interferometric configuration, corresponding to the geometry where the two wavepackets are held at fixed relative separation for the majority of the interferometric sequence (see Fig.~\ref{fig:paths}). 
In this case, 
\begin{equation}\label{app:eq:worldlines_single}
\overline{\boldsymbol{x}}_{L}(t) \approx 
\begin{cases} - \boldsymbol{\Delta x}/2  ~~\text{for $|t|\lesssim T/2$}  \\
0 ~~\text{otherwise}\
\end{cases}\, , \, \overline{\boldsymbol{x}}_{R}(t) \approx 
\begin{cases} + \boldsymbol{\Delta x}/2 ~~\text{for $|t|\lesssim T/2$}  \\
0 ~~\text{otherwise}\
\end{cases}\, .
\end{equation}
where $|\boldsymbol{\Delta x}|$ is the maximal distance between the two paths. 
Hence, $\widetilde{\mathcal{W}}(q;\overline{\boldsymbol{x}}_{L})$ can be obtained by splitting the path into its in-flight ($|t|\lesssim T/2$) and asymptotic ($|t|\gtrsim T/2$) segments: 
\begin{equation}\label{app:eq:stepW1}
\begin{aligned}
\widetilde{\mathcal{W}}(q;\overline{\boldsymbol{x}}_{L}) &= \int_{-\infty}^{\infty} \mathrm{d}t\, e^{\mi q_0 t}\, e^{-\mi\mathbf{q}\cdot \overline{\boldsymbol{x}}_{L}(t)} \\ 
& \approx  \int_{-\infty}^{-T/2} \mathrm{d}t\, e^{\mi q_0 t}\, + \int_{-T/2}^{T/2} \mathrm{d}t\, e^{\mi q_0 t} e^{\mi \mathbf{q}\cdot \boldsymbol{\Delta x}/2} + \int_{T/2}^{\infty} \mathrm{d}t\, e^{\mi q_0 t}\\
& = \deltabar(q_0) + T \mathrm{sinc}\left ( \tfrac{q_0 T}{2}\right ) \left(e^{\mi \mathbf{q}\cdot\boldsymbol{\Delta x}/2}-1\right) \, ,
\end{aligned}
\end{equation}
where we made use of $\int_{-T/2}^{T/2}dt\,e^{\mi q_0t} = T\,\mathrm{sinc}(q_0T/2)$. Hence, one finally obtains
\begin{equation}\label{app:eq:result}
\int\!\mathrm{d}^4x\;
  \varrho(\mathbf{x} -\overline{\boldsymbol{x}}_{L}(t))\,e^{\mi q\cdot x} \approx F_A(\mathbf{q})
  \left [ \deltabar(q_0)
    +T\,\mathrm{sinc} \Bigl(\tfrac{q_0 T}{2}\Bigr)
    \bigl(e^{\mi\mathbf{q}\cdot\boldsymbol{\Delta x}/2}-1\bigr) \right ]\, ,
\end{equation}
where $\widetilde{\mathcal{W}}(q;\overline{\boldsymbol{x}}_{R})=\widetilde{\mathcal{W}}(q;-\overline{\boldsymbol{x}}_{L})$.

\subsection{Mach--Zehnder sequences}  

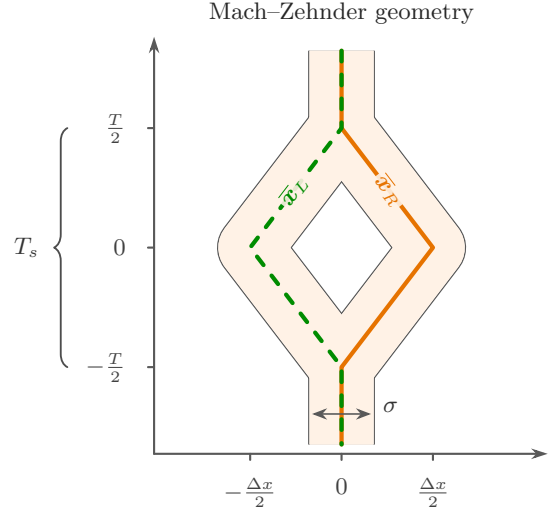
\begin{figure}
\centering
\begin{tikzpicture}[
  >=Stealth,
  line cap=round,
  line join=round,
  font=\footnotesize,
  x=0.62cm,
  y=0.62cm,
  axis/.style={line width=0.8pt, draw=black!65},
  tubeOuter/.style={draw=black!65, line width=25pt, line cap=butt},
  tubeInner/.style={draw=white, line width=24pt, line cap=butt},
  mainR/.style={line width=1.6pt, draw=orange!90!black},
  mainL/.style={line width=1.6pt, draw=green!55!black, dash pattern=on 4pt off 5pt},
  pathLabel/.style={
    sloped,
    fill=orange!10!white,
    fill opacity=0.9,
    text opacity=1,
    inner sep=1.2pt
  }
]
\def\TopY{2.55}
\def\BotY{-2.55}
\def\ArmX{1.95}
\def\AxisBot{-4.4}
\def\AxisTop{4.5}
\node[text=black!85] at (0,5.0) {Mach--Zehnder geometry};
\draw[axis,-{Stealth[length=2.2mm,width=1.4mm]}]
  (-4.0,\AxisBot) -- (-4.0,\AxisTop);
\draw[axis,-{Stealth[length=2.2mm,width=1.4mm]}]
  (-4.0,\AxisBot) -- (4.4,\AxisBot);
\draw[axis] (-4.0,\TopY) -- ++(-0.13,0);
\node[left=4pt,text=black!75] at (-4.18,\TopY)
  {$\frac{T}{2}$};
\draw[axis] (-4.0,0) -- ++(-0.13,0);
\node[left=4pt,text=black!75] at (-4.18,0)
  {$0$};
\draw[axis] (-4.0,\BotY) -- ++(-0.13,0);
\node[left=4pt,text=black!75] at (-4.18,\BotY)
  {$-\frac{T}{2}$};
\foreach \x/\lbl in {
  -1.95/{$-\frac{\Delta x}{2}$},
   0/{$0$},
   1.95/{$\frac{\Delta x}{2}$}
}{
  \draw[axis] (\x,\AxisBot) -- ++(0,-0.13);
  \node[below=3pt,text=black!75] at (\x,-4.55) {\lbl};
}
\coordinate (Top) at (0,\TopY);
\coordinate (UR) at (\ArmX,0);
\coordinate (LR) at (\ArmX,0);
\coordinate (Bot) at (0,\BotY);
\coordinate (LL) at (-\ArmX,0);
\coordinate (UL) at (-\ArmX,0);
\draw[
  decorate,
  decoration={brace, amplitude=5pt},
  line width=0.7pt,
  draw=black!70
]
  (-5.85,\BotY) -- (-5.85,\TopY)
  node[midway,left=7pt,text=black!80] {$T_s$};
\draw[tubeOuter]
  (0,4.2) -- (Top) -- (UR) -- (Bot) -- (0,-4.2);
\draw[tubeOuter]
  (0,4.2) -- (Top) -- (UL) -- (Bot) -- (0,-4.2);
\draw[tubeInner]
  (0,4.2) -- (Top) -- (UR) -- (Bot) -- (0,-4.2);
\draw[tubeInner]
  (0,4.2) -- (Top) -- (UL) -- (Bot) -- (0,-4.2);
\draw[orange!10!white,line width=24pt,line cap=butt]
  (0,4.2) -- (Top) -- (UR) -- (Bot) -- (0,-4.2);
\draw[orange!10!white,line width=24pt,line cap=butt]
  (0,4.2) -- (Top) -- (UL) -- (Bot) -- (0,-4.2);
\draw[mainR]
  (0,4.2) -- (Top) -- (UR) -- (Bot) -- (0,-4.2);
\draw[mainL]
  (0,4.2) -- (Top) -- (UL) -- (Bot) -- (0,-4.2);
\path (UL) -- (Top)
  node[pos=0.48,pathLabel,text=green!55!black]
  {$\overline{\boldsymbol{x}}_L$};
\path (Top) -- (UR)
  node[pos=0.52,pathLabel,text=orange!90!black]
  {$\overline{\boldsymbol{x}}_R$};
\draw[axis,<->,line width=0.6pt]
  (-0.63,-3.55) -- (0.63,-3.55);
\node[text=black!80] at (1.05,-3.4) {$\sigma$};
\end{tikzpicture}
\caption{Spacetime diagram for the Mach--Zehnder configuration in the mid-point trajectory frame, which is a limiting case of the split and hold geometry for $T=0$ and $T'\to T$. 
The total interferometric time is $T_s$ and the maximal spatial separation between the two paths is $\boldsymbol{\Delta x}$. 
The unperturbed classical worldlines $\overline{\boldsymbol{x}}_{L,R}(t)$, which correspond to the particular solutions of Eq.~\eqref{eq:unperturbed} and are defined in terms of external currents in Eq.~\eqref{eq:Xbar}, are shown in solid orange and dashed green.}\label{app:fig:paths}
\end{figure}

Mach--Zehnder sequences~\cite{Kasevich:1991zz} as viewed in the mid-point trajectory frame~\cite{Overstreet:2020ftt} are depicted in Fig.~\ref{app:fig:paths}. These can be derived from Eq.~\eqref{eq:current} by setting $T=0$ and mapping $T' \rightarrow T$ under the assumption that laboratory forces used to manipulate the atomic wavepackets act as impulses. The worldlines can then be modeled as
\begin{equation}\label{app:eq:worldlines_single}
\begin{aligned}
\overline{\boldsymbol{x}}_{L}(t) &=
\begin{cases} 
-\mathbf{v}(t+T/2)/2  ~~\text{for $- T/2 \leq t \leq 0$}  \\
- \mathbf{v}(T/2-t)/2  ~~\text{for $0 \leq t \leq T/2$}  \\
0 ~~\text{otherwise}\
\end{cases}\, , \\
\overline{\boldsymbol{x}}_{R}(t) &=
\begin{cases} 
\mathbf{v}(t+T/2)/2  ~~\text{for $- T/2 \leq t \leq 0$}  \\
\mathbf{v}(T/2-t)/2  ~~\text{for $0 \leq t \leq T/2$}  \\
0 ~~\text{otherwise}\
\end{cases} \, , 
\end{aligned}
\end{equation}
where $|\boldsymbol{\Delta x}| = |\mathbf{v}|T/2$ is the maximal distance between the two paths. 

With these paths, the worldline-dependent part of Eq.~\eqref{app:eq:FT_sys} can be evaluated using the same steps as above. 
Introducing the variable
\begin{equation}
\alpha_\pm \equiv q_0 \pm \frac{\mathbf{q}\cdot\mathbf{v}}{2} \, ,
\end{equation}
the left path contribution may be expressed as
\begin{equation}\label{app:eq:stepW}
\begin{aligned}
\widetilde{\mathcal{W}}(q;\overline{\boldsymbol{x}}_L) &= \int_{-\infty}^{\infty} \mathrm{d}t\, e^{\mi q_0 t}\, e^{-\mi \mathbf{q}\cdot \overline{\boldsymbol{x}}_{L}(t)} \\  
& = \deltabar(q_0) - \int_{-T/2}^{T/2} \mathrm{d}t e^{\mi q_0 t} +  e^{\mi\mathbf{q}\cdot\mathbf{v}T/4}
\left[\int_{-T/2}^{0} \mathrm{d}t e^{\mi\alpha_+ t}
+\int_{0}^{T/2} \mathrm{d}t e^{\mi\alpha_- t}\right] \\
& = \deltabar(q_0) - T \mathrm{sinc}\left ( \tfrac{q_0 T}{2}\right ) +  \tfrac{T}{2} e^{\mi\mathbf{q}\cdot\boldsymbol{\Delta x}/4}
\left[e^{-\mi q_0T/4}\,\mathrm{sinc}\!\left(\tfrac{\alpha_+ T}{4}\right)
+e^{\mi q_0T/4}\,\mathrm{sinc}\!\left(\tfrac{\alpha_- T}{4}\right)\right] \, .
\end{aligned}
\end{equation}
Hence, in the Markovian regime, Eq.~\eqref{app:eq:stepW} implies the appearance of energy conserving delta functions $\delta \left (q_0 \pm \tfrac{\mathbf{q}\cdot\mathbf{v}}{2} \right)$, which instead do not appear in the split-and-hold configuration.

\section{Higgs-portal dark matter constraints}\label{app:Higgs}

In this appendix, we briefly summarize the constraints shown in Fig.~\ref{fig:contact_reach}, which we collated from Refs.~\cite{VanTilburg:2024xib, Day:2023mkb}. 
\begin{itemize}
\item \textit{SN1987A and stellar cooling}.---The effective DM-nucleon coupling provides an additional cooling mechanism in stars and supernovae via $NN\to NN\chi\chi$. 
Because this interaction is irrelevant, the strongest astrophysical cooling constraints come from SN1987A~\cite{Olive:2007aj}, for which 
\begin{equation}
\frac{g}{\Lambda^2} \lesssim 4.5~\mathrm{GeV}^{-2} \times \left (\frac{\mathrm{eV}}{m_\chi} \right ) \, .
\end{equation}
\item \textit{Invisible Higgs decay}.---In the broken phase, $\mathcal{L} \supset -g_H \chi^2 H^\dagger H \to -g_H v h \chi^2$, where $v$ is the electroweak vev. 
Hence, this model predicts invisible Higgs decay at collider experiments, which bound the branching ratio $\Gamma_{h\rightarrow \chi\chi}/(\Gamma_{h\rightarrow \mathrm{SM}} + \Gamma_{h\rightarrow \chi\chi}) < 0.11$ at a 95\% C.L.~\cite{ATLAS:2023tkt}, where $\Gamma_\mathrm{SM} = 4.07$~MeV~\cite{Denner:2011mq} and $ \Gamma_{h\rightarrow \chi\chi} = g_H^2 v^2/8\pi m_h$ for $m_\chi \ll m_h$~\cite{Cline:2013gha}. 
This implies $|g_H| < 5.1\times 10^{-3}$, and 
\begin{equation}
\frac{g}{\Lambda^2} \lesssim 70 ~\mathrm{GeV}^{-2} \times \left (\frac{\mathrm{eV}}{m_\chi} \right ) \, .
\end{equation}
\item \textit{Bullet-cluster}.---In the broken phase, $\mathcal{L} \supset -g_H \chi^2 H^\dagger H \to -g_H v h \chi^2-g_H h^2 \chi^2/2$.
For $m_\chi \ll m_h$, by integrating out the Higgs one obtains the effective quartic coupling $-\lambda_\mathrm{eff} \chi^4/4!$. 
Assuming that the bare coupling cancels the tree-level Higgs exchange contribution, $\lambda_\mathrm{eff}\sim g_H^2/4\pi^2$.
Astrophysical observations of the Bullet cluster require that the cross section per unit mass satisfy $(1+n_{\chi,\mathrm{BC}})\sigma_{\chi\chi\to\chi\chi}/m_\chi \lesssim 1~\mathrm{cm}^2/\mathrm{g}$~\cite{Tulin:2017ara}, where $\sigma_{\chi\chi\to\chi\chi}=\lambda_\mathrm{eff}^2/128 \pi m_\chi^2$ is the single-particle self-interaction cross section~\cite{Egana-Ugrinovic:2021gnu}. Here, ${(1+n_{\chi,\mathrm{BC}})}$ accounts for Bose enhancements when the DM occupation number in the Bullet Cluster $n_{\chi,\mathrm{BC}} \simeq 23\times(1~\mathrm{eV}/m_\chi)^4$ exceeds unity~\cite{VanTilburg:2024xib}. Hence, 
\begin{equation}
\frac{g}{\Lambda^2} \lesssim \begin{cases} 
0.25~\mathrm{GeV}^{-2} \left (\frac{m_\chi}{\mathrm{eV}}\right )^{3/4} & \text{for $m_\chi\lesssim 2$~eV,}\\
0.55~\mathrm{GeV}^{-2} \left (\frac{\mathrm{eV}}{m_\chi}\right )^{1/4} & \text{for $m_\chi\gtrsim 2$~eV.}
\end{cases}
\end{equation}
\item \textit{Large-scale structure formation}.---In the early universe, the quartic self-interaction renders the oscillating $\chi$ condensate anharmonic. 
This generates non-vanishing pressure averaged over an oscillation cycle and hence a scale-independent sound speed, which can affect structure formation on length scales smaller than the associated Jeans scale $L_J \simeq (\pi|\lambda_\mathrm{eff}|M_\mathrm{Pl}^2/16m_\chi^4)^{1/2}$~\cite{Turner:1983he,Arvanitaki:2014faa}.
Requiring that self-interactions not substantially modify the growth of perturbations with present-day comoving wavelengths $\gtrsim100$~kpc between matter–radiation equality and today yields the approximate bound $|\lambda_\mathrm{eff}| \lesssim 1.8\times10^{-6}\,(m_\chi/\mathrm{eV})^4$~\cite{Arvanitaki:2014faa}, or equivalently 
\begin{equation}
\frac{g}{\Lambda^2} \lesssim 10^2~\mathrm{GeV}^{-2} \left (\frac{m_\chi}{\mathrm{eV}}\right ) \, .
\end{equation}
\end{itemize}

\section{Modifications to the DM distribution due to the Earth and the atmosphere}\label{app:DM_mod}

In this appendix, we compute the impact of the Earth and the atmosphere on the DM phase space distribution. Regions of parameter space affected lie above the black dot-dashed lines in Figs.~\ref{fig:contact_reach}--\ref{fig:sigma_reach_tot} .
We will specifically distinguish between: (i) contact and repulsive interactions, already considered in Refs.~\cite{Day:2023mkb,VanTilburg:2024xib}, and (ii) long-range attractive interactions. 
Here, we also restrict our attention to the regime where the DM de Broglie wavelength is much shorter than the size and thickness of the Earth and atmosphere.

\subsection{Contact repulsive interactions}

For the interaction $\mathcal L\supset-(g/\Lambda^2)m_\chi\chi^2\overline{N}N$, one can compute the impact of the Earth and atmosphere on the propagation of DM by solving the equations of motion of $\chi$ in matter. 
Replacing $\overline{N}N$ by the medium's nucleon number density $\rho_N=\langle\overline{N}N\rangle$ gives
\begin{equation}
\left[\partial_t^2-\nabla^2+m_\chi^2 +2\frac{g}{\Lambda^2}m_\chi \rho_N \right]\chi=0 \, .
\label{app:eq:contact_eom}
\end{equation}
As a result, DM acquires an effective mass in a medium of nucleons given by 
\begin{equation}
m_{\chi,\mathrm{eff}}^2=m_\chi^2+ \frac{2g}{\Lambda^2} m_\chi \rho_N \, .
\end{equation}

Imposing DM energy conservation inside and outside the medium then implies that, for repulsive interactions (i.e., $g>0$) and a homogeneous medium, DM modes with typical three-momentum $|\mathbf{p}| = m_\chi v_0$ incident on the medium become evanescent when~\cite{VanTilburg:2024xib}
\begin{equation}
\frac{g}{\Lambda^2} \gtrsim \frac{m_\chi v_0^2}{2 \rho_N} \, . 
\label{app:eq:evanescence}
\end{equation}
In the limit that the DM de Broglie wavelength is much shorter than the thickness and size of the shield, strong attenuation simply requires satisfying Eq.~\eqref{app:eq:evanescence}~\cite{Day:2023mkb}. 
Hence, using the typical number density of nucleons in the atmosphere, we find that atmospheric screening for an experiment on Earth becomes significant for 
\begin{equation}
\frac{g}{\Lambda^2} \gtrsim5\times10^4~\mathrm{GeV}^{-2} \left(\frac{m_\chi}{1~\mathrm{eV}}\right) \, ,
\end{equation}
throughout the mass range shown in Fig.~\ref{fig:contact_reach}.

Similarly, the Earth can screen DM. 
Approximating the Earth as a homogeneous solid sphere with mean nucleon number density $\rho_N \simeq3.3\times10^{24}~\mathrm{cm}^{-3}$, the Earth can significantly affect the reach of terrestrial experiments for
\begin{equation}
\frac{g}{\Lambda^2} \gtrsim 12~\mathrm{GeV}^{-2} \left(\frac{m_\chi}{1~\mathrm{eV}}\right).
\end{equation}

Space-based experiments can avoid atmospheric screening, but can also be sensitive to the Earth's DM ``shadow" during their orbit~\cite{Day:2023mkb}. 
Furthermore, surrounding spacecraft material may still attenuate the signal. 
This additional shielding depends on the material density, thickness, and geometry, and is left to future work.

\subsection{Long-range attractive interactions}

We now consider the case of DM-nucleon interactions via a hadrophilic light scalar mediator. 
In this case, interactions are attractive and the atmosphere and Earth can give rise to DM refraction, similar to the effect of gravitational focusing due to the Sun~\cite{Griest:1987vc,Kim:2021yyo}.
Neglecting the DM back-reaction and assuming a static and homogeneous distribution of nucleons, the equations of motion for the matter-sourced static field $\bar\phi$ and $\chi$ for the interaction $\mathcal L\supset-y_\chi m_\chi\phi\chi^2-y_n\phi\overline{N}N$ read as
\begin{align}
\label{app:eq:phi_eom}
(-\nabla^2 + m_\phi^2)\bar\phi&=- y_n\rho_N \, , \\
(\partial_t^2-\nabla^2+m_\chi^2+2y_\chi m_\chi\bar\phi)\chi&=0 \, .
\label{app:eq:chi_eom}
\end{align}

Imposing the boundary condition $\bar{\phi} = 0$ far from the medium, the solution to Eq.~\eqref{app:eq:phi_eom} is given by 
\begin{equation}
\begin{aligned}
\bar{\phi}(\mathbf{x}) &= -y_n \int d^3 x' G^{\phi}_\mathrm{r}(\mathbf{x}-\mathbf{x'})\rho_N (\mathbf{x'})  \, ,
\end{aligned}
\end{equation}
where $G^{\phi}_\mathrm{r}(\mathbf{x}-\mathbf{x'})$ is the position-space representation of the Green's function of $\phi$.
Using the decomposition $\chi=e^{-iEt}\chi'(\mathbf{x})+\mathrm{c.c.}$, Eq.~\eqref{app:eq:chi_eom} can be expressed as
\begin{equation}
\left[-\frac{\nabla^2}{2m_\chi}+V(\mathbf{x})\right]\chi'(\mathbf{x}) =\frac{E^2-m_\chi^2}{2m_\chi}\chi'(\mathbf{x}), \qquad V(\mathbf{x})=y_\chi\bar\phi(\mathbf{x}).
\end{equation}
where $V(\mathbf{x})$ is the in-medium potential seen by the DM.
Because $m_\phi^{-1}$ is much shorter than the Earth's radius and atmosphere's thickness, we compute $V(\mathbf{x})$ at the Earth's surface by approximating the environment as a half-space, such that
\begin{equation}
V(R_\oplus) \simeq - \frac{y_\chi y_n \rho_N}{2m_\phi^2} \, .
\end{equation}

Imposing energy conservation, we find that an $\mathcal{O}(1)$ fractional change in the DM speed due to the medium requires that the in-medium potential be similar to the typical kinetic energy of DM far from the medium, i.e., 
\begin{equation}
|V(R_\oplus) | \sim \frac{1}{2}m_\chi v_0^2 \implies y_\chi y_n \sim \frac{m_\phi^2 m_\chi v_0^2}{\rho_N} \, .
\end{equation}
Hence, in the limit $m_\chi \ll 1$~GeV and using the definition of the reference cross section (see Eq.~\eqref{eq:ref_cs}), we find that appreciable corrections to the DM phase space properties on the surface of the Earth occur for
\begin{align}
\overline{\sigma} & \gtrsim 5 \times 10^{-40}~\mathrm{cm}^2 \, \left ( \frac{m_\chi}{1~\mathrm{keV}}\right )^4 \left ( \frac{R_{\phi,\chi}}{10^{-5}}\right )^4 \, , \\
\overline{\sigma} & \gtrsim 10^{-32}~\mathrm{cm}^2 \, \left ( \frac{m_\chi}{1~\mathrm{keV}}\right )^4 \left ( \frac{R_{\phi,\chi}}{10^{-5}}\right )^4 \, ,
\end{align}
due to the Earth and the atmosphere, respectively.

The effect of laboratory shield, which could be relevant for certain space-based proposals and could be smaller in size than $m_\phi^{-1}$, is left to future work.

\bibliography{bib}{}
\bibliographystyle{apsrev4-2}

\end{document}